\documentclass[11pt,letterpaper]{article}
\pdfoutput=1

\usepackage[utf8x]{inputenc}%
\usepackage{tcolorbox,fancybox}%
\usepackage{physics}
\usepackage{cite}
\usepackage{moresize}
\usepackage{ntheorem}
\usepackage{mathtools}

\usepackage[english]{babel}
\usepackage{amsmath,amssymb,amsfonts}
\usepackage{graphicx}
\usepackage{booktabs}
\usepackage{color,xcolor}
\usepackage[numbers,sort&compress]{natbib}

\usepackage{tikz}
\usepackage[compat=1.1.0]{tikz-feynman}
\usetikzlibrary{positioning}
\usetikzlibrary{decorations.text}
\usetikzlibrary{decorations.pathmorphing}

\usetikzlibrary{calc}
\usetikzlibrary{shapes.misc}
\tikzset{
        >=latex,
    photon/.style={decorate, decoration={snake}, draw=black, thick},
    fermionnoarrow/.style={draw=black, postaction={decorate}, thick},
    scalar/.style={draw=black, postaction={decorate}, decoration={markings,mark=at position .55 with {\arrow{>}}}, thick, dashed},
    scalarnoarrow/.style={draw=black, postaction={decorate},  thick, dashed},
    fermion/.style={draw=black, postaction={decorate},decoration={markings,mark=at position .55 with {\arrow{>}}}, thick},
    gluon/.style={decorate, draw=black, decoration={coil,amplitude=4pt, segment length=5pt}, thick},
    vertex/.style={draw,shape=circle,fill=black,minimum size=3pt,inner sep=0pt},
    fillvertex/.style={draw,shape=circle,fill=black,minimum size=5pt,inner sep=0pt},
    openvertex/.style={draw,shape=circle,minimum size=5pt,inner sep=0pt},
    blob/.style={draw=red,shape=circle,fill=red,minimum size=6pt,inner sep=0pt},
    redvertex/.style={draw=red,shape=circle,fill=red,minimum size=3pt,inner sep=0pt},
    cross/.style={cross out, draw=black,thick, minimum size=5pt, inner sep=0pt, outer sep=0pt}
}

\usepackage{physics}
\usepackage{orcidlink}
\usepackage{slashed}
\usepackage{mathrsfs}
\usepackage{enumerate}
\usepackage{mdframed}
\usepackage{url}
\usepackage[makeroom]{cancel}
\usepackage{geometry}

\usepackage{pifont}
\newcommand{\cmark}{\ding{51}}%
\newcommand{\xmark}{\ding{55}}%

\newtheorem*{thm-non}{Theorem}

\newtheorem*{define}[thm-non]{Definition}

\usepackage{hyperref} 
\hypersetup{
    colorlinks=true,       
    linkcolor=red,          
    citecolor=blue,        
    filecolor=magenta,      
    urlcolor=purple           
}
\usepackage[all]{hypcap} 

\def\beqn{\begin{eqnarray}}
\def\eeqn{\end{eqnarray}}
\def\beqs{\begin{subequations}}
\def\eeqs{\end{subequations}}
\def\beq{\begin{equation}}
\def\eeq{\end{equation}}
\def\ba{\begin{array}}
\def\ea{\end{array}}

\def\non{\nonumber\\}

\def\hf{\frac{1}{2}}
\def\[{\left[}
\def\]{\right]}
\def\({\left(}
\def\){\right)}
\newcommand\para{\paragraph{}}

\newcommand{\Dslash}{D\hspace{-0.1in}\slash}

\def\gSU{\rm SU}

\def\gSO{\rm SO}

\newcommand{\rep}[1]{\mathbf{#1}}
\newcommand{\repb}[1]{\mathbf{\overline{#1}}}

\def\Bc{\mathcal{B}}

\def\Dc{\mathcal{D}}
\def\Ec{\mathcal{E}}
\def\Fc{\mathcal{F}}
\def\Gc{\mathcal{G}}

\def\Lc{\mathcal{L}}
\def\Mc{\mathcal{M}}
\def\Nc{\mathcal{N}}
\def\Oc{\mathcal{O}}

\def\Rc{\mathcal{R}}
\def\Sc{\mathcal{S}}
\def\Tc{\mathcal{T}}

\def\Xc{\mathcal{X}}
\def\Yc{\mathcal{Y}}

\def\AG{\mathfrak{A}}  \def\aG{\mathfrak{a}}
\def\BG{\mathfrak{B}}  \def\bG{\mathfrak{b}}
\def\CG{\mathfrak{C}}  \def\cG{\mathfrak{c}}
\def\DG{\mathfrak{D}}  \def\dG{\mathfrak{d}}
\def\EG{\mathfrak{E}}  \def\eG{\mathfrak{e}}

  \def\nG{\mathfrak{n}}

\def\UG{\mathfrak{U}}  \def\uG{\mathfrak{u}}

\title{
{\bf The gauge coupling evolutions of an ${\rm SU}(8)$ theory \\ with the maximally symmetry breaking pattern } \\
\author{\large Ning Chen$^{\,\heartsuit}$\,\orcidlink{0000-0002-0032-9012}, Zhanpeng Hou$^{\,\spadesuit}$\,\orcidlink{0000-0002-6035-368X}, Ying-nan Mao$^{\,\diamondsuit}$\,\orcidlink{0000-0001-8063-8968}, Zhaolong Teng$^{\,\clubsuit}$\,\orcidlink{0000-0002-7141-2331}}
\date{\small \it
$^\heartsuit\, ^\spadesuit \, ^\clubsuit$School of Physics, Nankai University, Tianjin, 300071, China \\
$^\diamondsuit$ Department of Physics, School of Physics and Mechanics, Wuhan University of Technology, \\ Wuhan, 430070, Hubei, China \\
}
}

\begin{document}

\maketitle
\setlength{\parskip}{0.2ex}

\begin{abstract}
\bigskip
We study the renormalizable group equations (RGEs) of the extended strong and weak gauge couplings in an ${\rm SU}(8)$ theory, where three-generational SM fermions are non-trivially embedded.
This framework was previously found to generate the observed SM quark/lepton mass hierarchies and the Cabibbo-Kobayashi-Maskawa mixing pattern through its maximally breaking pattern.
The field theoretical two-loop RGEs can not achieve the gauge coupling unification with the minimal setup, unless additional adjoint Higgs fields as well as the gravity-induced $d=5$ term to the ${\rm SU}(8)$ field strength term are included. 
\end{abstract}

\vspace{9.5cm}
{\emph{Emails:}\\  
$^{\,\heartsuit}$\url{chenning_symmetry@nankai.edu.cn},\\
$^{\,\spadesuit}$\url{houzhanpeng@mail.nankai.edu.cn},\\
$^{\,\diamondsuit}$\url{ynmao@whut.edu.cn},\\
$^{\,\clubsuit}$ \url{tengcl@mail.nankai.edu.cn}
 }

\thispagestyle{empty}  
\newpage  
 
\setcounter{page}{1}  

\vspace{1.0cm}
\eject
\tableofcontents

\section{Introduction}
\label{section:intro}
%
%

\para
Grand Unified Theories (GUTs), with their original formulations based on the gauge groups of $\gSU(5)$~\cite{Georgi:1974sy} and $\gSO(10)$~\cite{Fritzsch:1974nn}, were proposed to not only unify all three fundamental symmetries into one simple Lie group of $\Gc_U$, but to unify all Standard Model (SM) fermions into some anomaly-free irreps of the $\Gc_U$ as well.
The Georgi-Glashow $\gSU(5)$ GUT~\cite{Georgi:1974sy} contains chiral fermions of $3\times \[ \repb{5_F} \oplus \rep{10_F} \]$, and the Fritzsch-Minkowski $\gSO(10)$ GUT~\cite{Fritzsch:1974nn} contains chiral fermions of $3\times \rep{16_F}$.
Meanwhile, these $\gSU(5)$ and/or $\gSO(10)$ GUTs, with or without their supersymmetric (SUSY) extensions, are far from satisfactory, given that three-generational SM fermion masses and the Cabibbo-Kobayashi-Maskawa (CKM) mixing pattern~\cite{Cabibbo:1963yz,Kobayashi:1973fv} were not ultimately explained.
A common feature is that the fermion contents of these minimal GUTs follow the trivially repetitive flavor structure as we have observed in the SM.

\para
In a pioneering work by Georgi~\cite{Georgi:1979md}, an extension to the gauge group of ${\rm SU}(N >5)$ was proposed to embed three-generational SM fermions non-trivially.
In his original proposal, any anti-symmetric rank-$k$ fermion irrep can appear at most once with the anomaly-free condition.
Physically, this leads to a non-repetitive structure for both left-handed and right-handed components of chiral fermions based on an ${\rm SU}(11)$ gauge group.
Given the current LHC measurements of one single SM Higgs boson so far, which have already confirmed the hierarchical Yukawa couplings of the third-generational SM quarks/leptons and displayed the evidence for the muon~\cite{CMS:2022dwd,ATLAS:2022vkf}, Georgi's original proposal is quite insightful and can be further relaxed by requiring distinctive symmetry properties of three generations in the UV theory.
This relies on the concept of chiral \underline{ir}reducible \underline{a}nomaly-\underline{f}ree \underline{f}ermion \underline{s}ets (IRAFFSs) defined in Ref.~\cite{Chen:2023qxi}, which reads
\begin{define}\label{def:IRAFFS}

A chiral IRAFFS is a set of left-handed anti-symmetric fermions of $\sum_\Rc m_\Rc \, \Fc_L(\Rc)$, with $m_\Rc$ being the multiplicities of a particular fermion representation of $\Rc$.
Obviously, the anomaly-free condition reads $\sum_\Rc m_\Rc \, {\rm Anom}(  \Fc_L(\Rc) ) =0$.
We also require the following conditions to be satisfied for a chiral IRAFFS:
\begin{itemize}

\item the greatest common divisor (GCD) of the $\{ m_\Rc \}$ should satisfy that ${\rm GCD} \{  m_\Rc \} =1$;

\item the fermions in a chiral IRAFFS can no longer be removed, which would otherwise bring non-vanishing gauge anomalies;

\item there should not be any singlet, self-conjugate, adjoint fermions, or vectorial fermion pairs in a chiral IRAFFS.

\end{itemize}

\end{define}
Obviously, one SM generational fermions formulate one chiral IRAFFS, and we found that a minimal ${\rm SU}(8)$ theory~\cite{Barr:2008pn,Chen:2023qxi,Chen:2024cht} is sufficient to have all three-generational SM quarks/leptons transform differently with two distinctive chiral IRAFFSs in the UV setup~\footnote{To count the SM generations by decomposing the ${\rm SU}(8)$ fermion irreps into the ${\rm SU}(5)$ irreps~\cite{Georgi:1979md}, one finds three identical SM $\repb{5_F}$'s and three distinctive $\rep{10_F}$'s. Hence, it is sufficient to obtain three distinctive SM generations in the UV setup of the ${\rm SU}(8)$ theory.}.

\para
The ${\rm SU}(8)$ symmetry breaking pattern includes three intermediate scales above the EW scale by counting its rank.
By analyzing the emergent non-anomalous global $\widetilde{ {\rm U}}(1)_T$ symmetry in the UV, we derived the non-anomalous global $B-L$ symmetry in the SM, and one unique SM Higgs doublet from the $\rep{70_H}$ of the minimal setup was conjectured based on the requirement of the vanishing global $B-L$ charge~\cite{Chen:2023qxi}.
This SM Higgs doublet was found to give natural top quark mass with the renormalizable Yukawa coupling at the tree level.
Due to the distinctive symmetry properties of three-generational SM fermions in the UV, we found in Ref.~\cite{Chen:2024cht} that all SM quark/lepton masses as well as the CKM mixing pattern can be explained with (i) both $d=5$ direct Yukawa couplings and indirect Yukawa couplings to the $d=5$ irreducible Higgs mixing operators induced by the inevitable gravitational effects, (ii) three reasonable intermediate symmetry-breaking scales, and (iii) precise identifications of the SM flavors in their UV irreps.

\para
A central issue of a unified theory is to achieve the gauge coupling unification in terms of their renormalization group equations (RGEs)~\footnote{Some early studies of the RGEs based on the ${\rm SU}(5)$ and ${\rm SO}(10)$ groups include Refs.~\cite{Georgi:1974yf,Hall:1980kf,Dimopoulos:1981zb}.}.
A strong-weak-weak (SWW) symmetry breaking pattern of 
\beqn\label{eq:Pattern} 
&& {\rm SU}(8) \xrightarrow{ v_U } \Gc_{441} \xrightarrow{ v_{441} } \Gc_{341} \xrightarrow{v_{341} } \Gc_{331} \xrightarrow{ v_{331} } \Gc_{\rm SM} \xrightarrow{ v_{\rm EW} } {\rm SU}(3)_{c}  \otimes  {\rm U}(1)_{\rm EM} \,, \non
&&\Gc_{441} \equiv {\rm SU}(4)_{s} \otimes {\rm SU}(4)_W \otimes  {\rm U}(1)_{X_0 } \,, ~ \Gc_{341} \equiv {\rm SU}(3)_{c} \otimes {\rm SU}(4)_W \otimes  {\rm U}(1)_{X_1 } \,,\non
&&\Gc_{331} \equiv {\rm SU}(3)_{c} \otimes {\rm SU}(3)_W \otimes  {\rm U}(1)_{X_2 } \,,~ \Gc_{\rm SM} \equiv  {\rm SU}(3)_{c} \otimes {\rm SU}(2)_W \otimes  {\rm U}(1)_{Y } \,,
\eeqn
was previously considered in Ref.~\cite{Chen:2024cht}, and three intermediate scales of
\beqn\label{eq:benchmark}
&& v_{441}\simeq 1.4 \times 10^{17 }\,{\rm GeV}  \,,~ v_{341} \simeq  4.8\times 10^{15} \,{\rm GeV} \,, ~ v_{331} \simeq 4.8\times 10^{13} \,{\rm GeV} \,,
\eeqn
were predicted with the reduced Planck scale of $M_{\rm pl}= ( 8 \pi G_N)^{-1/2}= 2.4 \times 10^{18}\, {\rm GeV}$.
The maximally broken subgroup of $\Gc_{441}$ further admits the $\Gc_{\rm SM}$ as its subgroup~\footnote{Other different symmetry breaking patterns of the ${\rm SU}(8)$ group were previously described in Refs.~\cite{Chakrabarti:1980bn,Ma:1981pr}.}.
These scales were obtained by the reasonable predictions of three-generational SM quark/lepton masses and the CKM mixing pattern, together with the precise identifications of the SM flavors in their UV irreps.
In this paper, we perform the RG analysis of the gauge couplings in the extended strong and weak sectors based on the SWW symmetry breaking pattern in Eq.~\eqref{eq:Pattern} and the benchmark point in Eq.~\eqref{eq:benchmark}.

\para
The rest of the paper is organized as follows.
We review the ${\rm SU}(8)$ theory in section~\ref{section:pattern}, with the focus on the SWW symmetry breaking pattern.
Most of the contents in this section have been obtained in a previous Ref.~\cite{Chen:2024cht}.
In section~\ref{section:SU8_gauge}, we define the gauge sector of the ${\rm SU}(8)$ theory along the SWW symmetry breaking pattern defined in section~\ref{section:pattern}.
The gauge couplings at each symmetry breaking stage will be defined.
Due to the intrinsic symmetry properties, we explicitly display the flavor non-universalities of the $\Gc_{441}$ and $\Gc_{341}$ theories as well as the flavor universality of the $\Gc_{331}$ theory through their neutral currents.
In section~\ref{section:SU8_RGEs}, we study the gauge coupling evolutions in terms of the two-loop RGEs.
The ${\rm SU}(8)$ theory with the minimal fermion/Higgs contents as well as the intermediate symmetry breaking scales in Refs.~\cite{Chen:2023qxi,Chen:2024cht} cannot achieve the gauge coupling unification in terms of the conventional field theoretical analyses.
Instead, inclusion of additional ${\rm SU}(8)$ adjoint Higgs fields and the $d=5$ Hill-Shafi-Wetterich (HSW) operator~\cite{Hill:1983xh,Shafi:1983gz} are suggested to achieve the gauge coupling unification within the field theoretical framework.
We summarize and make future perspective in section~\ref{section:conclusion}, such as the possible issues in the supersymmetric extensions.
Appendix~\ref{section:Br} is given to define the gauge group indices and the decomposition rules according to the SWW symmetry breaking pattern in Eq.~\eqref{eq:Pattern}.
Appendix~\ref{section:proof} is added to prove a unique and elegant feature of the ${\rm U}(1)$ charge quantization in the ${\rm SU}(N)$ maximally symmetry breaking pattern.

\section{The ${\rm SU}(8)$ theory and its SWW symmetry breaking pattern}
\label{section:pattern}

\subsection{The setup of the ${\rm SU}(8)$ theory}

\para
The ${\rm SU}(8)$ theory contains two following chiral IRAFFSs at the GUT scale~\cite{Chen:2023qxi,Chen:2024cht}
\beqn\label{eq:SU8_3gen_fermions}
\{ f_L \}_{ {\rm SU}(8)}^{n_g=3}&=& \Big[ \repb{8_F}^\omega \oplus \rep{28_F} \Big] \bigoplus \Big[ \repb{8_F}^{ \dot \omega } \oplus \rep{56_F} \Big] \,,~ {\rm dim}_{ \mathbf{F}}= 156\,, \non
&& \Omega \equiv ( \omega \,, \dot \omega ) \,, ~ \omega = ( 3\,, {\rm IV}\,, {\rm V}\,, {\rm VI}) \,, ~  \dot \omega = (\dot 1\,, \dot 2\,, \dot {\rm VII}\,, \dot {\rm VIII}\,, \dot {\rm IX} ) \,,
\eeqn
with undotted/dotted indices for the $\repb{8_F}$'s in the rank-$2$ IRAFFS and the rank-$3$ IRAFFS, respectively.
The Roman numbers and the Arabic numbers are used for the heavy partner fermions and the SM fermions, respectively.
Three $\rep{10_F}$'s from the $\rep{28_F}$ and $\rep{56_F}$ are transforming differently, while three identical $\repb{5_F}$'s come from nine $\repb{8_F}$'s, as one can find in Tabs.~\ref{tab:SU8_8barferm}, \ref{tab:SU8_28ferm}, and \ref{tab:SU8_56ferm}.
These facts are sufficient to guarantee that three-generational SM quarks and leptons must transform differently in the UV theory.

\para
The non-anomalous global symmetries observed by Dimopoulos-Raby-Susskind (DRS)~\cite{Dimopoulos:1980hn} from chiral fermions in Eq.~\eqref{eq:SU8_3gen_fermions} are
\beqn\label{eq:DRS_SU8}
\widetilde{ \Gc}_{\rm DRS} \[{\rm SU}(8)\,, n_g=3 \]&=& \Big[ \widetilde{ {\rm SU}}(4)_\omega  \otimes \widetilde{ {\rm U}}(1)_{T_2} \Big]  \bigotimes \Big[ \widetilde{ {\rm SU}}(5)_{\dot \omega } \otimes \widetilde{ {\rm U}}(1)_{T_3}   \Big]  \,,
\eeqn
and we also denote the anomalous global Peccei-Quinn (PQ) symmetries~\cite{Peccei:1977hh} as
\beqn\label{eq:PQ_SU8}
\widetilde{ \Gc}_{\rm PQ} \[{\rm SU}(8)\,, n_g=3 \]&=& \widetilde{ {\rm U}}(1)_{{\rm PQ}_2} \bigotimes \widetilde{ {\rm U}}(1)_{ {\rm PQ}_3}  \,.
\eeqn
Both $\widetilde{ {\rm U}}(1)_{T_2}$ and $\widetilde{ {\rm U}}(1)_{{\rm PQ}_2}$ can be thought as the linear combinations of two global $\widetilde{ {\rm U}}(1)$'s of the $\repb{8_F}^\omega$ and the $\rep{28_F}$ in the rank-2 IRAFFS, and similarly for the $\widetilde{ {\rm U}}(1)_{T_3}$ and $\widetilde{ {\rm U}}(1)_{{\rm PQ}_3}$ symmetries in the rank-3 IRAFFS.
Since the global $B-L$ symmetry should be identical for all three generations, we further require a common non-anomalous $\widetilde{ {\rm U}}(1)_{T}\equiv \widetilde{ {\rm U}}(1)_{T_2} = \widetilde{ {\rm U}}(1)_{T_3}$ between two IRAFFSs.
Likewise, we also assume a common anomalous $\widetilde{ {\rm U}}(1)_{\rm PQ}\equiv \widetilde{ {\rm U}}(1)_{{\rm PQ}_2} = \widetilde{ {\rm U}}(1)_{ {\rm PQ}_3}$ between two IRAFFSs in Eq.~\eqref{eq:PQ_SU8}.
Accordingly, the non-anomalous global $\widetilde{ {\rm U}}(1)_T$ charges and the anomalous global $\widetilde{ {\rm U}}(1)_{\rm PQ}$ charges for fermions are assigned in Tab.~\ref{tab:U1TU1PQ}, where the anomalous global $\widetilde{ {\rm U}}(1)_{\rm PQ}$ charges are assigned such that
\beqn\label{eq:PQcharges_SU8}
&&  p : q_2  \neq -3 : +2  \,, \quad p  : q_3 \neq -3  : +1  \,.
\eeqn

\begin{table}[htp]
\begin{center}
\begin{tabular}{c|cccc}
\hline\hline
 Fermions &  $\repb{8_F}^\Omega$ &  $\rep{28_F}$  &  $\rep{56_F}$  &     \\[1mm]
\hline
$\widetilde{ {\rm U}}(1)_T$ &  $-3t$  &  $+2t$  & $+t$ &      \\[1mm]
$\widetilde{ {\rm U}}(1)_{\rm PQ}$ &  $p$  &  $q_2$  & $q_3$ &      \\[1mm]
\hline
Higgs  &  $\repb{8_H}_{\,, \omega }$  & $\repb{28_H}_{\,, \dot \omega }$   & $\rep{70_H}$ &  $\rep{63_H}$    \\[1mm]
\hline
$\widetilde{ {\rm U}}(1)_T$ &  $+t$  &  $+2t$   &  $-4t$  &  $0$    \\[1mm]
$\widetilde{ {\rm U}}(1)_{\rm PQ}$ &  $-(p+q_2)$  &  $-(p+q_3 )$  & $-2q_2$ &  $0$ \\[1mm]
\hline\hline
\end{tabular}
\end{center}
\caption{The non-anomalous $\widetilde{ {\rm U}}(1)_T$ charges and the anomalous global $\widetilde{ {\rm U}}(1)_{\rm PQ}$ charges for the $\gSU(8)$ fermions and Higgs fields.}
\label{tab:U1TU1PQ}
\end{table}%

\para
The most general gauge-invariant Yukawa couplings at least include the following renormalizable and non-renormalizable terms~\footnote{The term of $\rep{56_F} \rep{56_F} \rep{28_H}  + H.c.$ vanishes due to the anti-symmetric property~\cite{Barr:2008pn}. Instead, only a $d=5$ non-renormalizable term of $\frac{1}{ M_{\rm pl} } \rep{56_F}  \rep{56_F}  \repb{28_{H}}_{\,,\dot \omega }^\dag  \rep{63_{H}} $ is possible to generate masses for vectorlike fermions in the $\rep{56_F}$. Since it transforms as an $\widetilde{ {\rm SU}}(5)_{\dot \omega }$ vector and carries non-vanishing $\widetilde {\rm U}(1)_{\rm PQ}$ charge of $p+3q_3 \neq 0$ from Eq.~\eqref{eq:PQcharges_SU8}, it is only possible due to the gravitational effect.}
\beqn\label{eq:Yukawa_SU8}
-\Lc_Y&=& Y_\Bc  \repb{8_F}^\omega  \rep{28_F}  \repb{8_{H}}_{\,,\omega }  +  Y_\Tc \rep{28_F} \rep{28_F} \rep{70_H} \non
&+&  Y_\Dc \repb{8_F}^{\dot \omega  } \rep{56_F}  \repb{28_{H}}_{\,,\dot \omega }   + \frac{ c_4 }{ M_{\rm pl} } \rep{56_F}  \rep{56_F}  \repb{28_{H}}_{\,,\dot \omega }^\dag  \rep{63_{H}} + H.c.\,.
\eeqn
All renormalizable Yukawa couplings are assumed to be $(Y_\Bc \,, Y_\Tc\,, Y_\Dc)\sim\Oc(1)$, and the Wilson coefficient $c_4$ of the non-renormalizable term was assigned in Ref.~\cite{Chen:2024cht}.
Altogether, we collect the ${\rm SU}(8)$ Higgs fields as follows
\beqn\label{eq:SU8_Higgs}
 \{ H \}_{ {\rm SU}(8)}^{n_g=3} &=& \repb{8_H}_{ \,, \omega}  \oplus \repb{28_H}_{ \,, \dot \omega}  \oplus \rep{70_H}  \oplus \underline{ \rep{63_H} } \,,~ {\rm dim}_{ \mathbf{H}}= 547 \,,
\eeqn
where the adjoint Higgs field of $\rep{63_H}$ is real while all others are complex.
The adjoint Higgs field of $\rep{63_H}$ will maximally break the GUT symmetry as ${\rm SU}(8)\to \Gc_{441}$ through its VEV of
\beqn\label{eq:63H_VEV}
\langle  \rep{63_H}\rangle &=&\frac{1}{ 4 } {\rm diag}(- \mathbb{I}_{4\times 4} \,, +\mathbb{I}_{4\times 4} ) v_U \,.
\eeqn
We prove in App.~\ref{section:proof} that the maximal symmetry breaking pattern of ${\rm SU}(8)\to \Gc_{441}$ is the unique one where an additional normalization of the ${\rm U}(1)$ charge is unnecessary.
The $\rep{63_H}$ can be decomposed under the $\Gc_{441}$ as follows
\beqn\label{eq:63H_decomp}
&& \rep{63_H} = ( \rep{1} \,, \rep{1} \,, 0 )_{ \mathbf{H}} \oplus ( \rep{15} \,, \rep{1} \,, 0 )_{ \mathbf{H}} \oplus ( \rep{1} \,, \rep{15} \,, 0 )_{ \mathbf{H}} \oplus ( \repb{4} \,, \rep{4} \,, +\frac{1 }{2} )_{ \mathbf{H}} \oplus ( \rep{4} \,, \repb{4 } \,, -\frac{1 }{2 } )_{ \mathbf{H}}  \,,
\eeqn
where the VEV is developed by the $( \rep{1} \,, \rep{1} \,, 0 )_{ \mathbf{H}}$ component, and the $( \repb{4} \,, \rep{4} \,, +\frac{1 }{2} )_{ \mathbf{H}} \oplus ( \rep{4} \,, \repb{4 } \,, -\frac{1 }{2 } )_{ \mathbf{H}}$ represent the Nambu-Goldstone bosons.
We assign the non-anomalous global $\widetilde{ {\rm U}}(1)_T$ charges and the anomalous global $\widetilde{ {\rm U}}(1)_{\rm PQ}$ charges for Higgs fields in Tab.~\ref{tab:U1TU1PQ} according to the renormalizable Yukawa couplings in Eq.~\eqref{eq:Yukawa_SU8}. 
Obviously, the only non-renormalizable term in Eq.~\eqref{eq:Yukawa_SU8} is explicitly $\widetilde {\rm SU}(5)_{\dot \omega}$-breaking, and also has a non-vanishing global $\widetilde{ {\rm U}}(1)_{\rm PQ}$ charge according to Tab.~\ref{tab:U1TU1PQ} and Eq.~\eqref{eq:PQcharges_SU8}.
Thus, it can be explicitly broken by the gravitational effect~\cite{Barr:1992qq,Kamionkowski:1992mf,Holman:1992us,Harlow:2018jwu,Harlow:2018tng}.

\subsection{The SWW symmetry breaking pattern of the ${\rm SU}(8)$ theory}

\para
According to Ref.~\cite{Li:1973mq}, only the adjoint Higgs field can achieve the maximal symmetry breaking pattern of ${\rm SU}(N )\to {\rm SU}(N-k) \otimes {\rm SU}(k)\otimes {\rm U}(1)$ (with $k=[ \frac{N}{2}]$) in a non-SUSY gauge theory.
The SWW symmetry breaking pattern in Eq.~\eqref{eq:Pattern} follows from Ref.~\cite{Chen:2024cht} will be further studied in the following discussions.
The GUT scale symmetry breaking will be due to the VEV of the ${\rm SU}(8)$ adjoint Higgs field of $\rep{63_H}$.
In the subgroup of $\Gc_{441}$, both the strong and the weak symmetries are extended beyond the SM gauge symmetries. 
All intermediate symmetry breaking stages will be due to the VEVs of the Higgs fields from the gauge-invariant Yukawa couplings in Eq.~\eqref{eq:Yukawa_SU8}.
After the first stage of symmetry breaking pattern of $\Gc_{441} \to \Gc_{341}$ in Eq.~\eqref{eq:Pattern}, the sequential symmetry breaking stages are uniquely determined.
Along this SWW symmetry breaking pattern, the non-anomalous global $\widetilde{ {\rm U}}(1)_T$ symmetry will evolve to the global $\widetilde{ {\rm U}}(1)_{B-L}$ at the EW scale according to the following sequence~\cite{Chen:2023qxi}
\beqn\label{eq:U1T_def}
&& \Gc_{441}~:~ \Tc^\prime \equiv \Tc - 4t \Xc_0 \,, \quad  \Gc_{341}~:~ \Tc^{ \prime \prime} \equiv \Tc^\prime + 8 t \Xc_1 \,, \non
&& \Gc_{331}~:~   \Tc^{ \prime \prime \prime} \equiv  \Tc^{ \prime \prime} \,, \quad  \Gc_{\rm SM}~:~  \Bc- \Lc \equiv  \Tc^{ \prime \prime \prime} \,.
\eeqn
With these definitions, we have also checked that all Higgs components that can develop the VEVs at each symmetry breaking stage are neutral under the corresponding global $\widetilde{ {\rm U}}(1)_{T}$ symmetries~\cite{Chen:2023qxi}.

\subsection{Decompositions of the ${\rm SU}(8)$ fermions}
\label{section:SU8_fermions}

\begin{table}[htp] {\small
\begin{center}
\begin{tabular}{c|c|c|c|c}
\hline \hline
   $\gSU(8)$   &  $\Gc_{441}$  & $\Gc_{341}$  & $\Gc_{331}$  &  $\Gc_{\rm SM}$  \\
\hline \hline
 $\repb{ 8_F}^\Omega $   & $( \repb{4} \,, \rep{1}\,,  +\frac{1}{4} )_{ \mathbf{F} }^\Omega$  & $(\repb{3} \,, \rep{1} \,, +\frac{1}{3} )_{ \mathbf{F} }^\Omega $  & $(\repb{3} \,, \rep{1} \,, +\frac{1}{3} )_{ \mathbf{F} }^\Omega $  &  $( \repb{3} \,, \rep{ 1}  \,, +\frac{1}{3} )_{ \mathbf{F} }^{\Omega }~:~ { \Dc_R^\Omega }^c$  \\[1mm]
 &  &  $( \rep{1} \,, \rep{1} \,, 0)_{ \mathbf{F} }^{\Omega }$  &  $( \rep{1} \,, \rep{1} \,, 0)_{ \mathbf{F} }^{\Omega }$ &  $( \rep{1} \,, \rep{1} \,, 0)_{ \mathbf{F} }^{\Omega } ~:~ \check \Nc_L^{\Omega }$  \\[1.5mm]
 & $(\rep{1}\,, \repb{4}  \,,  -\frac{1}{4})_{ \mathbf{F} }^\Omega $  &  $(\rep{1}\,, \repb{4}  \,,  -\frac{1}{4})_{ \mathbf{F} }^\Omega$  &  $( \rep{1} \,, \repb{3} \,,  -\frac{1}{3})_{ \mathbf{F} }^{\Omega }$  &  $( \rep{1} \,, \repb{2} \,,  -\frac{1}{2})_{ \mathbf{F} }^{\Omega } ~:~\Lc_L^\Omega =( \Ec_L^\Omega \,, - \Nc_L^\Omega )^T$   \\[1mm]
 &   &   &   &  $( \rep{1} \,, \rep{1} \,,  0)_{ \mathbf{F} }^{\Omega^\prime} ~:~ \check \Nc_L^{\Omega^\prime }$  \\[1mm]
  &   &  &   $( \rep{1} \,, \rep{1} \,, 0)_{ \mathbf{F} }^{\Omega^{\prime\prime} }$ &   $( \rep{1} \,, \rep{1} \,, 0)_{ \mathbf{F} }^{\Omega^{\prime\prime} } ~:~ \check \Nc_L^{\Omega^{\prime \prime} }$   \\[1mm]   
\hline\hline
\end{tabular}
\caption{
The $\gSU(8)$ fermion representation of $\repb{8_F}^\Omega$ under the $\Gc_{441}\,,\Gc_{341}\,, \Gc_{331}\,, \Gc_{\rm SM}$ subgroups for the three-generational ${\rm SU}(8)$ theory, with $\Omega\equiv (\omega \,, \dot \omega )$.
Here, we denote $\underline{ {\Dc_R^\Omega}^c={d_R^\Omega}^c}$ for the SM right-handed down-type quarks, and ${\Dc_R^\Omega}^c={\DG_R^\Omega}^c$ for the right-handed down-type heavy partner quarks.
Similarly, we denote $\underline{ \Lc_L^\Omega = ( \ell_L^\Omega \,, - \nu_L^\Omega)^T}$ for the left-handed SM lepton doublets, and $\Lc_L^\Omega =( \eG_L^\Omega \,, - \nG_L^\Omega )^T$ for the left-handed heavy lepton doublets.
All left-handed neutrinos of $\check \Nc_L$ are sterile neutrinos, which are $\Gc_{\rm SM}$-singlets and do not couple to the EW gauge bosons.
}
\label{tab:SU8_8barferm}
\end{center} 
}
\end{table}%

\begin{table}[htp] {\small
\begin{center}
\begin{tabular}{c|c|c|c|c}
\hline \hline
   $\gSU(8)$   &  $\Gc_{441}$  & $\Gc_{341}$  & $\Gc_{331}$  &  $\Gc_{\rm SM}$  \\
\hline \hline 
 $\rep{28_F}$   & $( \rep{6}\,, \rep{ 1} \,, - \frac{1}{2})_{ \mathbf{F}}$ &  $ ( \rep{3}\,, \rep{ 1} \,, - \frac{1}{3})_{ \mathbf{F}}$   & $( \rep{3}\,, \rep{ 1} \,, - \frac{1}{3})_{ \mathbf{F}}$  & $( \rep{3}\,, \rep{ 1} \,, - \frac{1}{3})_{ \mathbf{F}} ~:~\DG_L$  \\[1mm]
                        &   & $( \repb{3}\,, \rep{ 1} \,, - \frac{2}{3})_{ \mathbf{F}}$  & $( \repb{3}\,, \rep{ 1} \,, - \frac{2}{3})_{ \mathbf{F}}$  & $\underline{( \repb{3}\,, \rep{ 1} \,, - \frac{2}{3})_{ \mathbf{F}}~:~ {t_R }^c }$   \\[1.5mm]
                        & $( \rep{1}\,, \rep{ 6} \,, +\frac{1}{2})_{ \mathbf{F}}$ & $( \rep{1}\,, \rep{ 6} \,, +\frac{1}{2})_{ \mathbf{F}}$   &  $( \rep{1}\,, \rep{ 3} \,, +\frac{1}{3})_{ \mathbf{F}}$ & $( \rep{1}\,, \rep{2} \,, +\frac{1}{2})_{ \mathbf{F}} ~:~( {\eG_R }^c \,, { \nG_R }^c)^T$  \\[1mm]
                       &   &   &   & $( \rep{1}\,, \rep{1} \,, 0 )_{ \mathbf{F}} ~:~ \check \nG_R^c $ \\[1mm]
                       &   &   & $( \rep{1}\,, \repb{ 3} \,, +\frac{2}{3})_{ \mathbf{F}}$  & $( \rep{1}\,, \repb{2} \,, +\frac{1}{2})_{ \mathbf{F}}^\prime ~:~( { \nG_R^{\prime} }^c\,, - {\eG_R^{\prime} }^c  )^T$   \\[1mm]
                       &   &   &   & $\underline{ ( \rep{1}\,, \rep{1} \,, +1 )_{ \mathbf{F}} ~:~ {\tau_R}^c}$ \\[1.5mm]
                        & $( \rep{4}\,, \rep{4} \,,  0)_{ \mathbf{F}}$ &  $( \rep{3}\,, \rep{4} \,,  -\frac{1}{12})_{ \mathbf{F}}$   & $( \rep{3}\,, \rep{3} \,,  0)_{ \mathbf{F}}$  & $\underline{ ( \rep{3}\,, \rep{2} \,,  +\frac{1}{6})_{ \mathbf{F}}~:~ (t_L\,, b_L)^T}$  \\[1mm]
                        &   &   &   & $( \rep{3}\,, \rep{1} \,,  -\frac{1}{3})_{ \mathbf{F}}^{\prime} ~:~\DG_L^\prime$  \\[1mm]
                        &   &   & $( \rep{3}\,, \rep{1} \,,  -\frac{1}{3})_{ \mathbf{F}}^{\prime\prime}$  & $( \rep{3}\,, \rep{1} \,,  -\frac{1}{3})_{ \mathbf{F}}^{\prime\prime} ~:~\DG_L^{\prime \prime}$ \\[1mm]
                        &   & $ ( \rep{1}\,, \rep{4} \,,  +\frac{1}{4} )_{ \mathbf{F}}$  & $( \rep{1}\,, \rep{3} \,,  +\frac{1}{3} )_{ \mathbf{F}}^{\prime\prime}$  & $( \rep{1}\,, \rep{2} \,,  +\frac{1}{2} )_{ \mathbf{F}}^{\prime\prime} ~:~( {\eG_R^{\prime\prime} }^c \,, { \nG_R^{\prime\prime}}^c )^T$  \\[1mm] 
                        &   &   &   & $( \rep{1}\,, \rep{1}\,, 0)_{ \mathbf{F}}^{\prime} ~:~ \check \nG_R^{\prime\,c}$ \\[1mm]  
                        &   &   & $( \rep{1}\,, \rep{1}\,, 0)_{ \mathbf{F}}^{\prime\prime}$ & $( \rep{1}\,, \rep{1}\,, 0)_{ \mathbf{F}}^{\prime\prime} ~:~\check \nG_R^{\prime \prime \,c}$ \\[1mm]  
\hline\hline
\end{tabular}
\caption{
The $\gSU(8)$ fermion representation of $\rep{28_F}$ under the $\Gc_{441}\,,\Gc_{341}\,, \Gc_{331}\,, \Gc_{\rm SM}$ subgroups for the three-generational ${\rm SU}(8)$ theory.
All SM fermions are marked with underlines.}
\label{tab:SU8_28ferm}
\end{center}
}
\end{table}%

\begin{table}[htp] {\small
\begin{center}
\begin{tabular}{c|c|c|c|c}
\hline \hline
   $\gSU(8)$   &  $\Gc_{441}$  & $\Gc_{341}$  & $\Gc_{331}$  &  $\Gc_{\rm SM}$  \\
\hline \hline
     $\rep{56_F}$   & $( \rep{ 1}\,, \repb{4} \,, +\frac{3}{4})_{ \mathbf{F}}$  &  $( \rep{ 1}\,, \repb{4} \,, +\frac{3}{4})_{ \mathbf{F}}$ & $( \rep{ 1}\,, \repb{3} \,, +\frac{2}{3})_{ \mathbf{F}}^\prime$   &  $( \rep{ 1}\,, \repb{2} \,, +\frac{1}{2})_{ \mathbf{F}}^{\prime\prime\prime} ~:~( {\nG_R^{\prime\prime\prime }}^c \,, -{\eG_R^{\prime\prime\prime } }^c )^T$  \\[1mm]
                            &   &   &   & $\underline{( \rep{ 1}\,, \rep{1} \,, +1)_{ \mathbf{F}}^{\prime} ~:~ {\mu_R}^c}$ \\[1mm]
                            &   &   & $( \rep{ 1}\,, \rep{1} \,, +1)_{ \mathbf{F}}^{\prime\prime}$  & $( \rep{ 1}\,, \rep{1} \,, +1)_{ \mathbf{F}}^{\prime \prime} ~:~{\EG_R}^c$   \\[1.5mm]
                       & $( \repb{ 4}\,, \rep{1} \,, -\frac{3}{4})_{ \mathbf{F}}$  &  $( \repb{3}\,, \rep{1} \,, -\frac{2}{3})_{ \mathbf{F}}^{\prime}$ & $( \repb{3}\,, \rep{1} \,, -\frac{2}{3})_{ \mathbf{F}}^\prime$  & $\underline{ ( \repb{3}\,, \rep{1} \,, -\frac{2}{3})_{ \mathbf{F}}^{\prime} ~:~{u_R}^c }$ \\[1mm]
                       &   &  $( \rep{1}\,, \rep{1} \,, -1)_{ \mathbf{F}}$ & $( \rep{1}\,, \rep{1} \,, -1)_{ \mathbf{F}}$  &  $( \rep{1}\,, \rep{1} \,, -1)_{ \mathbf{F}} ~:~\EG_L$  \\[1.5mm]
                       & $( \rep{ 4}\,, \rep{6} \,, +\frac{1}{4})_{ \mathbf{F}}$  &  $( \rep{3}\,, \rep{6} \,, +\frac{1}{6})_{ \mathbf{F}}$ & $( \rep{3}\,, \rep{3} \,, 0 )_{ \mathbf{F}}^\prime$ & $\underline{ ( \rep{3}\,, \rep{2} \,, +\frac{1}{6} )_{ \mathbf{F}}^{\prime} ~:~ ( c_L \,, s_L)^T} $  \\[1mm]
                       &   &   &   & $( \rep{3}\,, \rep{1} \,, -\frac{1}{3})_{ \mathbf{F}}^{\prime\prime \prime } ~:~\DG_L^{\prime \prime \prime}$ \\[1mm]
                       &   &   & $( \rep{3}\,, \repb{3} \,, +\frac{1}{3})_{ \mathbf{F}}$ & $( \rep{3}\,, \repb{2} \,, +\frac{1}{6})_{ \mathbf{F}}^{\prime\prime} ~:~ (\dG_L \,, - \uG_L )^T$   \\[1mm]
                       &   &   &   & $( \rep{3}\,, \rep{1} \,, +\frac{2}{3})_{ \mathbf{F}} ~:~\UG_L$  \\[1mm]
                       &   & $( \rep{1}\,, \rep{6} \,, +\frac{1}{2})_{ \mathbf{F}}^\prime$ & $( \rep{1}\,, \rep{3} \,, +\frac{1}{3})_{ \mathbf{F}}^\prime $ & $( \rep{1}\,, \rep{2} \,, +\frac{1}{2})_{ \mathbf{F}}^{\prime\prime \prime \prime} ~:~ ( {\eG_R^{\prime\prime\prime\prime }}^c \,, {\nG_R^{\prime\prime\prime\prime } }^c )^T$ \\[1mm]
                       &   &   &   & $( \rep{1}\,, \rep{1} \,, 0 )_{ \mathbf{F}}^{\prime\prime \prime} ~:~ {\check \nG_R}^{\prime \prime\prime \,c}$ \\[1mm]
                       &   &   & $( \rep{1}\,, \repb{3} \,, +\frac{2}{3})_{ \mathbf{F}}^{\prime\prime}$  & $( \rep{1}\,, \repb{2} \,, +\frac{1}{2})_{ \mathbf{F}}^{\prime\prime \prime \prime \prime} ~:~( {\nG_R^{\prime\prime\prime\prime\prime }}^c \,, -{\eG_R^{\prime\prime\prime\prime\prime } }^c )^T$  \\[1mm]
                       &   &   &   & $\underline{ ( \rep{1}\,, \rep{1} \,, +1 )_{ \mathbf{F}}^{\prime\prime \prime } ~:~ {e_R}^c }$ \\[1.5mm]
                       & $( \rep{ 6}\,, \rep{4} \,, -\frac{1}{4})_{ \mathbf{F}}$  & $( \rep{3}\,, \rep{4} \,, -\frac{1}{12})_{ \mathbf{F}}^\prime$ & $( \rep{3}\,, \rep{3} \,, 0)_{ \mathbf{F}}^{\prime\prime}$ & $\underline{ ( \rep{3}\,, \rep{2} \,, +\frac{1}{6})_{ \mathbf{F}}^{\prime\prime \prime } ~:~  (u_L \,, d_L)^T} $ \\[1mm]
                       &   &   &   &  $( \rep{3}\,, \rep{1} \,, -\frac{1}{3})_{ \mathbf{F}}^{\prime \prime \prime \prime} ~:~\DG_L^{\prime \prime \prime\prime}$ \\[1mm]
                       &   &   &  $( \rep{3}\,, \rep{1} \,, -\frac{1}{3})_{ \mathbf{F}}^{\prime \prime \prime\prime \prime}$ & $( \rep{3}\,, \rep{1} \,, -\frac{1}{3})_{ \mathbf{F}}^{\prime \prime \prime \prime \prime} ~:~ \DG_L^{\prime \prime \prime\prime \prime}$ \\[1mm]
                       &   & $( \repb{3}\,, \rep{4} \,, -\frac{5}{12})_{ \mathbf{F}}$ & $( \repb{3}\,, \rep{3} \,, -\frac{1}{3})_{ \mathbf{F}}$ & $( \repb{3}\,, \rep{2} \,, -\frac{1}{6})_{ \mathbf{F}} ~:~ ( {\dG_R}^c \,,{\uG_R}^c )^T$  \\[1mm]
                       &   &   &   & $( \repb{3}\,, \rep{1} \,, -\frac{2}{3})_{ \mathbf{F}}^{\prime \prime} ~:~{\UG_R}^c$  \\[1mm]
                       &   &   & $( \repb{3}\,, \rep{1} \,, -\frac{2}{3})_{ \mathbf{F}}^{\prime \prime \prime}$ & $\underline{ ( \repb{3}\,, \rep{1} \,, -\frac{2}{3})_{ \mathbf{F}}^{\prime \prime \prime} ~:~{c_R}^c }$  \\[1mm]
\hline\hline
\end{tabular}
\caption{
The $\gSU(8)$ fermion representation of $\rep{56_F}$ under the $\Gc_{441}\,,\Gc_{341}\,, \Gc_{331}\,, \Gc_{\rm SM}$ subgroups for the three-generational ${\rm SU}(8)$ theory. 
All SM fermions are marked with underlines.
}
\label{tab:SU8_56ferm}
\end{center}
}
\end{table}%

\para
By following the SWW symmetry breaking pattern in Eq.~\eqref{eq:Pattern}, we tabulate the fermion representations at various stages of the ${\rm SU}(8)$ model in Tabs.~\ref{tab:SU8_8barferm}, \ref{tab:SU8_28ferm}, and \ref{tab:SU8_56ferm}. 
All ${\rm U}(1)$ charges at different stages are obtained according to Eqs.~\eqref{eq:X1charge_4sfund}, \eqref{eq:X2charge_4Wfund}, \eqref{eq:Ycharge_4Wfund}, and~\eqref{eq:Qcharge_4Wfund}.
According to the rule by Georgi~\cite{Georgi:1979md}, it is straightforward to find $n_g=3$ in the current setup.
All chiral fermions are named by their $\Gc_{\rm SM}$ irreps.
For the right-handed down-type quarks of ${\Dc_R^\Omega}^c$, they are named as follows
\beqn\label{eq:DR_names}
&& {\Dc_R^{ \dot 1} }^c \equiv {d_R}^c \,, ~ {\Dc_R^{  \dot 2} }^c \equiv {s_R}^c \,,~ {\Dc_R^{\dot {\rm VII} } }^c \equiv {\DG_R^{\prime\prime\prime\prime \prime }}^c \,, ~  {\Dc_R^{\dot {\rm VIII} } }^c \equiv {\DG_R^{\prime\prime \prime }}^c  \,,~ {\Dc_R^{  \dot {\rm IX} } }^c \equiv {\DG_R^{\prime\prime\prime \prime }}^c \,, \non
&&{\Dc_R^{3} }^c \equiv {b_R}^c \,,~  {\Dc_R^{\rm IV } }^c \equiv {\DG_R}^c \,, ~  {\Dc_R^{\rm V } }^c \equiv {\DG_R^{\prime\prime }}^c  \,,~ {\Dc_R^{\rm VI } }^c \equiv {\DG_R^{\prime }}^c  \,.
\eeqn
For the left-handed ${\rm SU}(2)_W$ lepton doublets of $(\Ec_L^\Omega \,, - \Nc_L^\Omega )$, they are named as follows
\beqn\label{eq:ELNL_names}
&&   ( \Ec_L^{ \dot 1} \,,  - \Nc_L^{\dot 1})  \equiv (e_L\,, - \nu_{e\,L} ) \,, ~( \Ec_L^{  \dot 2} \,,   - \Nc_L^{\dot 2})  \equiv( \mu_L \,, - \nu_{\mu\,L} )  \,, \non
&&  ( \Ec_L^{ \dot {\rm VII} } \,,  - \Nc_L^{ \dot {\rm VII} })  \equiv ( \eG_L^{ \prime\prime \prime \prime} \,, - \nG_L^{\prime\prime \prime \prime } )  \,, ~ ( \Ec_L^{ \dot {\rm VIII} } \,, - \Nc_L^{ \dot {\rm VIII} }  )  \equiv ( \eG_L^{ \prime\prime  \prime} \,, - \nG_L^{\prime\prime \prime} ) \,,~  ( \Ec_L^{\dot {\rm IX} } \,,  - \Nc_L^{ \dot {\rm IX} } ) \equiv ( \eG_L^{ \prime\prime \prime \prime \prime } \,,  - \nG_L^{\prime\prime \prime\prime \prime} )  \,,\non
&&  ( \Ec_L^{ 3} \,, - \Nc_L^{3}) \equiv ( \tau_L \,, - \nu_{\tau\,L})\,,~  ( \Ec_L^{\rm IV } \,, - \Nc_L^{\rm IV }) \equiv ( \eG_L^{\prime\prime} \,, - \nG_L^{\prime\prime} ) \,, \non
&& ( \Ec_L^{\rm V } \,, -  \Nc_L^{\rm V }) \equiv ( \eG_L \,, - \nG_L )  \,,~ ( \Ec_L^{\rm VI } \,, - \Nc_L^{\rm VI } ) \equiv  ( \eG_L^\prime\,, - \nG_L^\prime ) \,.
\eeqn
Through the analysis in Sec.~\ref{section:pattern}, we shall see that all heavy $(\Dc^\Omega\,, \Ec^\Omega\,, \Nc^\Omega)$ (with $\Omega={\rm IV}\,, \ldots \,,\dot {\rm IX}$) acquire vectorlike masses during the intermediate symmetry breaking stages.
For the remaining left-handed sterile neutrinos of $(  \check \Nc_L^\Omega \,, \check \Nc_L^{\Omega^\prime } \,, \check \Nc_L^{\Omega^{ \prime\prime} } )$, several of them are massive and they are named as follows
\beqn\label{eq:stNL_names}
&& \check \Nc_L^{{\rm IV}^\prime }  \equiv \check \nG_L^\prime \,,~   \check \Nc_L^{{\rm IV}^{\prime \prime} }  \equiv \check \nG_L^{\prime \prime} \,, ~ \check \Nc_L^{{\rm V}^{\prime } }  \equiv \check \nG_L \,,~  \check \Nc_L^{\dot {\rm VII}^{\prime} }  \equiv \check \nG_L^{\prime \prime \prime } \,.
\eeqn
%
%
%
%
%
%
The flavor indices in Eqs.~\eqref{eq:DR_names} and \eqref{eq:ELNL_names} are chosen in accordance with the previous conventions in Ref.~\cite{Chen:2024cht}.
The SM flavor names in Tabs.~\ref{tab:SU8_8barferm}, \ref{tab:SU8_28ferm}, and~\ref{tab:SU8_56ferm} have been determined according to the consistent mass hierarchies and the CKM mixing patterns.

\para
Through Tabs.~\ref{tab:SU8_8barferm}, \ref{tab:SU8_28ferm}, and \ref{tab:SU8_56ferm}, it is straightforward to find that all three-generational SM fermions transform differently in the UV theories.
For example, three left-handed quark doublets belong to $(\rep{3}\,, \rep{2}\,, +\frac{1}{6} )_{\mathbf{F}}\subset (\rep{4}\,, \rep{4}\,, 0 )_{\mathbf{F}}$, $(\rep{3}\,, \rep{2}\,, +\frac{1}{6} )_{\mathbf{F}}^\prime \subset (\rep{4}\,, \rep{6}\,, +\frac{1}{4} )_{\mathbf{F}}$, and $(\rep{3}\,, \rep{2}\,, +\frac{1}{6} )_{\mathbf{F}}^{\prime\prime } \subset (\rep{6}\,, \rep{4}\,, -\frac{1}{4} )_{\mathbf{F}}$, when one traces back their irreps in the $\Gc_{441}$ effective theory.
Likewise, three $(\repb{3}\,, \rep{1}\,, -\frac{2}{3} )_{\mathbf{F}}$'s and three $(\rep{1}\,, \rep{1}\,, +1 )_{\mathbf{F}}$'s all stem from distinctive $\Gc_{441}$ irreps.
Thus, the flavor non-universality under the flavor-conserving neutral currents of the $\Gc_{441}$ and $\Gc_{341}$ theories can be expected.
When the theory flows to the $\Gc_{331}$ along the symmetry breaking pattern in Eq.~\eqref{eq:Pattern}, all three-generational SM fermions together with their heavy partner fermions, have already transformed the same.
Therefore, the flavor universality under the flavor-conserving neutral currents of the $\Gc_{331}$ theory can be expected.
These features will be explicitly demonstrated in terms of the neutral currents at each symmetry breaking stage in Sec.~\ref{section:SU8_gauge}.

\subsection{Decompositions of the ${\rm SU}(8)$ Higgs fields}
\label{section:SU8_Higgs}

\para
We decompose the Higgs fields in Eq.~\eqref{eq:Yukawa_SU8} into components that can be responsible for the sequential SWW symmetry breaking pattern in Eq.~\eqref{eq:Pattern}.
For Higgs fields of $\repb{8_H}_{\,,\omega }$, they read
\beqn\label{eq:SU8_Higgs_Br01}
\repb{8_H}_{\,,\omega }  &\supset&  \boxed{  ( \repb{4} \,, \rep{1} \,, +\frac{1}{4} )_{\mathbf{H}\,, \omega }  } \oplus  \underline{ ( \rep{1} \,, \repb{4} \,, -\frac{1}{4} )_{\mathbf{H}\,, \omega } }\non
&\supset&  \boxed{ ( \rep{1} \,, \repb{4} \,, -\frac{1}{4} )_{\mathbf{H}\,, \omega }  } \supset  \boxed{ ( \rep{1} \,, \repb{3} \,, -\frac{1}{3} )_{\mathbf{H}\,, \omega } } \supset \boxed{ ( \rep{1} \,, \repb{2} \,, -\frac{1}{2} )_{\mathbf{H}\,, \omega } } \,.
\eeqn
For Higgs fields of $\repb{28_H}_{\,,\dot \omega } $, they read
\beqn\label{eq:SU8_Higgs_Br02}
\repb{28_H}_{\,,\dot \omega } &\supset& ( \rep{6} \,, \rep{1} \,,  +\frac{1}{2} )_{\mathbf{H}\,, \dot\omega }  \oplus  \underline{ ( \rep{1} \,, \rep{6} \,, -\frac{1}{2} )_{\mathbf{H}\,, \dot\omega } } \oplus \underline{ ( \repb{4} \,, \repb{4} \,, 0 )_{\mathbf{H}\,, \dot\omega } }  \non
&\supset & \underline{  ( \rep{1} \,, \rep{6} \,, -\frac{1}{2} )_{\mathbf{H}\,, \dot\omega }  } \oplus \boxed{ ( \rep{1} \,, \repb{4} \,, -\frac{1}{4} )_{\mathbf{H}\,, \dot\omega }  } \non
&\supset& \[ \boxed{ ( \rep{1} \,, \repb{3} \,, -\frac{1}{3} )_{\mathbf{H}\,, \dot\omega }^\prime }  \oplus \underline{ ( \rep{1} \,, \rep{3} \,, -\frac{2}{3} )_{\mathbf{H}\,, \dot\omega }}  \] \oplus \boxed{ ( \rep{1} \,, \repb{3} \,, -\frac{1}{3} )_{\mathbf{H}\,, \dot\omega } } \non
&\supset& \[ \boxed{ ( \rep{1} \,, \repb{2} \,, -\frac{1}{2} )_{\mathbf{H}\,, \dot\omega }^\prime }  \oplus \boxed{ (  \rep{1} \,, \rep{2} \,, -\frac{1}{2}  )_{\mathbf{H}\,, \dot\omega }  }  \] \oplus \boxed{ ( \rep{1} \,, \repb{2} \,, -\frac{1}{2} )_{\mathbf{H}\,, \dot\omega}  }  \,.
\eeqn
For Higgs field of $\rep{70_H}$, they read
\beqn\label{eq:SU8_Higgs_Br05} 
\rep{70_H} &\supset& ( \rep{1} \,, \rep{1 } \,, -1 )_{\mathbf{H}}^{ \prime \prime } \oplus ( \rep{1} \,, \rep{1 } \,, +1 )_{\mathbf{H}}^{ \prime \prime \prime \prime } \oplus \underline{ ( \rep{4} \,, \repb{4} \,, +\frac{1}{2} )_{\mathbf{H}} } \oplus  ( \repb{4} \,, \rep{4} \,, -\frac{1}{2} )_{\mathbf{H}}   \oplus ( \rep{6 } \,, \rep{6 } \,, 0 )_{\mathbf{H}}  \non
&\supset& \underline{ ( \rep{1} \,, \repb{4} \,, +\frac{3}{4} )_{\mathbf{H}}^\prime } \supset  \underline{ ( \rep{1} \,, \repb{3} \,, +\frac{2}{3} )_{\mathbf{H}}^{\prime\prime\prime} }  \supset  \boxed{ ( \rep{1} \,, \repb{2} \,, +\frac{1}{2} )_{\mathbf{H}}^{\prime \prime\prime } } \,. 
\eeqn
%
%

\begin{table}[htp]
\begin{center}
\begin{tabular}{c|cccc}
\hline \hline
Higgs & $\Gc_{441}\to \Gc_{341}$ & $\Gc_{341}\to \Gc_{331}$  &  $\Gc_{331}\to \Gc_{\rm SM}$ & $\Gc_{\rm SM} \to {\rm SU}(3)_{c}  \otimes  {\rm U}(1)_{\rm EM}$  \\
\hline
$\repb{8_H}_{\,,\omega }$ & \cmark$\,(\omega ={\rm IV})$  & \cmark$\,(\omega ={\rm V})$  &  \cmark$\,(\omega =3\,, {\rm VI} )$  & \cmark  \\
$\repb{28_H}_{\,,\dot \omega }$ &  \xmark  & \cmark$\,(\dot \omega = \dot 1\,, \dot {\rm VII} )$  &  \cmark$\,(\dot \omega =\dot 2 \,, \dot {\rm VIII}\,, \dot{\rm IX})$  & \cmark  \\
$\rep{70_H}$ & \xmark  & \xmark  & \xmark  & \cmark \\
\hline\hline
\end{tabular}
\end{center}
\caption{The Higgs fields and their SWW symmetry breaking directions in the ${\rm SU}(8)$ theory.
The \cmark and \xmark represent possible and impossible symmetry breaking directions for a given Higgs field.
The flavor indices for the Higgs fields that develop VEVs at different stages are also specified in the parentheses.
}
\label{tab:SU8Higgs_directions}
\end{table}%

\para
All components that are likely to develop VEVs for the sequential symmetry breaking stages are framed by boxes, and other ${\rm SU}(3)_c\otimes {\rm U}(1)_{\rm EM}$-exact components are neglected.
Accordingly, we mark the allowed/disallowed symmetry breaking directions for all Higgs fields in Tab.~\ref{tab:SU8Higgs_directions}.
Schematically, we assign the Higgs VEVs according to the symmetry breaking pattern in Eq.~\eqref{eq:Pattern} as follows
\beqs\label{eqs:SU8_Higgs_VEVs_mini}
\beqn
\Gc_{441} \to \Gc_{341} ~&:&~ \langle ( \repb{4} \,, \rep{1} \,, +\frac{1}{4} )_{\mathbf{H}\,, {\rm IV}} \rangle \equiv \frac{1 }{ \sqrt{2} } W_{ \repb{4}\,, {\rm IV}}\,, \label{eq:SU8_Higgs_VEVs_mini01}\\[1mm]
\Gc_{341} \to \Gc_{331} ~&:&~ \langle ( \rep{1} \,, \repb{4} \,, -\frac{1}{4} )_{\mathbf{H}\,, {\rm V} } \rangle \equiv \frac{1 }{ \sqrt{2} } w_{\repb{4}\,, {\rm V} }\,, ~   \langle ( \rep{1} \,, \repb{4} \,, -\frac{1}{4} )_{\mathbf{H}\,,  \dot 1\,, \dot {\rm VII} } \rangle \equiv \frac{1 }{ \sqrt{2} } w_{\repb{4}\,,  \dot 1\,, \dot {\rm VII} }  \,, \label{eq:SU8_Higgs_VEVs_mini02}\\[1mm]
\Gc_{331} \to \Gc_{\rm SM} ~&:&~  \langle ( \rep{1} \,, \repb{3} \,, -\frac{1}{3} )_{\mathbf{H}\,, 3\,, {\rm VI}} \rangle \equiv \frac{1 }{ \sqrt{2} } V_{ \repb{3}\,, 3\,, {\rm VI} } \non
&&   \langle ( \rep{1} \,, \repb{3} \,, -\frac{1}{3} )_{\mathbf{H}\,, \dot {\rm IX} } \rangle \equiv \frac{1 }{ \sqrt{2} }V_{ \repb{3}\,, \dot {\rm IX} }  \,, ~  \langle ( \rep{1} \,, \repb{3} \,, -\frac{1}{3} )_{\mathbf{H}\,, \dot 2\,, \dot {\rm VIII}  }^\prime \rangle \equiv \frac{1 }{ \sqrt{2} } V_{ \repb{3}\,, \dot 2\,, \dot {\rm VIII} }^\prime  \,, \label{eq:SU8_Higgs_VEVs_mini03} \\[1mm]
 {\rm EWSB} ~&:&~   \langle ( \rep{1} \,, \repb{2} \,, +\frac{1}{2} )_{\mathbf{H} }^{ \prime\prime \prime} \rangle \equiv \frac{1 }{ \sqrt{2} } v_{\rm EW}   \,.\label{eq:SU8_Higgs_VEVs_mini05}
\eeqn
\eeqs
These VEV assignments are made such that the three-generational SM fermions are likely to generate the observed mass hierarchies and the CKM mixing pattern~\cite{Chen:2024cht}.
Given that the $\rep{70_H}$ is self-conjugate, one may speculate two EWSB components of $( \rep{1} \,, \repb{2} \,, +\frac{1}{2} )_{\mathbf{H}}^{\prime \prime\prime }\subset( \rep{4} \,, \repb{4} \,, +\frac{1}{2} )_{\mathbf{H}}$ and $( \rep{1} \,, \rep{2} \,, -\frac{1}{2} )_{\mathbf{H}}^{\prime \prime\prime }\subset( \repb{4} \,, \rep{4} \,, -\frac{1}{2} )_{\mathbf{H}}$ from the $\rep{70_H}$.
In Ref.~\cite{Chen:2023qxi}, we have proved that only one component of $( \rep{1} \,, \rep{2} \,, +\frac{1}{2} )_{\mathbf{H}}^{\prime \prime\prime }\subset \rep{70_H}$ is $\widetilde{ {\rm U}}(1)_{B-L}$-neutral according to the definition in Eq.~\eqref{eq:U1T_def}, and thus can develop the EWSB VEV.
Furthermore, this component also gives the top quark mass through the Yukawa coupling of $Y_\Tc \rep{28_F} \rep{28_F} \rep{70_H} +H.c.$.
An immediate consequence of this result is that the third-generational SM $\rep{10_F}$ must reside in the $\rep{28_F}$, while the first- and second-generational SM $\rep{10_F}$'s must reside in the $\rep{56_F}$.
The same observation was also pointed out in Ref.~\cite{Barr:2008pn}.

\subsection{A summary of the vectorlike fermions of the SWW symmetry breaking pattern}

\begin{table}[htp] {\small
\begin{center}
\begin{tabular}{c|c|c|c|c}
\hline \hline
   stages &  $Q_e=- \frac{1}{3}$  & $Q_e=+ \frac{2}{3}$   & $Q_e= -1$ & $Q_e= 0$  \\
\hline \hline
  $v_{441}$   & $\DG$   &  -  &  $( \eG^{ \prime\prime}\,, \nG^{\prime\prime} )$  &  $\{ \check \nG^{ \prime}\,, \check \nG^{\prime\prime} \}$  \\
  $\{ \Omega \}$  &  ${\rm IV}$  &    & ${\rm IV}$ &  $\{ {\rm IV}^\prime \,, {\rm IV}^{\prime\prime} \}$  \\[1mm]
    \hline
  $v_{341}$   &  $\dG\,, \{  \DG^{ \prime\prime}\,,  \DG^{\prime\prime\prime\prime\prime} \}$  & $\uG\,, \UG$   &  $\EG\,,( \eG\,, \nG )\,,( \eG^{ \prime\prime\prime\prime}\,, \nG^{\prime\prime\prime\prime} )$  & $\{ \check \nG\,, \check \nG^{\prime\prime\prime} \}$   \\
   $\{ \Omega \}$   &  $\{ {\rm V} \,, \dot {\rm VII}  \}$  &    &  $\{ {\rm V} \,, \dot {\rm VII}  \}$  &  $\{ {\rm V}^\prime \,, \dot {\rm VII}^{\prime} \}$  \\[1mm]
    \hline
   $v_{331}$   &  $\{  \DG^{ \prime}\,,  \DG^{ \prime\prime\prime}\,, \DG^{\prime\prime\prime\prime} \}$  &  -  &  $( \eG^\prime \,, \nG^\prime )\,,( \eG^{ \prime\prime\prime}\,, \nG^{\prime\prime\prime} )\,, ( \eG^{ \prime\prime\prime\prime\prime}\,, \nG^{\prime\prime\prime\prime\prime} )$  &  -  \\
    $\{ \Omega\}$  &  $\{ {\rm VI} \,, \dot {\rm VIII}\,, \dot {\rm IX} \}$  &   &  $\{ {\rm VI} \,, \dot {\rm VIII}\,, \dot {\rm IX} \}$  &    \\[1mm]
\hline\hline
\end{tabular}
\caption{
The vectorlike fermions at different intermediate symmetry breaking scales in the ${\rm SU}(8)$ theory.}
\label{tab:SU8_vectorferm}
\end{center} 
}
\end{table}%

\para
Here, we summarize all vectorlike fermions that acquire masses above the EWSB scale in Tab.~\ref{tab:SU8_vectorferm}, where the conventions of flavor indices aligned with our previous derivations in Refs.~\cite{Chen:2023qxi,Chen:2024cht}.
In the language of Georgi's decomposition and count rule~\cite{Georgi:1979md}, the $(\rep{5_F}\,, \repb{5_F})$ vectorlike pairs appear in the form of $( \DG \,, \eG^{ \prime\prime}\,, \nG^{\prime\prime}  )$ at the first stage, and so on.
At the second symmetry breaking stage, there is also a $(\rep{10_F}\,, \repb{10_F})$ vectorlike pair, which appear in the form of $( \uG\,, \dG \,, \UG \,, \EG)$.
The numbers of vectorlike pairs that become massive at each symmetry breaking stage have been precisely determined by gauge invariance and the anomaly-free conditions~\cite{Chen:2023qxi,Chen:2024cht}.

\section{The gauge sector of the ${\rm SU}(8)$ theory and the flavor non-universality}
\label{section:SU8_gauge}

\para
In this section, we describe the gauge sector according to the SWW symmetry breaking pattern in Eq.~\eqref{eq:Pattern}.
The gauge couplings at each symmetry breaking stage will be defined for the derivation of RGEs.
Due to the intrinsic symmetry properties, the flavor non-universality of three-generational SM fermions in the $\Gc_{441}$ and the $\Gc_{341}$ theories will be explicitly displayed.
Some of the earlier and recent references addressing the flavor non-universality scenario with product gauge groups in the UV setup include Refs.~\cite{Li:1981nk,Ma:1987ds,Li:1992fi,Craig:2011yk,Bordone:2017bld,Fuentes-Martin:2020pww,Davighi:2022fer,Fuentes-Martin:2022xnb,Davighi:2022bqf,FernandezNavarro:2023rhv,FernandezNavarro:2023hrf}.

\subsection{The ${\rm SU}(4)_s \otimes {\rm U}(1)_{ X_0}$ gauge bosons and gauge couplings}
\label{section:441_bosons}

\subsubsection{Covariant derivatives and gauge boson masses}

\para
We express the ${\rm SU}(4)_s \otimes {\rm U}(1)_{X_0}$ covariant derivatives as follows
\beqn\label{eq:441_cov_fund}
i D_\mu \psi_{ \rep{4}} &\equiv& i \partial_\mu \psi_{ \rep{4}} + ( g_{4 s} G_\mu^{\bar A} T_{ {\rm SU}(4)}^{\bar A} + g_{X_0} \Xc_0 \mathbb{I}_4 X_{0\,\mu} ) \cdot \psi_{ \rep{4}} \,,
\eeqn
for the ${\rm SU}(4)_s$ fundamental representation of $\psi_{ \rep{4}}$, and
\beqn\label{eq:441_cov_antifund}
i D_\mu \psi_{ \repb{4}} &\equiv& i \partial_\mu \psi_{ \repb{4}} + (- g_{4 s} G_\mu^{\bar A} ( T_{ {\rm SU}(4)}^{\bar A} )^T + g_{X_0} \Xc_0 \mathbb{I}_4 X_{0\,\mu} ) \cdot \psi_{ \repb{4}}\,,
\eeqn
for the ${\rm SU}(4)_s$ anti-fundamental representation of $\psi_{ \repb{4}}$.
For the ${\rm SU}(4)_s$ rank-$2$ anti-symmetric field of $\psi_{ \rep{6}}$, the covariant derivative acts as follows
\beqn\label{eq:441_cov_r2antisym}
i D_\mu \psi_{ \rep{6}} &\equiv&i \partial_\mu \psi_{ \rep{6}}  + g_{4s} G_\mu^{\bar A} \Big( T_{ {\rm SU}(4)}^{\bar A} \cdot \psi_{ \rep{6}}  + \psi_{ \rep{6}}  \cdot (T_{ {\rm SU}(4)}^{\bar A} )^T  \Big) + g_{X_0} \Xc_0 \mathbb{I}_4 X_{0\,\mu} \cdot \psi_{ \rep{6}} \,.
\eeqn
The ${\rm SU}(4)_s$ generators of $T_{ {\rm SU}(4)}^{\bar A}$ are normalized such that $ {\rm Tr}\Big( T_{ {\rm SU}(4)}^{\bar A} T_{ {\rm SU}(4)}^{\bar B}  \Big)= \hf \delta^{\bar A \bar B}$, with $\bar A\,, \bar B=1\,,...\,,15$.
Specifically, we split the ${\rm SU}(4)_s$ adjoint indices into $\bar A=\underbrace{1\,,...\,,8}_{ A}\,,\underbrace{9\,,...\,,15}_{ \hat A }$.
All ${\rm SU}(4)_s$ gluons of $G_\mu^{\bar A}$ are decomposed into
\beqn
\underbrace{ ( \rep{15} \,, \rep{1} \,, 0 )_{ \rep{V}} }_{ G_\mu^{\bar A}}&=& \underbrace{ ( \rep{8 } \,, \rep{1} \,, 0 )_{ \rep{V}} }_{ G_\mu^{ A} }\oplus \underbrace{ ( \rep{3 } \,, \rep{1} \,, - \frac{1 }{3} )_{ \rep{V} } }_{  \tilde C_\mu^{ a} }\oplus \underbrace{ ( \repb{ 3} \,, \rep{1} \,, +\frac{1 }{3 } )_{ \rep{V} }  }_{  \tilde C_{\mu \,, a } }\oplus \underbrace{ ( \rep{1 } \,, \rep{1} \,, 0 )_{ \rep{V} }  }_{ M_\mu}\,. 
\eeqn
It is useful to define a strong mixing angle of
\beqn\label{eq:Chen_angle}
t_{\vartheta_C}&\equiv& \tan\vartheta_{C} = \frac{g_{X_0} }{ \sqrt{6} g_{4s } }\,.
\eeqn
The explicit form for the gauge fields of $g_{4s} G_\mu^{\bar A}  T_{ {\rm SU}(4)}^{\bar A} +g_{X_0} \Xc_0 \mathbb{I}_4 X_{0\,\mu}$ can be expressed in terms of a $4\times 4$ matrix in the gauge eigenbasis
\beqn\label{eq:441_fundconnection_gauge}
&& g_{4s } G^{\bar A}_\mu T_{ {\rm SU}(4)}^{\bar A} + g_{X_0} \Xc_0 \mathbb{I}_4 X_{0\,\mu}=  \non
&& \frac{ g_{4s} }{2} \left(
\begin{array}{c c c c}
 & & & G_\mu^{ 9} - i  G_\mu^{10} \\
\multicolumn{3}{c}{  G_\mu^A T_{ {\rm SU}(3)}^A}  & G_\mu^{ 11} - i  G_\mu^{ 12} \\
 &  &  &  G_\mu^{ 13 } - i  G_\mu^{14 } \\
 G_\mu^{ 9} + i  G_\mu^{  10 } &  G_\mu^{ 11 } + i  G_\mu^{ 12 } & G_\mu^{ 13} + i  G_\mu^{  14} & 0
\end{array}
\right)   \non
 &+& \frac{ g_{4s} }{2} {\rm diag}  \Big(   ( \frac{1}{\sqrt{6} }   G_\mu^{  15}  + 2 \sqrt{6} t_{\vartheta_C} \Xc_0 X_{0\,\mu} ) \mathbb{I}_{3\times 3} \,,  - \frac{3}{\sqrt{6} } G_\mu^{  15} + 2 \sqrt{6} t_{\vartheta_C} \Xc_0 X_{0\,\mu}  \Big) \,.
\eeqn
By using the ${\rm U}(1)_{X_1}$ charge matrix for the ${\rm SU}(4)_s$ adjoint in Eq.~\eqref{eq:X1charge_4sadj}, we find that
\beqn
&& \Xc_1 (   G_\mu^{ 9} \mp i  G_\mu^{10} ) =  \Xc_1 ( G_\mu^{ 11 } \mp i  G_\mu^{ 12 }  ) = \Xc_1 ( G_\mu^{ 13 } \mp i  G_\mu^{14 }  ) = \mp \frac{1}{3} \,,
\eeqn
through the relation of $[ \hat X_1 (\rep{15_s} ) \,, G^{\bar A}_\mu T_{ {\rm SU}(4)}^{\bar A} ] = \Xc_1\, G^{\bar A}_\mu T_{ {\rm SU}(4)}^{\bar A} $.
Apparently, they are ${\rm SU}(3)_c$ fundamental and anti-fundamental vector bosons with electric charges of $\mp \frac{1}{3}$, which are vectorial leptoquarks (LQs).
Accordingly, we name these vectorial LQs as
\beqn\label{eq:LQ_Name}
&& \frac{1}{\sqrt{2}} (  G_\mu^{ 9} - i  G_\mu^{10} ) \equiv \tilde C_\mu^\clubsuit\,, ~  \frac{1}{\sqrt{2}} ( G_\mu^{ 11 } - i  G_\mu^{ 12 }  ) \equiv \tilde C_\mu^\diamondsuit\,, ~ \frac{1}{\sqrt{2}} ( G_\mu^{ 13 } - i  G_\mu^{14 } ) \equiv \tilde C_\mu^\heartsuit \,, \non
&& \frac{1}{\sqrt{2}} (  G_\mu^{ 9} + i  G_\mu^{10} ) \equiv \tilde C_{\mu\,,\clubsuit}\,, ~  \frac{1}{\sqrt{2}} (G_\mu^{ 11 } + i  G_\mu^{ 12 }  ) \equiv \tilde C_{\mu\,,\diamondsuit }\,, ~ \frac{1}{\sqrt{2}} ( G_\mu^{ 13 } + i  G_\mu^{14 } ) \equiv \tilde C_{\mu\,,\heartsuit} \,.
\eeqn

\para
The first symmetry breaking stage is achieved by the Higgs VEV given in Eq.~\eqref{eq:SU8_Higgs_VEVs_mini01}, and the Higgs kinematic terms lead to the gauge boson masses of
\beqn
&& \frac{ 1 }{4} g_{4s}^2 v_{441}^2 \cdot \Big[  \sum_{ a = \clubsuit\,,\diamondsuit\,, \heartsuit  }  \tilde C_\mu^{ a} \tilde C_{\mu\,, a} +  \frac{3 }{4} \Big(   \ G_\mu^{15} + t_{ \vartheta_C } X_{0\,\mu}  \Big)^2 \Big] \,, \quad v_{441}^2 = W_{ \repb{4}\,, {\rm IV}}^2  \,.
\eeqn
The mass eigenstates are diagonalized in terms of the mixing angle $\vartheta_C$ in Eq.~\eqref{eq:Chen_angle} as follows
\beqn
 \left( \ba{c} M_\mu  \\ X_{1\,\mu}  \\    \ea \right) &=& \left( \ba{cc}
   c_{\vartheta_{C}}  &  s_{\vartheta_C} \\
 - s_{\vartheta_C} &  c_{\vartheta_C}  \\   \ea \right) \cdot  \left( \ba{c}   G^{15 }_\mu  \\ X_{0\,\mu} \\     \ea \right)\,,
\eeqn
with $(c_{\vartheta_C}\,,s_{\vartheta_C})\equiv (\cos \vartheta_C \,,\sin \vartheta_C)$.

\para
The ${\rm SU}(3)_c \otimes {\rm U}(1)_{X_1}$ couplings of $(\alpha_{3c}\,,\alpha_{X_1})$ are related to the ${\rm SU}(4)_s \otimes {\rm U}(1)_{X_0}$ couplings of $(\alpha_{4s}\,, \alpha_{X_0} )$ as
\beqn\label{eq:441_coupMatch}
&& \alpha_{3c }^{-1}(v_{441} ) = \alpha_{4s }^{-1} (v_{441} ) \,,~ \alpha_{X_1}^{-1} (v_{441} ) = \frac{1}{6} \alpha_{4s }^{-1}(v_{441} ) +  \alpha_{X_0}^{-1} (v_{441} ) \,, \non
&& \frac{1}{6} \alpha_{4s }^{-1} = \alpha_{X_1}^{-1}  s_{\vartheta_{C}}^2 \,,~ \alpha_{X_0}^{-1} = \alpha_{X_1}^{-1} c_{\vartheta_{C}}^2  \,.
\eeqn
Obviously, three LQs of $\tilde C_\mu^{ a}$ and one neutral $M_\mu$ obtain masses of
\beqn\label{eq:441_GB}
&& m_{ \tilde C_\mu^a }^2 =  \frac{ g_{X_1}^2}{  24 s_{\vartheta_{C}}^2   } v_{441}^2 \,, \quad  m_{ M_\mu }^2 =  \frac{ g_{ X_1 }^2 }{ 16 s_{\vartheta_{C}}^2 c_{\vartheta_{C}}^2  } v_{441}^2  \,,
\eeqn
at this stage of symmetry breaking.
The remaining massless gauge boson of $X_{1\, \mu}$ will further mix with the ${\rm SU}(4)_W$ gauge bosons afterwards.

\para
Expressed in terms of mass eigenstates, the gauge fields in Eq.~\eqref{eq:441_fundconnection_gauge} become
\beqn\label{eq:441_fundconnection_mass}
&& g_{4s } G^{\bar A}_\mu T_{ {\rm SU}(4)}^{\bar A} + g_{X_0} \Xc_0 \mathbb{I}_4 X_{0\,\mu}=  \non
&& \frac{ g_{4s} }{2} \left(
\begin{array}{c c c c}
 & & & 0 \\
\multicolumn{3}{c}{  G_\mu^A T_{ {\rm SU}(3)}^A}  & 0 \\
 &  &  &  0 \\
 0 &  0  & 0 & 0
\end{array}
\right)  + \underline{ \frac{ g_{4s} }{ \sqrt{2} } \left(
\begin{array}{c c c c}
 & & & \tilde C_\mu^\clubsuit \\
\multicolumn{3}{c}{  0_{3\times 3} }  & \tilde C_\mu^\diamondsuit \\
 &  &  &  \tilde C_\mu^\heartsuit \\
 \tilde C_{\mu\,,\clubsuit}  &  \tilde C_{\mu\,,\diamondsuit } &  \tilde C_{\mu\,,\heartsuit }  & 0
\ea  \right) }  \non
&+&  g_{X_1}  {\rm diag}\, \Big(  ( - \frac{1}{12} +  \Xc_0 ) \mathbb{I}_{3\times 3} \,,  \frac{1}{4} +  \Xc_0  \Big) X_{1\, \mu}  \non
&+& \underline{ \frac{g_{X_1} }{ s_{\vartheta_C} c_{\vartheta_C} } {\rm diag}\, \Big(  \[ \frac{1}{12} + (- \frac{1}{12} +  \Xc_0 ) s_{\vartheta_C}^2 \] \mathbb{I}_{3\times 3} \,,   - \frac{1}{4} + ( \frac{1}{4} +  \Xc_0 ) s_{\vartheta_C}^2   \Big)  M_\mu } \,,
\eeqn
for the ${\rm SU}(4)_s$ fundamental representation.
It is straightforward to find
\beqn\label{eq:441_antifundconnection_mass}
&& - g_{4s } G^{\bar A}_\mu ( T_{ {\rm SU}(4)}^{\bar A} )^T + g_{X_0} \Xc_0 \mathbb{I}_4 X_{0\,\mu}  \non
&=& -\frac{ g_{4s} }{2} \left(
\begin{array}{c c c c}
 & & & 0 \\
\multicolumn{3}{c}{  G_\mu^A ( T_{ {\rm SU}(3)}^A)^T  }  & 0 \\
 &  &  &  0 \\
 0 &  0  & 0 & 0
\end{array}
\right)  - \underline{ \frac{ g_{4s} }{ \sqrt{2} } \left(  \ba{cccc}
 & & & \tilde C_{\mu\,,\clubsuit}   \\
\multicolumn{3}{c}{  0_{3\times 3} }  & \tilde C_{\mu\,,\diamondsuit } \\
 &  &  & \tilde C_{\mu\,,\heartsuit } \\
\tilde C_\mu^\clubsuit   &  \tilde C_\mu^\diamondsuit  & \tilde C_\mu^\heartsuit  & 0
\ea  \right) } \non
&+&  g_{X_1}  {\rm diag}\, \Big(  ( \frac{1}{12} +  \Xc_0 ) \mathbb{I}_{3\times 3} \,,  - \frac{1}{4}  + \Xc_0  \Big) X_{1\, \mu}  \non
&+& \underline{ \frac{g_{X_1} }{ s_{\vartheta_C} c_{\vartheta_C} } {\rm diag}\, \Big(  \[ - \frac{1}{12 } + ( \frac{1}{12} +  \Xc_0 ) s_{\vartheta_C}^2 \] \mathbb{I}_{3\times 3} \,,   \frac{1}{4} + ( - \frac{1}{4} +  \Xc_0 ) s_{\vartheta_C}^2   \Big)  M_\mu } \,,
\eeqn
for the ${\rm SU}(4)_s$ anti-fundamental representation.
For the ${\rm SU}(4)_s$-singlet representations, we have
\beqn
&& g_{X_0} \Xc_0 X_{0\,\mu}  = g_{X_1} ( + \Xc_0 ) X_{1\, \mu} +  \frac{g_{X_1} }{ s_{\vartheta_C} c_{\vartheta_C} } ( + \Xc_0 s_{\vartheta_C}^2 ) M_\mu \,.
\eeqn

\subsubsection{Gauge couplings of the $\Gc_{441}$ fermions}

\para
We proceed to obtain the neutral currents mediated by the ${\rm SU}(3)_c$-singlet vector boson of $M_\mu$.
For the anti-fundamental fermions of $\repb{8_F}^\Omega$, we have the following covariant derivatives of
\beqs
\beqn
&& i \overline \psi_{ ( \repb{4}\,, \rep{1}\,, +\frac{1}{4} )^\Omega }  \Dslash \psi_{ ( \repb{4}\,, \rep{1}\,, +\frac{1}{4} )^\Omega } \supset \frac{ g_{X_1} }{ s_{ \vartheta_C } c_{ \vartheta_C } } \Big[  ( \frac{1}{12} - \frac{1}{3} s_{ \vartheta_C}^2 ) \overline{ \Dc_R^\Omega } \gamma^\mu \Dc_R^\Omega  + ( + \frac{1}{4} ) \overline{ \check \Nc_L^\Omega } \gamma^\mu \check \Nc_L^\Omega \Big] \cdot M_\mu \,,\\[1mm]
&& i \overline \psi_{ ( \rep{1}\,, \repb{4}\,, -\frac{1}{4} )^\Omega }   \Dslash \psi_{ ( \rep{1}\,, \repb{4}\,, -\frac{1}{4} )^\Omega } \supset \frac{ g_{X_1} }{ s_{ \vartheta_C } c_{ \vartheta_C } } ( - \frac{1}{4} s_{ \vartheta_C}^2 ) \Big[ \overline{ \Ec_L^\Omega } \gamma^\mu \Ec_L^\Omega + \overline{ \Nc_L^\Omega } \gamma^\mu \Nc_L^\Omega  \non
&+&  \overline{ \check \Nc_L^{\Omega^\prime } } \gamma^\mu \check \Nc_L^{\Omega^\prime } + \overline{\check \Nc_L^{\Omega^{\prime\prime} } } \gamma^\mu \check \Nc_L^{\Omega^{ \prime\prime } } \Big] \cdot M_\mu \,.
\eeqn
\eeqs
For the rank-2 anti-symmetric fermions of $\rep{28_F}$, we have the covariant derivatives of
\beqs
\beqn
&&\Tr \[ i \overline \psi_{ ( \rep{6}\,, \rep{1}\,, -\frac{1}{2} ) }  \Dslash \psi_{ ( \rep{6}\,, \rep{1}\,, -\frac{1}{2} ) } \]  \supset \frac{ g_{X_1} }{ s_{ \vartheta_C } c_{ \vartheta_C } }  \Big[  ( -\frac{1 }{6} + \frac{2 }{3 } s_{ \vartheta_C }^2 ) \overline{ {t_R} }  \gamma^\mu  t_{ R} \non
&+& (- \frac{1 }{6} - \frac{1}{3 } s_{ \vartheta_C }^2  ) \overline{ \DG_L } \gamma^\mu \DG_L  \Big] \cdot M_\mu 
 \,,\\[1mm]
&& i \overline \psi_{ ( \rep{1}\,, \rep{6}\,, +\frac{1}{2} ) }  \Dslash \psi_{ ( \rep{1}\,, \rep{6}\,, +\frac{1}{2} ) } \supset \frac{ g_{X_1} }{ s_{ \vartheta_C } c_{ \vartheta_C } } ( - \frac{1}{2} s_{ \vartheta_C}^2 ) \Big[ \overline{ \eG_R} \gamma^\mu \eG_R + \overline{ \nG_R } \gamma^\mu \nG_R + \overline{ \check \nG_R } \gamma^\mu \check \nG_R \non
&+& \overline{ \nG_R^\prime } \gamma^\mu \nG_R^\prime + \overline{ \eG_R^\prime } \gamma^\mu \eG_R^\prime + \overline{ \tau_R } \gamma^\mu \tau_R  \Big] \cdot M_\mu \,,\\[1mm]
&& i \overline \psi_{ ( \rep{4}\,, \rep{4}\,, 0 ) }  \Dslash \psi_{ ( \rep{4}\,, \rep{4}\,, 0 ) }  \supset \frac{ g_{X_1} }{ s_{ \vartheta_C } c_{ \vartheta_C } }  \Big[ (  \frac{1}{12 } - \frac{1}{12 } s_{ \vartheta_C }^2  ) ( \overline{ t_L} \gamma^\mu t_L + \overline{ b_L} \gamma^\mu b_L + \overline{ \DG_L^\prime } \gamma^\mu \DG_L^\prime + \overline{ \DG_L^{ \prime \prime} } \gamma^\mu \DG_L^{ \prime\prime} ) \non
&+& (  \frac{1}{4 } - \frac{1}{4 } s_{ \vartheta_C }^2  ) ( \overline{ \eG_R^{ \prime\prime}  } \gamma^\mu \eG_R^{ \prime\prime}  +  \overline{ \nG_R^{ \prime\prime}  } \gamma^\mu \nG_R^{ \prime\prime}  +  \overline{ \check \nG_R^{ \prime}  } \gamma^\mu \check \nG_R^{ \prime}  +  \overline{ \check \nG_R^{ \prime\prime}  } \gamma^\mu \check \nG_R^{ \prime\prime}  )   \Big] \cdot M_\mu \,.
%
\eeqn
\eeqs
For the rank-3 anti-symmetric fermions of $\rep{56_F}$, we have the covariant derivatives of
\beqs
\beqn
&& i \overline \psi_{ ( \rep{1}\,, \repb{4}\,, +\frac{3}{4} ) }  \Dslash \psi_{ ( \rep{1}\,, \repb{4}\,, +\frac{ 3}{4} ) }  \supset \frac{ g_{X_1} }{ s_{ \vartheta_C } c_{ \vartheta_C } } (  - \frac{3}{4} s_{ \vartheta_C}^2 ) \Big[   \overline{ \nG_R^{ \prime\prime \prime} } \gamma^\mu \nG_R^{ \prime\prime \prime} + \overline{ \eG_R^{ \prime\prime \prime} } \gamma^\mu \eG_R^{ \prime\prime \prime} + \overline{ \mu_R } \gamma^\mu \mu_R  \non
&+& \overline{ \EG_R } \gamma^\mu \EG_R \Big]  \cdot M_\mu \,,\\[1mm]
&& i \overline \psi_{ ( \repb{4}\,, \rep{1}\,, -\frac{3}{4} ) }  \Dslash \psi_{  ( \repb{4}\,, \rep{1}\,, -\frac{3}{4} )}  \supset  \frac{ g_{X_1} }{ s_{ \vartheta_C } c_{ \vartheta_C } }  \Big[ (\frac{1 }{12 } + \frac{2 }{3 } s_{ \vartheta_C }^2 ) \overline{ u_R } \gamma^\mu u_R  + (  \frac{1 }{4 } -  s_{ \vartheta_C }^2 ) \overline{  \EG_L } \gamma^\mu \EG_L  \Big] \cdot M_\mu \,,\\[1mm]
&& i \overline \psi_{ ( \rep{4}\,, \rep{6}\,, +\frac{1}{4} ) } \Dslash \psi_{ ( \rep{4}\,, \rep{6}\,, +\frac{1}{4} )} \non
&\supset& \frac{ g_{X_1} }{ s_{ \vartheta_C } c_{ \vartheta_C } } \Big[  ( \frac{1}{12} + \frac{1}{6} s_{ \vartheta_C }^2 )( \overline{ c_L } \gamma^\mu c_L + \overline{ s_L } \gamma^\mu s_L + \overline{ \DG_L^{ \prime\prime\prime }  } \gamma^\mu \DG_L^{ \prime\prime\prime } + \overline{ \dG_L } \gamma^\mu \dG_L + \overline{ \uG_L } \gamma^\mu \uG_L + \overline{ \UG_L } \gamma^\mu \UG_L   ) \non
&+& ( \frac{1}{4} - \frac{1}{2} s_{ \vartheta_C }^2 ) ( \overline{ \eG_R^{ \prime\prime \prime \prime} } \gamma^\mu \eG_R^{ \prime\prime \prime \prime} +  \overline{ \nG_R^{ \prime\prime \prime \prime} } \gamma^\mu \nG_R^{ \prime\prime \prime \prime} +  \overline{\check \nG_R^{ \prime \prime \prime} } \gamma^\mu \check \nG_R^{ \prime \prime \prime} +  \overline{ \nG_R^{ \prime \prime\prime \prime \prime} } \gamma^\mu \nG_R^{ \prime \prime\prime \prime \prime} +  \overline{ \eG_R^{ \prime\prime \prime \prime \prime } } \gamma^\mu \eG_R^{ \prime \prime\prime \prime \prime} +  \overline{ e_R } \gamma^\mu e_R )  \Big] \cdot M_\mu \,,\\[1mm]
%
&&\Tr\[ i \overline \psi_{ ( \rep{6}\,, \rep{4}\,, - \frac{1}{4} ) } \Dslash  \psi_{ ( \rep{6}\,, \rep{4}\,, - \frac{1}{4} )} \]  \supset  \frac{ g_{X_1} }{ s_{ \vartheta_C } c_{ \vartheta_C } }  \Big[ (- \frac{1}{6} + \frac{5}{12 } s_{ \vartheta_C }^2 ) ( \overline{ \dG_R } \gamma^\mu \dG_R + \overline{ \uG_R } \gamma^\mu \uG_R  + \overline{ \UG_R } \gamma^\mu \UG_R  + \overline{ c_R } \gamma^\mu c_R  ) \non
&+& ( - \frac{1}{6} - \frac{1}{12 } s_{ \vartheta_C }^2 ) (  \overline{ u_L } \gamma^\mu u_L + \overline{ d_L } \gamma^\mu d_L + \overline{ \DG_L^{ \prime\prime\prime \prime} } \gamma^\mu \DG_L^{ \prime\prime\prime \prime} + \overline{ \DG_L^{ \prime\prime\prime \prime \prime} } \gamma^\mu \DG_L^{ \prime\prime\prime \prime \prime} )  \Big] \cdot M_\mu \,.
%
\eeqn
\eeqs
%
%

\begin{table}[htp]
\begin{center}
\begin{tabular}{c|cccccc}
\hline \hline
  & $u$   & $c$ &  $t$  &  $\uG$  &  $\UG$    &   \\ 
\hline
$g_f^{V\, \Mc}$  &  $- \frac{1}{ 24} + \frac{7 }{ 24} s_{ \vartheta_C}^2 $  & $- \frac{1}{ 24} + \frac{7 }{ 24} s_{ \vartheta_C}^2 $ &  $- \frac{1}{ 24} + \frac{7 }{ 24} s_{ \vartheta_C}^2 $  &  $- \frac{1}{ 24} + \frac{7 }{ 24} s_{ \vartheta_C}^2 $  &  $- \frac{1}{ 24} + \frac{7 }{ 24} s_{ \vartheta_C}^2 $   &    \\ 
$g_f^{A\, \Mc}$  &  $  \frac{1}{ 8 } + \frac{3 }{ 8 } s_{ \vartheta_C}^2$  & $ - \frac{1}{ 8 } + \frac{1 }{ 8 } s_{ \vartheta_C}^2$ &  $ - \frac{1}{ 8 } + \frac{3 }{ 8 } s_{ \vartheta_C}^2$  &  $- \frac{1}{ 8 } + \frac{1 }{ 8 } s_{ \vartheta_C}^2$  &  $- \frac{1}{ 8 } + \frac{1 }{ 8 } s_{ \vartheta_C}^2$  &   \\ 
\hline
  & $d$  & $s$  &  $b$  &  $\dG$  &    &   \\ 
  \hline
$g_f^{V\, \Mc}$  &  $- \frac{1}{ 24} - \frac{5 }{ 24} s_{ \vartheta_C}^2$  & $ \frac{1}{ 12} - \frac{1 }{ 12 } s_{ \vartheta_C}^2$ &  $ \frac{1}{ 12 } - \frac{5 }{ 24 } s_{ \vartheta_C}^2$  &  $- \frac{1}{ 24} + \frac{7 }{ 24} s_{ \vartheta_C}^2 $  &    &    \\ 
$g_f^{A\, \Mc}$  &  $ \frac{1}{ 8 } - \frac{1 }{ 8 } s_{ \vartheta_C}^2$  & $ - \frac{1 }{ 4 } s_{ \vartheta_C}^2$ &  $ - \frac{1}{ 8 } s_{ \vartheta_C}^2 $  &  $- \frac{1}{ 8 } + \frac{1 }{ 8 } s_{ \vartheta_C}^2$  &    &    \\ 
\hline
  & $\DG$  & $\DG^{ \prime  }$  &  $\DG^{ \prime \prime }$  &  $\DG^{ \prime \prime\prime }$ &  $\DG^{ \prime \prime \prime \prime  }$   & $\DG^{ \prime \prime \prime \prime \prime }$  \\ 
  \hline
$g_f^{V\, \Mc}$  &  $ - \frac{1}{ 24} - \frac{1 }{ 3} s_{ \vartheta_C}^2 $  & $ \frac{1}{ 12 } - \frac{5 }{ 24 } s_{ \vartheta_C}^2 $ &  $ \frac{1}{ 12 } - \frac{5 }{ 24 } s_{ \vartheta_C}^2 $  &  $ \frac{1}{ 12 } - \frac{1 }{ 12 } s_{ \vartheta_C}^2 $ &  $ - \frac{1}{ 24} - \frac{5 }{ 24 } s_{ \vartheta_C}^2 $    &  $- \frac{1}{ 24} - \frac{5 }{ 24 } s_{ \vartheta_C}^2$  \\ 
$g_f^{A\, \Mc}$  &  $ \frac{1}{8} $  & $- \frac{1 }{ 8 } s_{ \vartheta_C}^2 $ &  $- \frac{1 }{ 8 } s_{ \vartheta_C}^2 $  &  $- \frac{1 }{ 4 } s_{ \vartheta_C}^2$  &  $ \frac{1}{ 8 } - \frac{1 }{ 8 } s_{ \vartheta_C}^2$   &   $ \frac{1}{ 8 } - \frac{1 }{ 8 } s_{ \vartheta_C}^2$   \\ 
\hline
  & $e$  & $\mu$   &  $\tau$    &  $\EG$  &   &     \\ 
  \hline
$g_f^{V\, \Mc}$  &  $ \frac{1}{ 8 } - \frac{3 }{ 8 } s_{ \vartheta_C}^2$  & $ -\frac{1 }{ 2 } s_{ \vartheta_C}^2$ &  $ -\frac{3 }{ 8 } s_{ \vartheta_C}^2$  &  $\frac{1}{ 8 } - \frac{7 }{ 8 } s_{ \vartheta_C}^2$  &    &    \\ 
$g_f^{A\, \Mc}$  &  $ \frac{1}{ 8 } - \frac{1 }{ 8 } s_{ \vartheta_C}^2$  & $-\frac{1 }{ 4 } s_{ \vartheta_C}^2$ &  $ -\frac{1 }{ 8 } s_{ \vartheta_C}^2$  &  $- \frac{1}{ 8 } + \frac{1 }{ 8 } s_{ \vartheta_C}^2$  &    &    \\ 
\hline
  & $(\eG\,,\nG)$  & $(\eG^{ \prime } \,, \nG^{ \prime })$  &  $(\eG^{ \prime \prime } \,, \nG^{ \prime \prime })$  &  $(\eG^{ \prime \prime\prime } \,, \nG^{ \prime \prime \prime })$  &  $(\eG^{ \prime \prime\prime \prime } \,,  \nG^{ \prime \prime \prime \prime })$ & $(\eG^{ \prime \prime\prime \prime \prime } \,, \nG^{ \prime \prime\prime \prime \prime })$  \\ 
  \hline
$g_f^{V\,\Mc }$  &  $-  \frac{3 }{ 8 } s_{ \vartheta_C}^2$  & $-  \frac{3 }{ 8 } s_{ \vartheta_C}^2$ &  $\frac{1 }{8 } -  \frac{1 }{ 4 } s_{ \vartheta_C}^2$  &  $-  \frac{1 }{ 2 } s_{ \vartheta_C}^2$  &  $\frac{1 }{ 8 } -  \frac{3 }{ 8 } s_{ \vartheta_C}^2$ &  $\frac{1 }{ 8 } -  \frac{3 }{ 8 } s_{ \vartheta_C}^2$   \\ 
$g_f^{A\, \Mc}$  &  $-  \frac{1 }{ 8 } s_{ \vartheta_C}^2$  & $-  \frac{1 }{ 8 } s_{ \vartheta_C}^2$ &  $\frac{1 }{8 }$  &  $-  \frac{1 }{ 4 } s_{ \vartheta_C}^2$  &  $\frac{1 }{ 8 } -  \frac{1 }{ 8 } s_{ \vartheta_C}^2$ & $\frac{1 }{ 8 } -  \frac{1 }{ 8 } s_{ \vartheta_C}^2$  \\  
\hline
  & $\nu_{e\, L}/\nu_{\mu\, L}/\nu_{\tau\, L}$  &  $ \check \Nc_L^{ \Omega \,, \Omega^\prime \,, \Omega^{\prime\prime} } $ & $\check \nG$  & $\check \nG^{ \prime  }$  &  $\check \nG^{ \prime\prime  }$  &  $\check \nG^{ \prime \prime \prime }$     \\ 
  \hline
$g_f^{V\, \Mc}$   &  $-\frac{1}{8} s_{ \vartheta_C}^2$   &  $- \frac{1}{8 } s_{ \vartheta_C}^2$ &  $-  \frac{3 }{ 8 } s_{ \vartheta_C}^2$  & $\frac{1 }{8 } -  \frac{1 }{ 4 } s_{ \vartheta_C}^2$ &  $\frac{1 }{8 } -  \frac{1 }{ 4 } s_{ \vartheta_C}^2$  &  $\frac{1 }{ 8 } -  \frac{3 }{ 8 } s_{ \vartheta_C}^2$      \\ 
$g_f^{A\, \Mc}$  &  $\frac{1}{8} s_{ \vartheta_C}^2$  &  $ \frac{1}{8} s_{ \vartheta_C}^2 $  &  $-  \frac{1 }{ 8 } s_{ \vartheta_C}^2$  & $\frac{1 }{8 } $  &  $\frac{1 }{8 } $ &  $\frac{1 }{ 8 } -  \frac{1 }{ 8 } s_{ \vartheta_C}^2$     \\ 
\hline\hline
\end{tabular}
\end{center}
\caption{
The couplings of the neutral currents mediated by $M_\mu$ in the $V-A$ basis.
The $ \check \Nc_L^{ \Omega \,, \Omega^\prime \,, \Omega^{\prime\prime} } $ represent all 23 left-handed massless sterile neutrinos expect four massive sterile neutrinos of $(\check \nG \,, \check \nG^{ \prime }\,, \check \nG^{ \prime \prime } \,, \check \nG^{ \prime \prime \prime })$.
}
\label{tab:Maon_VA}
\end{table}%

\para
The flavor-conserving neutral currents mediated by $M_\mu$ are expressed as follows in the $V-A$ basis
\beqn\label{eq:SU4c_MaoVA_coup}
\Lc_{ {\rm SU}(4)_s}^{\rm NC\,, F}&=& \frac{g_{X_1 }}{ s_{\vartheta_C } c_{\vartheta_C } } ( g_f^{V\,\Mc } \overline f \gamma^\mu f + g_f^{A\, \Mc } \overline f \gamma^\mu \gamma_5 f ) M_\mu \,,
\eeqn
and we tabulate the vectorial and axial couplings of $( g_f^{V\,\Mc } \,, g_f^{A\,\Mc } )$ in Tab.~\ref{tab:Maon_VA}. 
The flavor non-universality of the neutral currents mediated by $M_\mu$ among three generational SM quarks and leptons are displayed explicitly. 
Three generational SM left-handed neutrinos exhibit flavor universal coupling to the $M_\mu$, since they all originate from the identical irrep of $\repb{8_F}$.
This will always be true in their couplings to the neutral gauge bosons in the sequential symmetry breaking stages.
%
%

\subsection{The ${\rm SU}(4)_W \otimes {\rm U}(1)_{ X_1}$ gauge bosons and gauge couplings}
\label{section:341_bosons}

\subsubsection{Covariant derivatives and gauge boson masses}

\para
We express the ${\rm SU}(4)_W \otimes {\rm U}(1)_{X_1}$ covariant derivatives as follows
\beqn\label{eq:341_cov_fund}
i D_\mu \psi_{ \rep{4}} &\equiv& i \partial_\mu  \psi_{ \rep{4}}  +  (  g_{4W} A_\mu^{\bar I} T_{ {\rm SU}(4)}^{\bar I} +  g_{X_1} \Xc_1 \mathbb{I}_4 X_{1\,\mu} ) \cdot \psi_{ \rep{4}}  \,,
\eeqn
for the ${\rm SU}(4)_W $ fundamental representation, and
\beqn\label{eq:341_cov_antifund}
i D_\mu \psi_{ \repb{4}}&\equiv&  i \partial_\mu \psi_{ \repb{4}} +  ( - g_{4W} A_\mu^{\bar I} ( T_{ {\rm SU}(4)}^{\bar I} )^T + g_{X_1} \Xc_1 \mathbb{I}_4 X_{1\,\mu} ) \cdot \psi_{ \repb{4}} \,,
\eeqn
for the ${\rm SU}(4)_W $ anti-fundamental representation.
The ${\rm SU}(4)_W $ generators of $T_{ {\rm SU}(4)}^a$ are normalized such that $ \Tr\Big( T_{ {\rm SU}(4)}^{\bar I} T_{ {\rm SU}(4)}^{\bar J}  \Big)= \hf \delta^{\bar I \bar J}$.
For the ${\rm SU}(4)_W $ rank-$2$ anti-symmetric field of $\psi_6$, the covariant derivative acts as follows
\beqn\label{eq:341_cov_r2antisym}
i D_\mu \psi_{ \rep{6}} &\equiv& i \partial_\mu \psi_{ \rep{6}}  + g_{4W} A_\mu^{\bar I} ( T_{ {\rm SU}(4)}^{\bar I} \cdot \psi_{ \rep{6}}  + \psi_{ \rep{6}} \cdot T_{ {\rm SU}(4)}^{\bar I\,,T} )  +  g_{X_1} \Xc_1  X_{1\,\mu} \psi_{ \rep{6}} \,.
\eeqn
The corresponding kinematic term for the rank-$2$ anti-symmetric field of $\psi_6$ should be ${\rm Tr}( \bar \psi_{ \rep{6}}  i \Dslash \psi_{ \rep{6}} )$.
For the singlet of $\psi_{ \rep{1}}$, the covariant derivative is simply given by
\beqn\label{eq:341_cov_singlet}
i D_\mu \psi_{ \rep{1}} &\equiv& (  i \partial_\mu  +   g_{X_1} \Xc_1  X_{1\,\mu} ) \psi_{ \rep{1}}  \,,
\eeqn

\para
The explicit form for the gauge fields of $g_{4W} A_\mu^{\bar I}  T_{ {\rm SU}(4)}^{\bar I} +g_{X_1} \Xc_1 \mathbb{I}_4 X_{1\,\mu}$ can be expressed in terms of a $4\times 4$ matrix below
\beqn\label{eq:341_connection_fund}
&& g_{4W} A^{\bar I}_\mu T_{ {\rm SU}(4)}^{\bar I} + g_{X_1} \Xc_1 \mathbb{I}_4 X_{1\,\mu}  \non
&=& \hf g_{4W} \,  \left( 
\ba{cccc}  
     0  &   A_\mu^1 - i A_\mu^2  &   A_\mu^4 - i A_\mu^5  &  A_\mu^9 - i A_\mu^{10}   \\ 
 A_\mu^1 +  i A_\mu^2  &  0   &  A_\mu^6 - i A_\mu^7 &  A_\mu^{11} - i A_\mu^{12}    \\ 
 A_\mu^4 + i A_\mu^5  &  A_\mu^6 + i A_\mu^7  &  0 &  A_\mu^{13} - i  A_\mu^{14}   \\
 A_\mu^{9} + i A_\mu^{10}  & A_\mu^{11} + i A_\mu^{12} &  A_\mu^{13} + i A_\mu^{14}   &  0  \\   \ea  \right)  \non
 &+& \frac{ g_{4W}  }{2} {\rm diag}  \Big(   A_\mu^3 + \frac{1}{\sqrt{3} } A_\mu^8 \,, - A_\mu^3 + \frac{1}{\sqrt{3} } A_\mu^8  \,,  - \frac{ 2 }{\sqrt{3} }  A_\mu^8  \,,  0 \Big) \non
&+& \frac{ g_{4W} }{2 \sqrt{6}} {\rm diag}  \Big(  \[   A_\mu^{15}  + 12 t_{\vartheta_G } \Xc_1 X_{1\,\mu} \] \mathbb{I}_{3\times 3} \,,   - 3  A_\mu^{15} +  12 t_{\vartheta_G }  \Xc_1 X_{1\,\mu}  \Big) \,,
\eeqn
where we have defined an ${\rm SU}(4)_W$ mixing angle of
\beqn\label{eq:Glashow_angle}
 t_{\vartheta_G} &\equiv& \tan\vartheta_{G} = \frac{g_{X_1} }{ \sqrt{6} g_{4W } }\,.
\eeqn

\para
By using the relation of $\[ \hat Q_e( \rep{15_W}) \,, A^{\bar I}_\mu T_{ {\rm SU}(4)}^{\bar I}  \] = Q_e\, A^{\bar I}_\mu T_{ {\rm SU}(4)}^{\bar I}$ with $\hat Q_e( \rep{15_W})$ in Eq.~\eqref{eq:Qcharge_4Wadj}, we find the electrical charges of gauge bosons as follows
\beqn
&& Q_e ( A_\mu^1 \mp i A_\mu^2 ) =  Q_e ( A_\mu^4 \mp i A_\mu^5 ) =  Q_e ( A_\mu^9 \mp i A_\mu^{10} ) = \pm 1 \,,
\eeqn
while all other gauge bosons are electrically neutral.
Correspondingly, the gauge fields of $g_{4W} A_\mu^{\bar I}  T_{ {\rm SU}(4)}^{\bar I} +g_{X_1} \Xc_1 \mathbb{I}_4 X_{1\,\mu}$ can also be expressed as
\beqn\label{eq:341_connection_fund01}
&& g_{4W} A^{\bar I}_\mu T_{ {\rm SU}(4)}^{\bar I} + g_{X_1} \Xc_1 \mathbb{I}_4 X_{1\,\mu}=  \non
&& \frac{ g_{4W}}{\sqrt{2} }  \,  \left( 
\ba{cccc}  
     0  &   W_\mu^+  &   W_\mu^{\prime\, +}  &  W_\mu^{\prime\prime \, +}   \\ 
 W_\mu^-  &    &   &    \\ 
 W_\mu^{\prime\, -}   &     \multicolumn{3}{c}{  0_{3\times 3} }  \\
 W_\mu^{\prime\prime \, -}  &  &  &   \\   \ea  \right) + \frac{ g_{4W}}{\sqrt{2} } \,  \left( 
\ba{cccc}  
     0  & 0  &  0  &  0  \\ 
 0  &  0   &  N_\mu &  N_\mu^\prime  \\ 
0  &  \bar N_\mu   &  0 &  N_\mu^{\prime\prime}  \\
 0  &   \bar N_\mu^{\prime}  &  \bar N_\mu^{\prime\prime}  &  0  \\   \ea  \right)  \non
 &+& \frac{ g_{4W}  }{2} {\rm diag}  \Big(   A_\mu^3 + \frac{1}{\sqrt{3} } A_\mu^8 \,, - A_\mu^3 + \frac{1}{\sqrt{3} } A_\mu^8  \,,  - \frac{ 2 }{\sqrt{3} }  A_\mu^8  \,,  0 \Big) \non
&+& \frac{ g_{4W} }{2 \sqrt{6}} {\rm diag}  \Big(  \[   A_\mu^{15}  + 12 t_{\vartheta_G } \Xc_1 X_{1\,\mu} \] \mathbb{I}_{3\times 3} \,,   - 3  A_\mu^{15} +  12 t_{\vartheta_G }  \Xc_1 X_{1\,\mu}  \Big) \,.
\eeqn
Here, all diagonal gauge bosons are neutral, while the off-diagonal components contain both the electrically charged gauge bosons of $(W_\mu^\pm \,, W_\mu^{\prime\,\pm} \,, W_\mu^{ \prime\prime\,\pm} )$ as well as the neutral gauge bosons of $(N_\mu \,, N_\mu^{\prime} \,,N_\mu^{\prime\prime} )$.

\para
This stage of symmetry breaking is achieved by Higgs VEVs in Eq.~\eqref{eq:SU8_Higgs_VEVs_mini02}, and the Higgs kinematic terms lead to the $(W_\mu^{\prime\prime\, \pm} \,, N_\mu^\prime\,, N_\mu^{\prime\prime} )$ gauge boson masses of 
\beqn
&& \frac{1}{4} g_{4W}^2 v_{341}^2 \cdot   \Big( W_\mu^{\prime \prime +}  W^{\prime \prime -\, \mu} +N_\mu^\prime \bar N^{\prime \, \mu} +N_\mu^{\prime \prime } \bar N^{\prime \prime \, \mu}  \Big) \,, \quad v_{341}^2 = ( w_{ \repb{4} \,, {\rm V} } )^2 + ( w_{ \repb{4} \,, \dot 1 } )^2 + ( w_{ \repb{4} \,, \dot {\rm VII } } )^2  \,.
\eeqn
For the flavor-conserving neutral gauge bosons of $(A_\mu^{15}\,, X_{1\, \mu} )$, the mass squared matrix reads
\beqn
&& \frac{3 }{16  } g_{4W}^2 v_{341}^2  (   A^{15}_\mu \,, X_{1\,\mu}  )  \cdot \left( 
\ba{cc}  
 1   &  - t_{\vartheta_G}  \\ 
- t_{\vartheta_G}  & t_{\vartheta_G}^2   \\   \ea \right)  \cdot  \left( \ba{c}  A^{15\,\mu}  \\ X_{1}^\mu \\   \ea \right)\,.
\eeqn
Obviously it contains a zero eigenvalue which corresponds to the massless gauge boson of $X_{2\, \mu}$ after the ${\rm SU}(4)_W$ symmetry breaking.
The gauge eigenstates of $(  A^{15}_\mu  \,, X_{1\,\mu})$ are diagonalized as follows
\beqn
 \left( \ba{c}  Z^{\prime\prime}_\mu  \\ X_{2\,\mu}  \\    \ea \right) &=& \left( \ba{cc}  
   c_{\vartheta_{G}}  &   - s_{\vartheta_{G}} \\ 
  s_{\vartheta_{G}}  &  c_{\vartheta_{G}}  \\   \ea \right) \cdot  \left( \ba{c}  A^{15}_\mu  \\ X_{1\,\mu} \\     \ea \right)\,.
\eeqn

\para
The ${\rm SU}(4)_W \otimes {\rm U}(1)_{X_1}$ gauge couplings of $(\alpha_{4W}\,, \alpha_{X_1} )$ match with the ${\rm SU}(3 )_W \otimes {\rm U}(1)_{X_2}$ gauge couplings as follows
\beqn\label{eq:341_coupMatch}
&& \alpha_{3W }^{-1} (v_{341} ) =  \alpha_{4W }^{-1} (v_{341} )  \,,~ \alpha_{X_2}^{-1} (v_{341} )  = \frac{1}{6} \alpha_{4W }^{-1} (v_{341} )  +  \alpha_{X_1}^{-1} (v_{341} )  \,, \non
&& \frac{1}{6} \alpha_{4W }^{-1} = \alpha_{X_2}^{-1} s_{\vartheta_{G}}^2 \,,~ \alpha_{X_1}^{-1} = \alpha_{X_2}^{-1} c_{\vartheta_{G}}^2  \,.
\eeqn
According to the definitions of Eqs.~\eqref{eq:Glashow_angle} and \eqref{eq:Salam_angle}, we find a relation of
\beqn
&&\boxed{ \sin\vartheta_G = \frac{1}{ \sqrt{2} } \tan\vartheta_S } \,.
\eeqn
The tree-level masses for seven gauge bosons at this stage read
\beqn\label{eq:341_GB}
&& m_{W_\mu^{\prime\prime\, \pm} }^2 =  m_{ N_\mu^\prime\,, \bar N_\mu^\prime }^2   = m_{ N_\mu^{\prime\prime}\,, \bar N_\mu^{\prime\prime}  }^2   =  \frac{g_{X_2}^2}{ 4 s_{\vartheta_G}^2 }   v_{341}^2 \,, \quad  m_{Z_\mu^{\prime\prime} }^2 =  \frac{g_{X_2}^2}{16 s_{\vartheta_G}^2 c_{\vartheta_G}^2  }   v_{341}^2 \,.
\eeqn
The remaining massless gauge bosons are $(W_\mu^\pm\,, W_\mu^{\prime\, \pm}\,, N_\mu\,,\bar N_\mu )$ and $(A_\mu^3\,, A_\mu^8\,, X_{2\, \mu})$.

\para
In terms of mass eigenstates, the vector bosons from the covariant derivative in Eq.~\eqref{eq:341_connection_fund} becomes
\beqn\label{eq:341_connection_fundmass}
&& g_{4W} A^{\bar I}_\mu T_{ {\rm SU}(4)}^{\bar I} + g_{X_1} \Xc_1 \mathbb{I}_4 X_{1\,\mu}  \non
&=& \frac{ g_{4W}}{\sqrt{2} }  \,  \left( 
\ba{cccc}  
     0  &   W_\mu^+  &   W_\mu^{\prime\, +}  &  0   \\ 
 W_\mu^-  &  0  &  N_\mu  & 0   \\ 
 W_\mu^{\prime\, -}   &  \bar N_\mu  &  0 & 0  \\
 0  &  0 &  0  & 0  \\   \ea  \right) + \underline{  \frac{ g_{4W}}{\sqrt{2} } \,  \left( 
\ba{cccc}  
     0  & 0  &  0  &  W_\mu^{\prime\prime \, +}  \\ 
 0  &  0   &  0 &  N_\mu^\prime  \\ 
0  &  0   &  0 &  N_\mu^{\prime\prime}  \\
 W_\mu^{\prime\prime \, -}  &   \bar N_\mu^{\prime}  &  \bar N_\mu^{\prime\prime}  &  0  \\   \ea  \right) } \non
 &+& \frac{ g_{4W}  }{2} {\rm diag}  \Big(   A_\mu^3 + \frac{1}{\sqrt{3} } A_\mu^8 \,, - A_\mu^3 + \frac{1}{\sqrt{3} } A_\mu^8  \,,  - \frac{ 2 }{\sqrt{3} }  A_\mu^8  \,,  0 \Big) \non
&+&   g_{X_2}  {\rm diag}  \Big(   ( \frac{1}{12 } +  \Xc_1 ) \mathbb{I}_{3\times 3}  \,,  - \frac{1}{4} +  \Xc_1  \Big) X_{2\,\mu}  \non
&+&  \underline{ \frac{ g_{X_2}}{  s_{\vartheta_G} c_{\vartheta_G}  } {\rm diag}  \Big(  \[   \frac{1}{12} - (  \frac{1}{12} +  \Xc_1) s_{\vartheta_G}^2 \] \mathbb{I}_{3\times 3} \,,    - \frac{1}{4} +  ( \frac{1}{4} -  \Xc_1 ) s_{\vartheta_G}^2  \Big) Z_\mu^{\prime\prime } } \,.
\eeqn
We split terms with massless vector bosons and massive bosons (marked with underlines), respectively.
As a consistent check, the $(44)$-component in Eq.~\eqref{eq:341_connection_fundmass} is reduced to $-\frac{ g_{X_2} }{ 2 s_{\vartheta_G} c_{\vartheta_G} }  Z_\mu^{\prime\prime }$ when setting $\Xc_1=+\frac{1}{4}$ for the fundamental representation.
Likewise, we find the explicit form the gauge fields of $- g_{4W} A_\mu^{\bar I}  (T_{ {\rm SU}(4)}^{\bar I} )^T +g_{X_1} \Xc_1 \mathbb{I}_4 X_{1\,\mu}$ as follows
\beqn\label{eq:341_connection_antifundmass}
&& - g_{4W} A^{\bar I}_\mu ( T_{ {\rm SU}(4)}^{\bar I} )^T + g_{X_1} \Xc_1 \mathbb{I}_4 X_{1\,\mu}=  \non
&& - \frac{ g_{4W}}{\sqrt{2} }  \,  \left( 
\ba{cccc}  
     0  &   W_\mu^-  &   W_\mu^{\prime\, -}  &  0   \\ 
 W_\mu^+  &  0  & \bar N_\mu  & 0   \\ 
 W_\mu^{\prime\, +}   &   N_\mu  &  0 & 0  \\
 0  &  0 &  0  & 0  \\   \ea  \right) - \underline{  \frac{ g_{4W}}{\sqrt{2} } \,  \left( 
\ba{cccc}  
     0  & 0  &  0  &  W_\mu^{\prime\prime \, -}  \\ 
 0  &  0   &  0 & \bar N_\mu^\prime  \\ 
0  &  0   &  0 & \bar  N_\mu^{\prime\prime}  \\
 W_\mu^{\prime\prime \, +}  &   N_\mu^{\prime}  &  N_\mu^{\prime\prime}  &  0  \\   \ea  \right) } \non
 &-& \frac{g_{4W}}{2} {\rm diag}\,  \Big(    A_\mu^3 + \frac{1}{\sqrt{3} } A_\mu^8  \,,   - A_\mu^3 + \frac{1}{\sqrt{3} } A_\mu^8  \,, - \frac{ 2 }{\sqrt{3} }  A_\mu^8 \,, 0 \Big) \non
&+&  g_{X_2} {\rm diag}  \Big(   ( - \frac{1}{12 } +  \Xc_1 ) \mathbb{I}_{3\times 3}  \,,   \frac{1}{4} +  \Xc_1  \Big) X_{2\,\mu}  \non
&+&\underline{  \frac{ g_{X_2}}{  s_{\vartheta_G} c_{\vartheta_G}  } {\rm diag}  \Big(  \[  - \frac{1}{12 } + ( \frac{1}{12 } -  \Xc_1 ) s_{\vartheta_G}^2 \] \mathbb{I}_{3\times 3} \,,    \frac{1}{4} -  ( \frac{1}{4} +  \Xc_1 ) s_{\vartheta_G}^2  \Big) Z_\mu^{\prime\prime } } \,.
\eeqn

\subsubsection{Gauge couplings of the $\Gc_{341}$ fermions}

\para
For the anti-fundamental fermions of $\repb{8_F}^\Omega$, the covariant derivatives lead to the neutral currents of
\beqs
\beqn
&& i \overline \psi_{ ( \rep{3}\,, \rep{1}\,, -\frac{1}{3} )^\Omega }  \Dslash \psi_{ ( \rep{3}\,, \rep{1}\,, -\frac{1}{3} )^\Omega } \supset   \frac{ g_{X_2} }{ s_{\vartheta_G} c_{\vartheta_G}  } ( + \frac{1}{3} s_{\vartheta_G}^2 ) \overline{ \Dc_R^\Omega } \gamma^\mu \Dc_R^\Omega \cdot Z_\mu^{\prime \prime }  \,,\\[1mm]
&& i \overline \psi_{ ( \rep{1}\,, \repb{4}\,, -\frac{1}{4} )^\Omega }  \Dslash \psi_{ ( \rep{1}\,, \repb{4}\,, -\frac{1}{4} )^\Omega } \non
&\supset& \frac{ g_{X_2} }{ s_{\vartheta_G} c_{\vartheta_G}  } \Big[  (+ \frac{1}{4} )  \overline{ \check \Nc_L^{\Omega^{\prime \prime}} } \gamma^\mu \check \Nc_L^{ \Omega^{\prime\prime } }  + (- \frac{1}{12 } +  \frac{1}{3} s_{\vartheta_G}^2 ) ( \overline{ \Ec_L^\Omega} \gamma^\mu \Ec_L^\Omega + \overline{ \Nc_L^\Omega} \gamma^\mu \Nc_L^\Omega + \overline{ \check \Nc_L^{\Omega^\prime} } \gamma^\mu \check \Nc_L^{ \Omega^\prime } ) \Big]  \cdot Z_\mu^{\prime \prime }   \,.
\eeqn
\eeqs
For the rank-2 anti-symmetric fermionic representation of $( \rep{1}\,, \rep{6}\,, +\frac{1}{2} )_{ \mathbf{F} }$ from the $\rep{28_F}$, we first transform to its conjugate of $( \rep{1}\,, \rep{6}\,, -\frac{1}{2} )_{ \mathbf{F} }= ( \rep{1}\,, \rep{3}\,, -\frac{2}{3 } )_{ \mathbf{F} } \oplus ( \rep{1}\,, \repb{3}\,, -\frac{1}{3 } )_{ \mathbf{F} }$, with $( \rep{1}\,, \rep{3}\,, -\frac{2}{3 } )_{ \mathbf{F} } \equiv ( \nG_R^\prime \,, - \eG_R^\prime \,, \tau_R)^T$ and $( \rep{1}\,, \repb{3}\,, -\frac{1}{3 } )_{ \mathbf{F} } \equiv ( \eG_R \,, \nG_R \,, \check \nG_R)^T$.
Then, we denote the rank-$2$ anti-symmetric fermion of $( \rep{1}\,, \rep{6}\,, -\frac{1}{2} )_{ \mathbf{F} }$ as follows
\beqn\label{eq:16F_matrix}
\psi_{ ( \rep{1}\,, \rep{6}\,, -\frac{1}{2} ) } &\equiv& \frac{1}{ \sqrt{2} } \left( \ba{cccc}
 0  & \check \nG_R & -\nG_R & \nG_R^\prime  \\
 -\check \nG_R &  0  & \eG_R  & -\eG_R^\prime  \\
 \nG_R  & -\eG_R  &  0   & \tau_R  \\
 -\nG_R^\prime & \eG_R^\prime  & -\tau_R  & 0  \\
\ea \right) \,,
\eeqn
and its covariant derivative should be obtained according to Eq.~\eqref{eq:341_cov_r2antisym}.
The covariant derivatives lead to the neutral currents of
\beqs
\beqn
&& i \overline \psi_{ ( \rep{3}\,, \rep{1}\,, -\frac{1}{3} ) }   \Dslash \psi_{ ( \rep{3}\,, \rep{1}\,, -\frac{1}{3} ) }  \supset \frac{ g_{X_2} }{ s_{\vartheta_G} c_{\vartheta_G}  } ( + \frac{1}{3} s_{\vartheta_G}^2 ) \overline{ \DG_L } \gamma^\mu \DG_L \cdot Z_\mu^{\prime \prime }  \,,\\[1mm] 
&& i \overline \psi_{ ( \rep{3}\,, \rep{1}\,, +\frac{2}{3} ) }  \Dslash \psi_{ ( \rep{3}\,, \rep{1}\,, +\frac{2}{3} ) } \supset   \frac{ g_{X_2} }{ s_{\vartheta_G} c_{\vartheta_G}  }  ( - \frac{2}{3} s_{\vartheta_G}^2) \overline{ t_R} \gamma^\mu t_R  \cdot Z_\mu^{\prime \prime }  \,,\\[1mm]
&&\Tr \[  i \overline \psi_{ ( \rep{1}\,, \rep{6}\,, -\frac{1}{2} ) }  \Dslash \psi_{ ( \rep{1}\,, \rep{6}\,, -\frac{1}{2} ) } \] \supset \frac{ g_{X_2} }{ s_{\vartheta_G} c_{\vartheta_G}  } \Big[  ( \frac{1}{6} + \frac{1}{3}  s_{\vartheta_G}^2 ) ( \overline{ \check \nG_R  }  \gamma^\mu \check \nG_R +   \overline{ \nG_R  }  \gamma^\mu  \nG_R  +  \overline{ \eG_R  }  \gamma^\mu \eG_R  ) \non
&& + ( - \frac{1}{6} + \frac{2}{3}  s_{\vartheta_G}^2  ) (  \overline{ \nG_R^\prime  }  \gamma^\mu  \nG_R^\prime  +  \overline{ \eG_R^\prime  }  \gamma^\mu \eG_R^\prime +  \overline{ \tau_R  }  \gamma^\mu  \tau_R  )   \Big] \cdot Z_\mu^{\prime \prime } \,,\\[1mm]
&& i \overline \psi_{ ( \rep{3}\,, \rep{4}\,, -\frac{1}{12} ) }   \Dslash \psi_{ ( \rep{3}\,, \rep{4}\,, -\frac{1}{12} ) }  \supset   \frac{ g_{X_2} }{ s_{\vartheta_G} c_{\vartheta_G}  }  \Big[ (+\frac{1}{12} ) ( \overline{t_L } \gamma^\mu t_L + \overline{b_L } \gamma^\mu b_L + \overline{ \DG_L^\prime } \gamma^\mu \DG_L^\prime  )  \non
&+&  ( - \frac{1}{4}  + \frac{1}{3}  s_{\vartheta_G}^2  ) \overline{ \DG_L^{\prime \prime} } \gamma^\mu \DG_L^{\prime \prime}   \Big] \cdot Z_\mu^{\prime \prime }  \,,\\[1mm]
&& i \overline \psi_{ ( \rep{1}\,, \repb{4}\,, -\frac{1}{4} ) }  \Dslash \psi_{ ( \rep{1}\,, \repb{4}\,, -\frac{1}{ 4} ) } \non
&\supset &  \frac{ g_{X_2} }{ s_{\vartheta_G} c_{\vartheta_G}  }  \Big[   ( - \frac{1}{12 } +  \frac{1}{3} s_{\vartheta_G}^2 ) ( \overline{ \eG_R^{ \prime\prime} } \gamma^\mu \eG_R^{ \prime\prime} +  \overline{ \nG_R^{ \prime\prime} } \gamma^\mu \nG_R^{ \prime\prime}  + \overline{ \check \nG_R^{ \prime} } \gamma^\mu \check \nG_R^{ \prime}   )  +  (+ \frac{1}{4} )  \overline{ \check \nG_R^{\prime \prime} } \gamma^\mu \check \nG_R^{ \prime\prime  }  \Big] \cdot Z_\mu^{\prime \prime }   \,.
\eeqn
\eeqs

\para
For the rank-3 anti-symmetric fermions of $\rep{56_F}$, we denote the rank-$2$ anti-symmetric fermion of $( \rep{3}\,, \rep{6}\,, +\frac{1}{6} )_{ \mathbf{F} }$ and $( \rep{1}\,, \rep{6}\,, - \frac{1}{2} )_{ \mathbf{F} }^\prime$ as follows
\beqs
\beqn
\psi_{ ( \rep{3}\,, \rep{6}\,, +\frac{1}{6} ) } &\equiv& \frac{1}{ \sqrt{2} } \left( \ba{cccc}
 0  & \UG_L &  \uG_L & c_L  \\
- \UG_L &  0  & \dG_L  & s_L  \\
 - \uG_L &  -\dG_L &  0   & \DG_L^{ \prime \prime \prime}  \\
 -c_L & -s_L & -  \DG_L^{ \prime \prime \prime}  & 0  \\
\ea \right) \,,\label{eq:36F_matrix} \\[1mm]
\psi_{ ( \rep{1}\,, \rep{6}\,, -\frac{1}{2} )^\prime } &\equiv& \frac{1}{ \sqrt{2} }  \left( \ba{cccc}
 0  & \check \nG_R^{ \prime\prime \prime} & - \nG_R^{ \prime\prime \prime \prime} & \nG_R^{ \prime\prime \prime \prime\prime }  \\
 -\check \nG_R^{ \prime\prime \prime} &  0  & \eG_R^{ \prime\prime \prime \prime}  & -\eG_R^{ \prime\prime \prime \prime\prime }  \\
 \nG_R^{ \prime\prime \prime \prime}  & -\eG_R^{ \prime\prime \prime \prime}  &  0   & e_R  \\
 -\nG_R^{ \prime\prime \prime \prime\prime } & \eG_R^{ \prime\prime \prime \prime\prime }  & - e_R  & 0  \\
\ea \right) \,.\label{eq:16primeF_matrix}
\eeqn
\eeqs
Their covariant derivatives should be obtained according to Eq.~\eqref{eq:341_cov_r2antisym}.
The covariant derivatives from the $\rep{56_F}$ lead to the neutral currents of
\beqs
\beqn
&& i \overline \psi_{ ( \rep{1}\,, \rep{4}\,, -\frac{3}{4} ) }  \Dslash \psi_{ ( \rep{1}\,, \rep{4}\,, -\frac{3}{4} ) }  \supset \frac{ g_{X_2} }{ s_{\vartheta_G} c_{\vartheta_G}  } \Big[  ( - \frac{1}{4} +  s_{\vartheta_G}^2 ) \overline{ \EG_R } \gamma^\mu \EG_R  \non
&+& ( \frac{1}{12} + \frac{2}{3} s_{\vartheta_G}^2 ) (  \overline{ \nG_R^{ \prime\prime\prime} } \gamma^\mu \nG_R^{ \prime\prime\prime} +  \overline{ \eG_R^{ \prime\prime\prime} } \gamma^\mu \eG_R^{ \prime\prime\prime} + \overline{\mu_R} \gamma^\mu \mu_R  )  \Big]  \cdot Z_\mu^{\prime \prime }  \,,\\[1mm]
&& i \overline \psi_{ ( \rep{3}\,, \rep{1}\,, +\frac{2}{3} )^\prime }  \Dslash \psi_{ ( \rep{3}\,, \rep{1}\,, +\frac{2}{3} )^\prime }  \supset  \frac{ g_{X_2} }{ s_{\vartheta_G} c_{\vartheta_G}  } (  - \frac{2}{3} s_{\vartheta_G}^2 ) \overline{ u_R } \gamma^\mu u_R  \cdot Z_\mu^{\prime \prime } \,,\\[1mm]
&& i \overline \psi_{ ( \rep{1}\,, \rep{1}\,, -1 ) }  \Dslash  \psi_{ ( \rep{1}\,, \rep{1}\,, -1 ) } \supset \frac{ g_{X_2} }{ s_{\vartheta_G} c_{\vartheta_G}  } (  + s_{\vartheta_G}^2 ) \overline{ \EG_L } \gamma^\mu \EG_L  \cdot Z_\mu^{\prime \prime } \,,\\[1mm] 
&&  \Tr \[ i \overline \psi_{ ( \rep{3}\,, \rep{6}\,, +\frac{1}{6} )  }  \Dslash  \psi_{ ( \rep{3}\,, \rep{6}\,, +\frac{1}{6} )  } \]  \supset \frac{ g_{X_2} }{ s_{\vartheta_G} c_{\vartheta_G}  }  \Big[   ( \frac{1}{6} - \frac{1}{3 } s_{\vartheta_G}^2 ) ( \overline{ \UG_L } \gamma^\mu \UG_L +  \overline{ \uG_L } \gamma^\mu \uG_L  + \overline{ \dG_L } \gamma^\mu \dG_L    ) \non
&+& ( - \frac{1}{6} ) ( \overline{ c_L} \gamma^\mu c_L + \overline{ s_L} \gamma^\mu s_L  + \overline{ \DG_L^{\prime\prime \prime } } \gamma^\mu \DG_L^{\prime\prime \prime }   ) \Big] \cdot Z_\mu^{\prime \prime }  \,,\\[1mm]
&& \Tr \[ i \overline \psi_{ ( \rep{1}\,, \rep{6}\,, -\frac{1}{2} )^\prime  }  \Dslash  \psi_{ ( \rep{1}\,, \rep{6}\,, -\frac{1}{2} )^\prime  } \]  \supset \frac{ g_{X_2} }{ s_{\vartheta_G} c_{\vartheta_G}  }  \Big[   ( \frac{1}{6} + \frac{1}{3} s_{\vartheta_G}^2   )  ( \overline{ \check \nG_R^{ \prime\prime\prime } } \gamma^\mu \check \nG_R^{ \prime\prime\prime } + \overline{ \nG_R^{ \prime \prime \prime \prime } } \gamma^\mu \nG_R^{ \prime \prime \prime \prime } +  \overline{ \eG_R^{ \prime \prime \prime \prime } } \gamma^\mu \eG_R^{ \prime \prime \prime \prime }  )  \non
&& + ( - \frac{1}{6} + \frac{2}{3} s_{\vartheta_G}^2  )  ( \overline{ \nG_R^{ \prime \prime \prime \prime \prime }  } \gamma^\mu \nG_R^{ \prime \prime \prime \prime \prime } + \overline{ \eG_R^{ \prime \prime \prime \prime \prime } } \gamma^\mu \eG_R^{ \prime \prime \prime \prime \prime }  + \overline{ e_R } \gamma^\mu e_R  )  \Big] \cdot Z_\mu^{\prime \prime } \,,\\[1mm]
&& i \overline \psi_{ ( \rep{3}\,, \rep{4}\,, -\frac{1}{12} )^\prime  }  \Dslash  \psi_{ ( \rep{3}\,, \rep{4}\,, -\frac{1}{12} )^\prime  } \supset  \frac{ g_{X_2} }{ s_{\vartheta_G} c_{\vartheta_G}  }  \Big[ (+ \frac{1}{12} ) ( \overline{u_L} \gamma^\mu u_L   + \overline{ d_L} \gamma^\mu d_L + \overline{ \DG_L^{\prime \prime \prime \prime } } \gamma^\mu  \DG_L^{\prime \prime \prime \prime }    )   \non
&+& ( - \frac{1}{4} + \frac{1}{3}  s_{\vartheta_G}^2) \overline{ \DG_L^{ \prime \prime \prime \prime \prime }} \gamma^\mu \DG_L^{ \prime \prime \prime \prime \prime }  \Big] \cdot Z_\mu^{\prime \prime }   \,,\\[1mm]
&& i \overline \psi_{ ( \rep{3}\,, \repb{4}\,, +\frac{5}{12} )  }  \Dslash \psi_{ ( \rep{3}\,, \repb{4}\,, +\frac{5}{12} )  }   \supset  \frac{ g_{X_2} }{ s_{\vartheta_G} c_{\vartheta_G}  }  \Big[   ( - \frac{1}{ 12} - \frac{1}{3}  s_{\vartheta_G}^2   ) ( \overline{ \dG_R} \gamma^\mu \dG_R  + \overline{ \uG_R} \gamma^\mu \uG_R +  \overline{ \UG_R} \gamma^\mu \UG_R  )   \non
&+&  (  \frac{1}{4} - \frac{2}{3}  s_{\vartheta_G}^2 ) \overline{ c_R} \gamma^\mu c_R \Big] \cdot Z_\mu^{\prime \prime } \,.
\eeqn
\eeqs

\begin{table}[htp]
\begin{center}
\begin{tabular}{c|cccccc}
\hline \hline
  & $u$   & $c$ &  $t$  &  $\uG$  &  $\UG$    &   \\ 
\hline
$g_f^{V\,\prime\prime }$  &  $ \frac{1}{24} - \frac{1}{3} s_{\vartheta_G}^2 $  &  $ \frac{1}{24} - \frac{1}{3} s_{\vartheta_G}^2 $  &  $ \frac{1}{24} - \frac{1}{3} s_{\vartheta_G}^2$  &  $ \frac{1}{24} - \frac{1}{3} s_{\vartheta_G}^2$  &  $ \frac{1}{24} - \frac{1}{3} s_{\vartheta_G}^2$   &     \\
$g_f^{A\,\prime\prime }$  & $ - \frac{1}{24} - \frac{1}{3} s_{\vartheta_G}^2$  & $ \frac{ 5 }{24} - \frac{1}{3} s_{\vartheta_G}^2 $  &  $-\frac{1}{24} - \frac{1}{3} s_{\vartheta_G}^2$  & $-\frac{1}{8} $  & $-\frac{1}{8} $   &    \\
\hline
  & $d$  & $s$  &  $b$  &  $\dG$  &    &   \\ 
  \hline
$g_f^{V\,\prime\prime }$  &  $ \frac{1}{24 } + \frac{1}{6} s_{\vartheta_G}^2 $  & $ - \frac{1}{12 } + \frac{1}{6} s_{\vartheta_G}^2$ &  $ \frac{1}{24 } + \frac{1}{6} s_{\vartheta_G}^2$  &  $ \frac{1}{24} - \frac{1}{3} s_{\vartheta_G}^2$  &   &   \\ 
$g_f^{A\,\prime\prime }$  &  $-\frac{1}{24 } + \frac{1}{6} s_{\vartheta_G}^2$  & $ \frac{1}{12 } + \frac{1}{6} s_{\vartheta_G}^2$  &  $ -\frac{1}{24 } + \frac{1}{6} s_{\vartheta_G}^2$ &  $ -\frac{1}{8}$ &    &   \\ 
\hline
  & $\DG$  & $\DG^{ \prime \prime }$  &  $\DG^{ \prime }$  &  $\DG^{ \prime \prime \prime \prime \prime }$  &  $\DG^{ \prime \prime\prime }$ & $\DG^{ \prime \prime \prime \prime }$  \\ 
  \hline
$g_f^{V\,\prime\prime }$  &  $ \frac{1}{3} s_{\vartheta_G}^2  $  & $ -\frac{1}{8 } + \frac{1}{3} s_{\vartheta_G}^2 $ &  $\frac{1}{24 } + \frac{1}{6} s_{\vartheta_G}^2 $  &  $-\frac{1}{8 } + \frac{1}{3} s_{\vartheta_G}^2  $   &  $-\frac{1}{12 } + \frac{1}{6} s_{\vartheta_G}^2 $  &  $ \frac{1}{24 } + \frac{1}{6} s_{\vartheta_G}^2$ \\ 
$g_f^{A\,\prime\prime }$  &  $ 0$  & $\frac{1}{8 } $  &  $ -\frac{1}{24 } + \frac{1}{6} s_{\vartheta_G}^2$ &  $ \frac{1}{8 } $  &  $ \frac{1}{12 } + \frac{1}{6} s_{\vartheta_G}^2 $  &  $ -\frac{1}{24 } + \frac{1}{6} s_{\vartheta_G}^2$ \\ 
\hline
  & $e$  & $\mu$   &  $\tau$    &  $\EG$  &   &     \\ 
  \hline
$g_f^{V\,\prime\prime }$  & $-\frac{1}{8 } + \frac{1}{2} s_{\vartheta_G}^2 $   & $ \frac{1}{2} s_{\vartheta_G}^2$   &  $ -\frac{1}{8 } + \frac{1}{2} s_{\vartheta_G}^2$   &  $ -\frac{1}{8 } +  s_{\vartheta_G}^2$  &    &   \\ 
$g_f^{A\,\prime\prime }$  & $ -\frac{1}{24 } + \frac{1}{6} s_{\vartheta_G}^2$   & $ \frac{1}{12 } + \frac{1}{6} s_{\vartheta_G}^2$   &  $ -\frac{1}{24 } + \frac{1}{6} s_{\vartheta_G}^2$   &  $ -\frac{1}{8 }$  &   &  \\ 
\hline
  & $(\eG\,,\nG)$  & $(\eG^{ \prime } \,, \nG^{ \prime })$  &  $(\eG^{ \prime \prime } \,, \nG^{ \prime \prime })$  &  $(\eG^{ \prime \prime\prime } \,, \nG^{ \prime \prime \prime })$  &  $(\eG^{ \prime \prime\prime \prime } \,,  \nG^{ \prime \prime \prime \prime })$ & $(\eG^{ \prime \prime\prime \prime \prime } \,, \nG^{ \prime \prime\prime \prime \prime })$  \\ 
  \hline
$g_f^{V\,\prime\prime }$  &  $\frac{1}{24 } + \frac{1}{3} s_{\vartheta_G}^2 $  & $-\frac{1}{8 } + \frac{1}{2} s_{\vartheta_G}^2 $ &  $-\frac{1}{12 } + \frac{1}{3} s_{\vartheta_G}^2 $  &  $  \frac{1}{2} s_{\vartheta_G}^2$   &  $ \frac{1}{24 } + \frac{1}{3} s_{\vartheta_G}^2$ &  $ -\frac{1}{8 } + \frac{1}{2} s_{\vartheta_G}^2$ \\ 
$g_f^{A\,\prime\prime }$  &  $\frac{1}{8 }$  & $ -\frac{1}{24 } + \frac{1}{6} s_{\vartheta_G}^2$  &  $ 0$ &  $ \frac{1}{12 } + \frac{1}{6} s_{\vartheta_G}^2$  &  $ \frac{1}{8 }$   &  $ -\frac{1}{24 } + \frac{1}{6} s_{\vartheta_G}^2$ \\ 
\hline
  & $\nu_{e\,L}/ \nu_{\mu\,L} /\nu_{\tau\,L}$ & $\check \Nc_L^{ \Omega\,, \Omega^\prime\,, \Omega^{ \prime \prime } }$ & $\check \nG$  & $\check \nG^{ \prime  }$  &  $\check \nG^{ \prime\prime  }$  &  $\check \nG^{ \prime \prime \prime }$    \\ 
  \hline
$g_f^{V\,\prime\prime }$  & $-\frac{1}{24 } + \frac{1}{6} s_{\vartheta_G}^2$ &  $-\frac{1}{24 } + \frac{1}{6} s_{\vartheta_G}^2$ & $ \frac{1}{24} + \frac{1}{3} s_{ \vartheta_G }^2 $  & $ - \frac{1}{12} + \frac{1}{3} s_{ \vartheta_G }^2 $ &  $ \frac{1}{4}$  &  $\frac{1}{24} + \frac{1}{3} s_{ \vartheta_G }^2 $    \\ 
$g_f^{A\,\prime\prime }$  & $\frac{1}{24 } - \frac{1}{6} s_{\vartheta_G}^2$ & $\frac{1}{24 } - \frac{1}{6} s_{\vartheta_G}^2$ & $ \frac{1}{8}$  & $0 $  &  $ 0$ &  $ \frac{1}{8}$   \\ 
\hline\hline
\end{tabular}
\end{center}
\caption{
The couplings of the neutral currents mediated by $Z_\mu^{\prime\prime}$ in the $V-A$ basis.
}
\label{tab:Zpp_VA}
\end{table}%

\para
The neutral currents are expressed as follows in the $V-A$ basis
\beqn\label{eq:SU4W_ZppVA_coup}
\Lc_{ {\rm SU}(4)_W}^{\rm NC\,, F}&=& \frac{g_{X_2}}{ s_{\vartheta_G } c_{\vartheta_G} } ( g_f^{V\,\prime\prime } \overline f \gamma^\mu f + g_f^{A\, \prime \prime } \overline f \gamma^\mu \gamma_5 f ) Z_\mu^{\prime\prime } \,,
\eeqn
and we tabulate the vectorial and axial couplings of $( g_f^{V\,\prime\prime } \,, g_f^{A\,\prime\prime } )$ in Tab.~\ref{tab:Zpp_VA}. 
The flavor non-universality of the neutral currents mediated by $Z_\mu^{\prime\prime}$ among three generational SM quarks and leptons are displayed explicitly. 
The heavy partner fermions of $( \DG \,, \eG^{ \prime \prime } \,, \nG^{ \prime \prime } \,, \check \nG^\prime \,, \check \nG^{ \prime \prime} )$ only exhibit vectorial couplings, since they have already obtained their masses at the first stage of symmetry breaking.

\subsection{The ${\rm SU}(3)_W \otimes {\rm U}(1)_{ X_2}$ gauge bosons}
\label{section:331_bosons}

\subsubsection{Covariant derivatives and gauge boson masses}

\para
We express the ${\rm SU}(3)_W \otimes {\rm U}(1)_{X_2}$ covariant derivatives for the ${\rm SU}(3)_W $ fundamental and anti-fundamental representations as follows
\beqs\label{eqs:331_covariant}
\beqn
i D_\mu \psi_{ \rep{3}} &\equiv& i \partial_\mu \psi_{ \rep{3}}  +  ( g_{3W} A_\mu^{\tilde I}  T_{ {\rm SU}(3)}^{\tilde I} +  g_{X_2} \Xc_2 \mathbb{I}_3  X_{2\,\mu} ) \cdot  \psi_{ \rep{3}} \,,\label{eq:331_covariant_fund} \\[1mm]
i D_\mu \psi_{ \repb{3}}  &\equiv&i \partial_\mu \psi_{ \repb{3}}  +( - g_{3W} A_\mu^{\tilde I} ( T_{ {\rm SU}(3)}^{\tilde I} )^T + g_{X_2} \Xc_2 \mathbb{I}_3  X_{2\,\mu} ) \cdot \psi_{\repb{3}} \,. \label{eq:331_covariant_antifund}
\eeqn
\eeqs
The ${\rm SU}(3)_W$ generators are normalized such that ${\rm Tr}\Big(  T_{ {\rm SU}(3)}^{\tilde I} T_{ {\rm SU}(3)}^{\tilde J }  \Big)= \hf \delta^{\tilde I  \tilde J}$.
The gauge fields from Eq.~\eqref{eq:331_covariant_fund} can be expressed in terms of a $3\times 3$ matrix 
\beqn\label{eq:331_connection_gauge}
&& g_{3W} A^{\tilde I}_\mu  T_{ {\rm SU}(3)}^{\tilde I} + g_{X_2} \Xc_2 \mathbb{I}_3  X_{2\,\mu}  \non
&=&  \frac{g_{3W}}{ \sqrt{2} }\left( 
\ba{ccc}  
     0  &   W_\mu^+  &   W_\mu^{\prime\, +}  \\ 
 W_\mu^- &  0   &  0  \\ 
 W_\mu^{\prime\, -}   &  0 &  0   \\  \ea  \right)  + \frac{g_{3W}}{ \sqrt{2} }\left( 
\ba{ccc}  
     0  &  0  &  0  \\ 
 0 &  0   &  N_\mu  \\ 
 0   &  \bar N_\mu  &  0   \\  \ea  \right)  \non
 &+& \frac{ g_{3W} }{2} {\rm diag}  \Big(    A_\mu^3 + \frac{1}{\sqrt{3} } A_\mu^8  + 2 \sqrt{3} t_{\theta_S} \Xc_2 X_{2\,\mu} \,,  \non
&&  - A_\mu^3 + \frac{1}{\sqrt{3} } A_\mu^8  + 2 \sqrt{3} t_{\vartheta_S} \Xc_2 X_{2\,\mu} \,, - \frac{ 2 }{\sqrt{3} }  A_\mu^8  + 2 \sqrt{3} t_{\vartheta_S} \Xc_2 X_{2\,\mu} \Big)  \,,
\eeqn
with an ${\rm SU}(3)_W$ mixing angle defined by
\beqn\label{eq:Salam_angle}
t_{\vartheta_S} &\equiv& \tan\vartheta_{S} = \frac{g_{X_2} }{ \sqrt{3} g_{3W } }\,.
\eeqn
One can identify that 
\beqn
&& W_\mu^\pm = \frac{1}{ \sqrt{2} } ( A_\mu^1 \mp i A_\mu^2 ) \,,\quad W_\mu^{\prime\, \pm} = \frac{1}{ \sqrt{2} } ( A_\mu^4 \mp i A_\mu^5 ) \,, \non
&& N_\mu = \frac{1}{ \sqrt{2} } ( A_\mu^6 - i A_\mu^7 ) \,,\quad \bar N_\mu = \frac{1}{ \sqrt{2} } ( A_\mu^6 + i A_\mu^7 ) \,.
\eeqn
%
%
%
%

\para
This stage of symmetry breaking is achieved by Higgs VEVs in Eq.~\eqref{eq:SU8_Higgs_VEVs_mini03}, and the corresponding gauge boson masses from the Higgs kinematic terms are
\beqn
&& \frac{1}{4} g_{3W}^2  v_{331}^2 \Big[   \Big(  W_\mu^{\prime\,+} W^{\prime\,-\,\mu} + N_\mu \bar N^\mu \Big) +  \frac{2}{3}  \Big( A_\mu^8 - t_{\vartheta_S } X_{2\,\mu} \Big)^2 \Big] \,, \non
&&  v_{331}^2 = ( V_{ \repb{3} \,, 3 } )^2 + ( V_{ \repb{3} \,, {\rm VI} } )^2 + ( V_{ \repb{3} \,, \dot {\rm IX} } )^2  + ( V_{ \repb{3} \,, \dot 2 }^\prime )^2 + ( V_{ \repb{3} \,, \dot {\rm VIII} }^\prime )^2  \,.
\eeqn
It is straightforward to obtain the mass eigenstates in terms of the mixing angle in Eq.~\eqref{eq:Salam_angle} for this case
\beqn
 \left( \ba{c}  Z^{\prime}_\mu  \\ B_{\mu}  \\  \ea \right) &=& \left( \ba{cc}  
   c_{\vartheta_{S}}  &  - s_{\vartheta_{S}}  \\ 
  s_{\vartheta_{S}}   &  c_{\vartheta_{S}}    \\   \ea \right) \cdot  \left( \ba{c}  A^{8}_\mu  \\ X_{2\,\mu} \\     \ea \right) \,.
\eeqn
The ${\rm SU}(3)_W \otimes {\rm U}(1)_{X_2}$ gauge  couplings of $(\alpha_{3W}\,, \alpha_{X_1} )$ match with the EW gauge couplings as follows
\beqn\label{eq:331_coupMatch}
&&  \alpha_{2W }^{-1} (v_{331} ) =  \alpha_{3W }^{-1} (v_{331} ) \,, ~ \alpha_Y^{-1} (v_{331} ) = \frac{1}{3} \alpha_{3W }^{-1} (v_{331} ) +  \alpha_{X_2}^{-1} ( v_{331} ) \,, \non
&& \frac{1}{3} \alpha_{3W}^{-1} = \alpha_Y^{-1} s_{\vartheta_{S} }^2 \,,~ \alpha_{X_2}^{-1} = \alpha_Y^{-1} c_{\vartheta_{S} }^2 \,.
\eeqn
From the definitions of two mixing angles in Eqs.~\eqref{eq:Salam_angle} and \eqref{eq:Weinberg_angle}, we find a relation of
\beqn
&& \boxed{ \sin \vartheta_S = \frac{1}{ \sqrt{3}} \tan\vartheta_W } \,,
\eeqn
from the gauge coupling matching in Eq.~\eqref{eq:331_coupMatch}.
The tree-level masses for five gauge bosons at this stage read
\beqn\label{eq:331_GB}
&& m_{W_\mu^{\prime\, \pm} }^2    = m_{ N_\mu\,, \bar N_\mu  }^2   =  \frac{ g_Y^2 }{ 12 s_{\vartheta_S}^2 }   v_{331}^2 \,, \quad   m_{Z_\mu^{\prime} }^2 = \frac{g_{X_1}^2}{ 9 s_{\vartheta_S}^2 c_{\vartheta_S}^2 }  v_{331}^2 \,.
\eeqn

\para
In terms of mass eigenstates, the gauge bosons from the covariant derivative in Eq.~\eqref{eq:331_connection_gauge} become
\beqn\label{eq:331_connection_fund}
&& \frac{g_{3W}}{ \sqrt{2} } \left( 
\ba{ccc}  
 0  & W_\mu^+ & 0  \\ 
W_\mu^-  &  0 & 0 \\ 
0  & 0 &  0  \\  \ea  \right)   + \underline{ \frac{g_{3W}}{ \sqrt{2} } \left( 
\ba{ccc}  
 0  & 0  & W_\mu^{\prime +}  \\ 
0  &  0 &  N_\mu  \\ 
W_\mu^{\prime\, -}   &  \bar N_\mu  &  0  \\  \ea  \right) } \non
&+&  g_{3W} {\rm diag}  \Big( \hf \,, - \hf \,,  0 \Big) A_\mu^3 + g_Y {\rm diag}  \Big( ( \frac{1}{6} + \Xc_2 ) \mathbb{I}_{2\times 2}  \,, -\frac{1}{3} +  \Xc_2 \Big) B_\mu \non
&+& \underline{ \frac{g_Y}{ s_{\vartheta_S}  c_{\vartheta_S}  }   {\rm diag}  \Big(  \[ \frac{1}{6} -  ( \frac{1}{6} +  \Xc_2 ) s_{\vartheta_S}^2 \] \mathbb{I}_{2\times 2}  \,, - \frac{1}{3} + (\frac{1}{3} -  \Xc_2 ) s_{ \vartheta_S }^2 \Big) Z_\mu^\prime } \,.
\eeqn
Likewise, the covariant derivative for the ${\rm SU}(3)_W$ anti-fundamental representation gives
\beqn\label{eq:331_connection_antifund}
&& - \frac{g_{3W}}{ \sqrt{2} } \left( 
\ba{ccc}  
 0  & W_\mu^- & 0  \\ 
W_\mu^+  &  0 & 0 \\ 
0 &  0  &  0  \\  \ea  \right)  - \underline{ \frac{g_{3W}}{ \sqrt{2} } \left( 
\ba{ccc}  
 0  & 0  & W_\mu^{\prime\,-}  \\ 
0  &  0 & \bar N_\mu  \\ 
W_\mu^{\prime\, +}  & N_\mu  &  0  \\  \ea  \right) } \non 
&-&   g_{3W}  {\rm diag}\, \Big( \hf \,, - \hf \,,  0 \Big) A_\mu^3  + g_Y {\rm diag}\,\Big( (-\frac{1}{6} + \Xc_1 ) \mathbb{I}_{2\times 2} \,,  \frac{1}{3} + \Xc_1 \Big) B_\mu \non
&+& \underline{ \frac{g_Y}{ s_{\vartheta_S}  c_{\vartheta_S}  } {\rm diag} \Big( \[ - \frac{1}{6} +  ( \frac{1}{6} - \Xc_1 ) s_{\vartheta_S}^2 \] \mathbb{I}_{2 \times 2}  \,, \frac{1}{3} - (\frac{1}{3} +  \Xc_1 ) s_{ \vartheta_S }^2 \Big) Z_\mu^\prime } \,.
\eeqn

\subsubsection{Gauge couplings of the $\Gc_{331}$ fermions}

\para
From the anti-fundamental fermions of $\repb{8_F}^\Omega$, we find the neutral currents of
\beqs
\beqn
&& i \overline \psi_{ ( \rep{3}\,, \rep{1}\,, -\frac{1}{3} )^\Omega }  \Dslash \psi_{ ( \rep{3}\,, \rep{1}\,, -\frac{1}{3} )^\Omega }  \supset  \frac{ g_Y }{ s_{ \vartheta_S} c_{ \vartheta_S} } ( + \frac{1}{3} s_{ \vartheta_S}^2 ) \overline{ \Dc_R^\Omega } \gamma^\mu \Dc_R^\Omega  \cdot Z_\mu^\prime \,, \\[1mm] 
&& i \overline \psi_{ ( \rep{1}\,, \repb{3}\,, -\frac{1}{3} )^\Omega }  \Dslash \psi_{ ( \rep{1}\,, \repb{3}\,, -\frac{1}{3} )^\Omega } \non
&\supset&  \frac{ g_Y }{ s_{ \vartheta_S} c_{ \vartheta_S} } \Big[ ( - \frac{1}{6} + \frac{1}{2}  s_{ \vartheta_S}^2 ) ( \overline{ \Ec_L^\Omega } \gamma^\mu \Ec_L^\Omega + \overline{ \Nc_L^\Omega } \gamma^\mu \Nc_L^\Omega )  +  (+ \frac{1}{3} ) \overline{ \check \Nc_L^{\Omega^\prime}  } \gamma^\mu \check  \Nc_L^{\Omega^\prime}  \Big] \cdot Z_\mu^\prime \,.
\eeqn
\eeqs
%
%
%
%
%
%
From the rank-$2$ anti-symmetric fermion of $\rep{28_F}$, we find the neutral currents of
\beqs
\beqn
&& i \overline \psi_{ ( \rep{3}\,, \rep{1}\,, -\frac{1}{3} ) } \Dslash \psi_{ ( \rep{3}\,, \rep{1}\,, -\frac{1}{3} ) }  \supset \frac{ g_Y }{ s_{ \vartheta_S} c_{ \vartheta_S} } ( + \frac{1}{3} s_{ \vartheta_S}^2 ) \overline{ \DG_L } \gamma^\mu \DG_L   \cdot Z_\mu^\prime \,, \\[1mm]
&& i \overline \psi_{ ( \rep{3}\,, \rep{1}\,, +\frac{2}{3} ) }  \Dslash \psi_{ ( \rep{3}\,, \rep{1}\,, +\frac{2}{3} ) }  \supset \frac{ g_Y }{ s_{ \vartheta_S} c_{ \vartheta_S} } ( - \frac{2}{3} s_{ \vartheta_S}^2 ) \overline{ t_R } \gamma^\mu t_R \cdot Z_\mu^\prime \,, \\[1mm]
&& i \overline \psi_{ ( \rep{1}\,, \repb{3}\,, -\frac{1}{3} ) } \Dslash \psi_{ ( \rep{1}\,, \repb{3}\,, - \frac{1}{3} ) }  \non
&\supset & \frac{ g_Y }{ s_{ \vartheta_S} c_{ \vartheta_S} } \Big[ ( - \frac{1}{6} + \frac{1}{2} s_{ \vartheta_S}^2   ) ( \overline{ \eG_R } \gamma^\mu  \eG_R +  \overline{ \nG_R } \gamma^\mu  \nG_R  ) + (+\frac{1}{3} ) \overline{\check \nG_R } \gamma^\mu \check \nG_R \Big] \cdot Z_\mu^\prime \,, \\[1mm]
&& i \overline \psi_{ ( \rep{1}\,, \rep{3}\,, -\frac{2}{3} ) }  \Dslash  \psi_{ ( \rep{1}\,, \rep{3}\,, - \frac{2}{3} ) }  \non
&\supset & \frac{ g_Y }{ s_{ \vartheta_S} c_{ \vartheta_S} }  \Big[  ( \frac{1}{6} + \frac{1}{2} s_{ \vartheta_S}^2) (  \overline{ \eG_R^\prime } \gamma^\mu  \eG_R^\prime +  \overline{ \nG_R^\prime } \gamma^\mu  \nG_R^\prime ) + (- \frac{1}{3} + s_{ \vartheta_S}^2) \overline{ \tau_R} \gamma^\mu \tau_R   \Big] \cdot Z_\mu^\prime \,, \\[1mm]
&& i \overline \psi_{ ( \rep{3}\,, \rep{3}\,, 0 ) }  \Dslash \psi_{ ( \rep{3}\,, \rep{3}\,, 0 ) } \non
& \supset & \frac{ g_Y }{ s_{ \vartheta_S} c_{ \vartheta_S} } \Big[  \frac{1}{6} ( 1 - s_{\vartheta_S}^2 ) ( \overline{t_L} \gamma^\mu t_L + \overline{b_L} \gamma^\mu b_L   ) + \frac{1}{3} ( -1 + s_{\vartheta_S}^2 ) \overline{ \DG_L^\prime } \gamma^\mu \DG_L^\prime \Big] \cdot Z_\mu^\prime \,, \\[1mm]
&& i \overline \psi_{ ( \rep{3}\,, \rep{1}\,, -\frac{1}{3} )^{ \prime \prime} } \Dslash \psi_{ ( \rep{3}\,, \rep{1}\,, -\frac{1}{3} )^{ \prime \prime} }  \supset \frac{ g_Y }{ s_{ \vartheta_S} c_{ \vartheta_S} } (+ \frac{1}{3} s_{ \vartheta_S }^2 )  \overline{ \DG_L^{\prime\prime } } \gamma^\mu \DG_L^{\prime\prime } \cdot Z_\mu^\prime \,, \\[1mm]
&& i \overline \psi_{ ( \rep{1}\,, \repb{3}\,, -\frac{1}{3} )^{ \prime \prime} }  \Dslash \psi_{ ( \rep{1}\,, \repb{3}\,, -\frac{1}{3} )^{ \prime \prime} } \non
& \supset& \frac{ g_Y }{ s_{ \vartheta_S} c_{ \vartheta_S} } \Big[ (- \frac{1}{6} + \frac{1}{2} s_{ \vartheta_S }^2  ) ( \overline{ \eG_R^{\prime\prime } } \gamma^\mu \eG_R^{\prime\prime } + \overline{ \nG_R^{\prime\prime } } \gamma^\mu \nG_R^{\prime\prime }  ) + (+ \frac{1}{3} ) \overline{  \check \nG_R^\prime } \gamma^\mu \check \nG_R^\prime   \Big] \cdot Z_\mu^\prime  \,.
\eeqn
\eeqs
From the rank-$3$ anti-symmetric fermion of $\rep{56_F}$, we find the neutral currents of
\beqs
\beqn
&& i \overline \psi_{ ( \rep{1}\,, \rep{3}\,, -\frac{2}{3} )^\prime } \Dslash \psi_{ ( \rep{1}\,, \rep{3}\,, -\frac{2}{3} )^\prime }  \non
&\supset& \frac{ g_Y }{ s_{ \vartheta_S} c_{ \vartheta_S} }  \Big[   ( \frac{1}{6} + \frac{1}{2} s_{ \vartheta_S }^2 ) (  \overline{\eG_R^{ \prime\prime \prime }  } \gamma^\mu \eG_R^{ \prime \prime\prime }  +  \overline{\nG_R^{ \prime\prime  \prime}  } \gamma^\mu \nG_R^{ \prime \prime \prime} ) + ( - \frac{1}{3} +  s_{ \vartheta_S }^2 ) \overline{ \mu_R } \gamma^\mu \mu_R \Big] \cdot Z_\mu^\prime \,,\\[1mm]
&& i \overline \psi_{ ( \rep{1}\,, \rep{1}\,, -1 )^{\prime \prime} } \Dslash \psi_{ ( \rep{1}\,, \rep{1}\,, -1 )^{\prime \prime} }  \supset \frac{ g_Y }{ s_{ \vartheta_S} c_{ \vartheta_S} }  (+ s_{ \vartheta_S}^2 ) \overline{ \EG_R } \gamma^\mu \EG_R \cdot Z_\mu^\prime \,,\\[1mm]
&& i \overline \psi_{ ( \rep{3}\,, \rep{1}\,, +\frac{2}{3} )^{\prime} } \Dslash \psi_{ ( \rep{3}\,, \rep{1}\,, +\frac{2}{3} )^{\prime} }  \supset \frac{ g_Y }{ s_{ \vartheta_S} c_{ \vartheta_S} } (-\frac{2}{3} s_{ \vartheta_S}^2 ) \overline{ u_R } \gamma^\mu u_R  \cdot Z_\mu^\prime \,,\\[1mm]
&& i \overline \psi_{ ( \rep{1}\,, \rep{1}\,, -1 ) } \Dslash  \psi_{ ( \rep{1}\,, \rep{1}\,, -1 ) }  \supset \frac{ g_Y }{ s_{ \vartheta_S} c_{ \vartheta_S} } (+ s_{ \vartheta_S}^2 ) \overline{ \EG_L} \gamma^\mu \EG_L  \cdot Z_\mu^\prime \,,\\[1mm]
&& i \overline \psi_{ ( \rep{3}\,, \rep{3}\,, 0 )^{ \prime} } \Dslash \psi_{ ( \rep{3}\,, \rep{3}\,, 0 )^{ \prime } }   \supset  \frac{ g_Y }{ s_{ \vartheta_S} c_{ \vartheta_S} }  \Big[ \frac{1}{6} ( 1 - s_{\vartheta_S}^2 ) ( \overline{c_L } \gamma^\mu c_L  + \overline{ s_L } \gamma^\mu s_L ) \non
&+&  \frac{1}{3} ( -1 + s_{\vartheta_S}^2 ) \overline{ \DG_L^{\prime\prime \prime } } \gamma^\mu  \DG_L^{\prime \prime \prime} \Big]   \cdot Z_\mu^\prime \,,\\[1mm]
&& i \overline \psi_{ ( \rep{3}\,, \repb{3}\,, +\frac{1}{3} ) } \Dslash \psi_{ ( \rep{3}\,, \repb{3}\,, +\frac{1}{3} ) }  \non
&\supset & \frac{ g_Y }{ s_{ \vartheta_S} c_{ \vartheta_S} }  \Big[  - \frac{1}{6}  ( 1+ s_{ \vartheta}^2 ) ( \overline{\uG_L } \gamma^\mu \uG_L  + \overline{\dG_L } \gamma^\mu \dG_L  )  + ( \frac{1}{3} - \frac{2}{3} s_{ \vartheta}^2 ) \overline{ \UG_L } \gamma^\mu \UG_L \Big]   \cdot Z_\mu^\prime \,,\\[1mm]
&& i \overline \psi_{ ( \rep{1}\,, \repb{3}\,, -\frac{1}{3} )^\prime } \Dslash \psi_{ ( \rep{1}\,, \repb{3}\,,  -\frac{1}{3} )^\prime }  \non
&\supset& \frac{ g_Y }{ s_{ \vartheta_S} c_{ \vartheta_S} }  \Big[  ( - \frac{1}{6} + \frac{1}{2} s_{ \vartheta_S }^2  ) ( \overline{\eG_R^{ \prime\prime \prime\prime }  } \gamma^\mu \eG_R^{ \prime \prime\prime \prime}  +  \overline{\nG_R^{ \prime\prime \prime \prime}  } \gamma^\mu \nG_R^{ \prime \prime\prime \prime}  ) + ( + \frac{1}{3}  ) \overline{\check \nG_R^{ \prime\prime \prime } } \gamma^\mu \check \nG_R^{ \prime\prime \prime }  \Big]  \cdot Z_\mu^\prime \,,\\[1mm]
&& i \overline \psi_{ ( \rep{1}\,, \rep{3}\,, -\frac{2}{3} )^{\prime\prime}  }  \Dslash \psi_{ ( \rep{1}\,, \rep{3}\,,  -\frac{2}{3} )^{\prime \prime} }  \non
&\supset& \frac{ g_Y }{ s_{ \vartheta_S} c_{ \vartheta_S} }  \Big[  (  \frac{1}{6} + \frac{1}{2} s_{ \vartheta_S }^2  ) ( \overline{\eG_R^{ \prime\prime \prime\prime \prime}  } \gamma^\mu \eG_R^{ \prime\prime \prime\prime \prime}  +  \overline{\nG_R^{ \prime\prime \prime\prime \prime}  } \gamma^\mu \nG_R^{ \prime\prime \prime\prime \prime}  ) + ( - \frac{1}{3} + s_{ \vartheta_S }^2 ) \overline{e_R } \gamma^\mu e_R  \Big]  \cdot Z_\mu^\prime \,,\\[1mm]
&& i \overline \psi_{ ( \rep{3}\,, \rep{3}\,, 0 )^{ \prime\prime } }  \Dslash \psi_{ ( \rep{3}\,, \rep{3}\,, 0 )^{ \prime\prime } }  \non
&\supset& \frac{ g_Y }{ s_{ \vartheta_S} c_{ \vartheta_S} }  \Big[ \frac{1}{6} ( 1 - s_{\vartheta_S}^2 ) ( \overline{u_L } \gamma^\mu u_L  + \overline{d_L } \gamma^\mu d_L ) + \frac{1}{3} ( -1 + s_{\vartheta_S}^2 ) \overline{ \DG_L^{\prime\prime \prime \prime} } \gamma^\mu  \DG_L^{\prime\prime \prime \prime} \Big]  \cdot Z_\mu^\prime \,,\\[1mm]
&& i \overline \psi_{ ( \rep{3}\,, \rep{1}\,, -\frac{1}{3} )^{ \prime\prime \prime\prime \prime } }  \Dslash \psi_{ ( \rep{3}\,, \rep{1}\,, -\frac{1}{3} )^{ \prime\prime \prime \prime \prime } }  \supset \frac{ g_Y }{ s_{ \vartheta_S} c_{ \vartheta_S} }  ( + \frac{1}{3} s_{ \vartheta_S}^2 ) \overline{ \DG_L^{ \prime\prime \prime\prime \prime } } \gamma^\mu  \DG_L^{ \prime\prime \prime\prime \prime } \cdot Z_\mu^\prime \,,\\[1mm]
&& i \overline \psi_{ ( \rep{3}\,, \repb{3}\,, +\frac{1}{3} ) }  \Dslash \psi_{ ( \rep{3}\,, \repb{3}\,, +\frac{1}{3} ) }  \supset  \frac{ g_Y }{ s_{ \vartheta_S} c_{ \vartheta_S} }  \Big[  - \frac{1}{6}  ( 1+ s_{ \vartheta}^2 ) ( \overline{\uG_R } \gamma^\mu \uG_R  + \overline{\dG_R } \gamma^\mu \dG_R  ) \non 
&+& ( \frac{1}{3} - \frac{2}{3} s_{ \vartheta}^2 ) \overline{ \UG_R} \gamma^\mu \UG_R \Big] \cdot Z_\mu^\prime \,,\\[1mm]
&& i \overline \psi_{ ( \rep{3}\,, \rep{1}\,, +\frac{2}{3} )^{ \prime\prime \prime } }  \Dslash \psi_{ ( \rep{3}\,, \rep{1}\,, +\frac{2}{3} )^{  \prime \prime \prime } }  \supset \frac{ g_Y }{ s_{ \vartheta_S} c_{ \vartheta_S} } ( - \frac{2}{3} s_{ \vartheta_S}^2 ) \overline{ c_R } \gamma^\mu c_R   \cdot Z_\mu^\prime \,.
\eeqn
\eeqs
%
%

\begin{table}[htp]
\begin{center}
\begin{tabular}{c|cccc}
\hline \hline
  & $u/c/t$   &     $\uG$  &  $\UG$  &     \\ 
\hline
$g_f^{V\,\prime }$  &  $\frac{1}{12} - \frac{5}{12}  s_{\vartheta_S}^2 $ &     $-\frac{1}{6}  -\frac{1}{6} s_{ \vartheta_S}^2  $   &  $\frac{1}{3}  - \frac{2}{3} s_{ \vartheta_S}^2 $   &   \\
$g_f^{A\,\prime }$  &  $ -\frac{1}{12}  - \frac{1}{4}  s_{\vartheta_S}^2$  &    $0$  & $0$  &     \\
\hline
  & $d/s/b$   &  $\dG$  &  $\DG/ \DG^{\prime \prime}/ \DG^{ \prime \prime\prime \prime \prime }$  &  $\DG^\prime / \DG^{\prime \prime \prime }/ \DG^{ \prime \prime \prime \prime }$      \\ 
  \hline
$g_f^{V\,\prime }$  &  $\frac{1}{12} + \frac{1}{12} s_{\vartheta_S}^2$  &  $-\frac{1}{6}  -\frac{1}{6} s_{ \vartheta_S}^2 $ & $\frac{1}{3} s_{ \vartheta_S}^2$ & $ - \frac{1}{6} +\frac{1}{3} s_{ \vartheta_S}^2$     \\ 
$g_f^{A\,\prime }$  & $- \frac{1}{12} + \frac{1}{4} s_{\vartheta_S}^2$ &  $0$ &  $0$  & $\frac{1}{6}$    \\ 
\hline
  & $e/\mu/\tau$   &  $\EG$  &  $(\eG \,, \nG) / (\eG^{ \prime \prime}\,, \nG^{ \prime \prime}) $  &  $(\eG^\prime \,, \nG^\prime )/ (\eG^{ \prime \prime \prime} \,, \nG^{ \prime \prime \prime}) $     \\ 
    & $$   &  $$    & $/ ( \eG^{ \prime\prime\prime \prime}\,, \nG^{ \prime\prime\prime \prime} ) $  &  $/ ( \eG^{ \prime\prime\prime \prime \prime } \,,  \nG^{ \prime\prime\prime \prime \prime }) $ \\
  \hline
$g_f^{V\,\prime }$  & $- \frac{1}{4} + \frac{3}{4} s_{\vartheta_S}^2$   & $s_{ \vartheta_S}^2$   & $-\frac{1}{6} + \frac{1}{2} s_{ \vartheta_S}^2$   & $\frac{1}{2} s_{ \vartheta_S}^2$   \\ 
$g_f^{A\,\prime }$  & $- \frac{1}{12} + \frac{1}{4} s_{\vartheta_S}^2$    & $0$  & $0$   & $\frac{1}{6}$   \\ 
\hline
& $\nu_e/ \nu_\mu / \nu_\tau $ & $\check \nG/ \check \nG^\prime / \check \nG^{ \prime \prime } / \check \nG^{ \prime \prime \prime}$ &   &      \\ 
\hline
$g_f^{V\,\prime }$  &  $- \frac{1}{12} + \frac{1}{4} s_{\vartheta_S}^2$  & $\frac{1}{3}$   &  &   \\
$g_f^{A\,\prime }$  &  $ \frac{1}{12} - \frac{1}{4} s_{\vartheta_S}^2$  & $0$ &  &    \\ 
\hline\hline
\end{tabular}
\end{center}
\caption{
The couplings of the neutral currents mediated by $Z_\mu^{\prime}$ in the $V-A$ basis.
}
\label{tab:Zp_VA}
\end{table}%

\para
The neutral currents are expressed as follows 
\beqn\label{eq:SU3W_ZpVA_coup}
\Lc_{ {\rm SU}(3)_W}^{\rm NC\,, F}&=& \frac{g_{Y}}{ s_{\vartheta_S} c_{\vartheta_S} } ( g_f^{V\,\prime } \overline f \gamma^\mu f + g_f^{A\, \prime } \overline f \gamma^\mu \gamma_5 f ) Z_\mu^{\prime } \,,
\eeqn
in the $V-A$ basis, and we tabulate the vectorial and axial couplings in Tab.~\ref{tab:Zp_VA}. 
Three generational SM fermions with the same electric charges display the universal vectorial and axial couplings to the $Z_\mu^{\prime}$ gauge boson.
Both of the SM fermion vectorial and axial couplings depend on the mixing angle of $\vartheta_S$.
Besides of vectorlike fermions of $( \DG \,, \eG^{ \prime \prime } \,, \nG^{ \prime \prime } \,, \check \nG^\prime \,, \check \nG^{ \prime \prime} )$, the $(\uG\,, \dG\,,  \UG\,, \DG^{ \prime \prime} \,, \DG^{ \prime \prime\prime \prime \prime} \,, \EG\,, \eG\,, \nG \,, \eG^{ \prime \prime \prime \prime} \,, \nG^{ \prime \prime \prime \prime} )$ also exhibit vectorial couplings to the $Z_\mu^{\prime}$ gauge boson, since they have obtained their masses at the second stage of symmetry breaking.
For vectorlike fermions of $(\DG^\prime \,, \DG^{ \prime\prime\prime } \,, \DG^{ \prime \prime\prime \prime} \,, \nG^\prime\,, \eG^\prime \,, \nG^{ \prime \prime\prime} \,, \eG^{ \prime \prime\prime } \,, \nG^{ \prime \prime\prime\prime}\,, \eG^{ \prime \prime\prime\prime})$ that obtain masses at the third stage of symmetry breaking, they exhibit both the vectorial and axial couplings of $(g_f^{V\, \prime}\,, g_f^{A\, \prime} )$. 
Different from the SM fermions, only their vectorial couplings depend on the mixing angle of $\vartheta_S$, while their axial couplings are independent of the mixing angle of $\vartheta_S$.

\subsection{The ${\rm SU}(2)_W \otimes {\rm U}(1)_{ Y}$ gauge bosons}
\label{section:EW_bosons}

\subsubsection{Covariant derivatives and gauge boson masses}

\para
We express the ${\rm SU}(2)_W \otimes {\rm U}(1)_{Y}$ covariant derivatives in the EW sector for the fundamental and anti-fundamental representations as follows
%
%
\beqs\label{eqs:EW_covariant}
\beqn
i D_\mu \psi_{\rep{2} } &\equiv& i  \partial_\mu \psi_{\rep{2} } + (  g_{2W} A_\mu^{ I} T_{ {\rm SU}(2 )}^{ I}  +  g_{Y} \Yc \mathbb{I}_2  B_{\mu} ) \cdot \psi_{\rep{2} } \,,\label{eq:EW_covariant_fund} \\[1mm]
i D_\mu \psi_{\repb{2} } &\equiv& i \partial_\mu \psi_{\repb{2} }  + (-  g_{2W} A_\mu^{ I} ( T_{ {\rm SU}(2 )}^{ I} )^T +   g_{Y} \Yc \mathbb{I}_2  B_{\mu} ) \cdot \psi_{\repb{2} } \,, \label{eq:EW_covariant_antifund}
\eeqn
\eeqs
%
%
with $T_{ {\rm SU}(2 )}^{ I} \equiv \hf \sigma^I$.
The gauge fields from Eqs.~\eqref{eqs:EW_covariant} can be expressed in terms of a $2\times 2$ matrix 
\beqs
\beqn
&& g_{2W} A^{ I}_\mu T_{ {\rm SU}(2 )}^{ I}  + g_{Y} \Yc \mathbb{I}_2  B_{\mu}  \non
&=&  \frac{g_{2W}}{ \sqrt{2} }\left( 
\ba{cc}  
     0  &   W_\mu^+    \\ 
 W_\mu^- &  0   \\   \ea  \right)  + \frac{ g_{2W} }{2} {\rm diag}  \Big(  A_\mu^3 +  2  t_{\vartheta_W} \Yc B_{ \mu} \,,   - A_\mu^3  + 2  t_{\vartheta_W} \Yc B_{ \mu}  \Big)  \,, \\[1mm]
 && - g_{2W} A^{ I}_\mu ( T_{ {\rm SU}(2 )}^{ I} )^T + g_{Y} \Yc \mathbb{I}_2  B_{\mu}  \non
&=&  - \frac{g_{2W}}{ \sqrt{2} }\left( 
\ba{cc}  
     0  &   W_\mu^-    \\ 
 W_\mu^+ &  0   \\   \ea  \right)  +  \frac{ g_{2W} }{2} {\rm diag}  \Big( - A_\mu^3 +  2  t_{\vartheta_W} \Yc B_{ \mu} \,,   A_\mu^3  + 2  t_{\vartheta_W} \Yc B_{ \mu}   \Big) \,,
\eeqn
\eeqs
with the ${\rm SU}(2)_W$ Weinberg angle defined by
\beqn\label{eq:Weinberg_angle}
t_{\vartheta_W} &\equiv& \tan\vartheta_{W} = \frac{g_{Y} }{ g_{2W } }\,.
\eeqn
One can identify that 
\beqn
&& W_\mu^\pm = \frac{1}{ \sqrt{2} } ( A_\mu^1 \mp i A_\mu^2 )  \,.
\eeqn

\subsubsection{Fermion $\Gc_{\rm SM}$ gauge couplings}

\begin{table}[htp]
\begin{center}
\begin{tabular}{c|cccc}
\hline \hline
  & $u/c/t$   &     $\uG$  &  $\UG$  &     \\ 
\hline
$g_f^{V }$  &  $ \frac{1 }{4 } - \frac{2}{3} s_{ \vartheta_W}^2$ &     $ \frac{1 }{2} -\frac{2}{3} s_{ \vartheta_W}^2 $   &  $-\frac{2}{3} s_{ \vartheta_W}^2 $   &     \\
$g_f^{A }$  &   $-\frac{1 }{4 }$  &    $0$  & $0$  &    \\
\hline
  & $d/s/b$   &  $\dG$  &  $\DG/ \DG^{\prime } /\DG^{\prime \prime}/   \DG^{\prime \prime \prime }/ \DG^{\prime \prime \prime \prime }/ \DG^{ \prime \prime\prime \prime \prime }$  &      \\ 
  \hline
$g_f^{V }$  &  $- \frac{1 }{4 } + \frac{1}{3} s_{ \vartheta_W}^2$  &  $- \frac{1 }{2} + \frac{1}{3} s_{ \vartheta_W}^2 $ & $+\frac{1}{3} s_{ \vartheta_W}^2$ &     \\ 
$g_f^{A }$  &  $+\frac{1 }{4 }$ &  $0$ &  $0$  &    \\ 
\hline
  & $e/\mu/\tau$   &  $\EG$  &  $\nu_e /\nu_\mu/ \nu_\tau$  &  $  $     \\ 
  \hline
$g_f^{V }$  & $ - \frac{1 }{4 } + s_{ \vartheta_W}^2$   & $+ s_{ \vartheta_W}^2 $   &  $+\frac{1 }{4 }$  & $$   \\ 
$g_f^{A }$  & $+\frac{1 }{4 }$    & $0$  & $-\frac{1 }{4 }$   & $$   \\ 
\hline\hline
\end{tabular}
\end{center}
\caption{
The couplings of the flavor-conserving neutral currents mediated by $Z_\mu$ in the $V-A$ basis.
}
\label{tab:Z_VA}
\end{table}%

\para
The flavor-conserving neutral currents are expressed as follows in the $V-A$ basis
\beqn\label{eq:SU2W_ZpVA_coup}
\Lc_{ {\rm SU}(2)_W}^{\rm NC\,, F}&=& \frac{ e }{ s_{\vartheta_W } c_{\vartheta_W} } ( g_f^{V } \overline f \gamma^\mu f + g_f^{A } \overline f \gamma^\mu \gamma_5 f ) Z_\mu \,,
\eeqn
and we recovered the SM vectorial and axial couplings in Tab.~\ref{tab:Z_VA}.

\section{The gauge coupling evolutions in the ${\rm SU}(8)$ theory}
\label{section:SU8_RGEs}

\para
The gauge coupling unification relies on the RGEs of the ${\rm SU}(8)$ theory.
The two-loop RGE of a gauge coupling of $\alpha_\Upsilon$ in the $\overline{\rm MS}$ scheme is given by~\cite{Machacek:1983tz}
\beqn\label{eq:gauge_RGE}
&& \frac{d \alpha_\Upsilon^{-1}( \mu ) }{ d\log \mu} = \frac{ b_\Upsilon^{ (1) } }{2\pi }  +  \sum_{ \Upsilon^\prime } \frac{b_{  \Upsilon \Upsilon^\prime }^{ ( 2 ) } }{ 8 \pi^2 \alpha_\Upsilon^{-1} ( \mu) } \,,
\eeqn
where the one-loop and two-loop $\beta$ coefficients for the non-Abelian groups are
\beqs
\beqn
b_\Upsilon^{ (1 ) } &=& - \frac{11 }{ 3 } C_2( \Gc_\Upsilon ) + \frac{2 }{3 } \sum_{ F  } T(  \Rc^F_\Upsilon ) +  \frac{1 }{3 } \sum_{ S   } T(  \Rc^S_\Upsilon ) \,, \\[1mm]
b_{ \Upsilon \Upsilon^\prime }^{ ( 2  ) }&=& - \frac{ 34}{3 }  C_2 ( \Gc_\Upsilon )^2 \alpha_\Upsilon - \sum_{F  } \[ 2 \sum_{ \Upsilon^\prime } C_2 ( \Rc^F_{\Upsilon^\prime} ) \alpha_{ \Upsilon^\prime } + \frac{ 10}{ 3} C_2 ( \Gc_\Upsilon ) \alpha_{ \Upsilon}  \] T( \Rc^F_\Upsilon )\non
&& - \sum_S \[ 4 \sum_{ \Upsilon^\prime }  C_2 ( \Rc^S_{\Upsilon^\prime} ) \alpha_{ \Upsilon^\prime } + \frac{2 }{ 3 } C_2 ( \Gc_\Upsilon )\alpha_{ \Upsilon}  \] T( \Rc^S_\Upsilon ) \,.
\eeqn
\eeqs
Here, $C_2( \Gc_\Upsilon )$ is the quadratic Casimir of the group $\Gc_\Upsilon$, $T(  \Rc^F_\Upsilon ) $ and $T(  \Rc^S_\Upsilon ) $ are trace invariants of the chiral fermions in the irrep of $\Rc^F_\Upsilon$ and complex scalars in the irrep of $\Rc^S_\Upsilon$.
For the ${\rm U}(1)$ Abelian groups with charges denoted as $\Xc^{F/S}$, the one-loop $\beta$ coefficients become
\beqn
b_\Upsilon^{ (1 ) } &=&  \frac{2 }{3 } \sum_F (  \Xc^F_\Upsilon )^2 +  \frac{1 }{3 } \sum_S ( \Xc^S_\Upsilon )^2 \,.
\eeqn

\subsection{RGEs of the minimal setup with the survival hypothesis}
\label{section:SU8_RGEs_survival}

\para 
By following the survival hypothesis~\cite{Georgi:1979md,Glashow:1979nm,Barbieri:1979ag,Barbieri:1980vc,Barbieri:1980tz,delAguila:1980qag}, we first list the minimal set of massless Higgs fields and the massless fermions between different symmetry breaking stages according the minimal Higgs VEVs in Eqs.~\eqref{eqs:SU8_Higgs_VEVs_mini}.
The two-loop RGEs within the minimal ${\rm SU}(8)$ setup are derived by using the PyR@TE~\cite{Sartore:2020gou}.

\para
Between the $v_{441}\leq \mu\leq v_U$, almost all massless Higgs fields are in the first lines of Eqs.~\eqref{eq:SU8_Higgs_Br01}, \eqref{eq:SU8_Higgs_Br02}, and \eqref{eq:SU8_Higgs_Br05}, which are
\beqn\label{eq:Higgs_441to8}
&& (  \repb{4} \,, \rep{1} \,, +\frac{1}{4} )_{\rep{H}\,,\omega }  \oplus ( \rep{1} \,, \repb{4} \,, -\frac{1}{4} )_{\rep{H}\,,\omega }  \subset  \repb{8_H}_{\,, \omega } \,, \non
&& (\rep{6 } \,,   \rep{1 } \,, +\frac{1}{2} )_{\rep{H}\,, \dot \omega } \oplus  ( \rep{1 } \,, \rep{6 } \,, -\frac{1}{2} )_{\rep{H}\,, \dot \omega } \oplus ( \repb{4 } \,, \repb{4 } \,, 0 )_{\rep{H}\,, \dot \omega }  \subset \repb{28_H}_{\,,\dot \omega } \,,\non
&&  ( \rep{1 } \,, \rep{1 } \,, - 1  )_{\rep{H} }^{ \prime \prime  }  \oplus ( \rep{1 } \,, \rep{1 } \,, + 1  )_{\rep{H} }^{ \prime \prime \prime \prime }  \oplus ( \rep{4 } \,, \repb{4} \,, +\frac{1 }{ 2 } )_{\rep{H} } \oplus  (\repb{4 } \,, \rep{4} \,, -\frac{1 }{ 2}  )_{\rep{H} }  \oplus (\rep{6 } \,, \rep{6} \,,  0  )_{\rep{H} }   \subset \rep{70_H} \,.
\eeqn
All ${\rm SU}(8)$ fermions in Eq.~\eqref{eq:SU8_3gen_fermions} remain massless after the decomposition into the $\Gc_{441}$ irreps.
Correspondingly, we have the $\Gc_{441}$ $\beta$ coefficients of
\beqn\label{eq:441_beta}
&& (  b_{{\rm SU}(4)_s}^{ (1 ) }  \,,b_{{\rm SU}(4)_W}^{ (1 ) }  \,, b_{{\rm U}(1)_{X_0} }^{ (1 ) } ) = (  + \frac{13 }{3} \,,  + \frac{13 }{3} \,,  + \frac{55}{3 }) \,, \non
&& b_{\Gc_{441} }^{ (2 ) } = \left( \ba{ccc}
 2299/6  & 405/2  & 12  \\
 405/2  & 2299/6 & 12  \\
  180 & 180 & 31  \\
\ea \right)   \,.
\eeqn
When the RGEs evolve down to $v_{441}$, the gauge couplings should match according to Eq.~\eqref{eq:441_coupMatch}.

\para
Between the $v_{341}\leq \mu\leq v_{441}$, the massless Higgs fields are
\beqn\label{eq:Higgs_341to441}
&&  ( \rep{1} \,, \repb{4} \,, -\frac{1}{4} )_{\rep{H}\,, 3\,, {\rm V} \,, {\rm VI}}  \subset  \repb{8_H}_{\,, 3\,, {\rm V} \,, {\rm VI}} \,, \non
&&    ( \rep{1} \,, \rep{6 } \,, -\frac{1}{2} )_{\rep{H}\,, \dot 1\,, \dot 2\,, \dot {\rm VIII} }  \subset ( \rep{1 } \,, \rep{6 } \,, -\frac{1}{2} )_{\rep{H}\,, \dot 1\,, \dot 2\,,  \dot {\rm VIII} } \subset \repb{28_H}_{\,, \dot 1\,, \dot 2\,, \dot {\rm VIII}} \,,\non
&&  ( \rep{1} \,, \repb{4} \,, -\frac{1}{4} )_{\rep{H}\,, \dot 1 \,, \dot 2\,, \dot {\rm VII} \,, \dot {\rm IX} }  \subset ( \repb{4 } \,, \repb{4} \,, 0 )_{\rep{H}\,, \dot 1\,,  \dot 2\,, \dot {\rm VII} \,, \dot {\rm IX} } \subset \repb{28_H}_{\,,\dot 1\,, \dot 2\,, \dot {\rm VII} \,, \dot {\rm IX} } \,,\non
&&  ( \rep{1 } \,, \repb{4} \,, +\frac{3 }{ 4} )_{\rep{H} }^{\prime  } \subset ( \rep{4 } \,, \repb{4} \,, +\frac{1 }{ 2 } )_{\rep{H} }  \subset \rep{70_H} \,.
\eeqn
The massless $\Gc_{341}$ fermions are listed as follows 
\beqn\label{eq:341_fermions}
&& \Big[ ( \repb{3}\,, \rep{1}\,, +\frac{1}{3} )_{ \mathbf{F}}^\Omega \oplus ( \rep{1}\,, \rep{1}\,, 0 )_{ \mathbf{F}}^\Omega \Big]  \oplus ( \rep{1}\,, \repb{4}\,, -\frac{1}{4} )_{ \mathbf{F}}^\Omega  \subset \repb{8_F}^\Omega \,, \non
&& \Omega = ( \omega \,, \dot \omega ) \,,\quad \omega = (3\,, {\rm V}\,, {\rm VI} ) \,,\quad \dot \omega = ( \dot 1\,, \dot 2\,, \dot {\rm VII} \,,\dot {\rm VIII} \,, \dot {\rm IX}) \,,\non
&& ( \rep{1}\,, \rep{1}\,, 0 )_{ \mathbf{F}}^{{\rm IV} } \subset \repb{8_F}^{ {\rm IV}} \,,\non
&&  ( \repb{3}\,, \rep{1}\,, -\frac{2}{3} )_{ \mathbf{F}}  \oplus ( \rep{1}\,, \rep{6}\,, +\frac{1}{2} )_{ \mathbf{F}}  \oplus ( \rep{3}\,, \rep{4}\,, -\frac{1}{12} )_{ \mathbf{F}}  \subset \rep{28_F}\,, \non
&& ( \rep{1}\,, \repb{4}\,, +\frac{3}{4} )_{ \mathbf{F}} \oplus  ( \repb{3}\,, \rep{1}\,, -\frac{2}{3} )_{ \mathbf{F}}^\prime \oplus ( \rep{1}\,, \rep{1}\,, -1 )_{ \mathbf{F}} \oplus  ( \rep{3}\,, \rep{6}\,, +\frac{1}{6} )_{ \mathbf{F}} \oplus ( \rep{1}\,, \rep{6}\,, +\frac{1}{2})_{ \mathbf{F}}^\prime  \non
&\oplus&  ( \rep{3}\,, \rep{4}\,, -\frac{1}{12} )_{ \mathbf{F}}^\prime \oplus ( \repb{3}\,, \rep{4}\,, -\frac{5}{12})_{ \mathbf{F}}  \subset \rep{56_F}\,.
\eeqn
Correspondingly, we have the $\Gc_{341}$ $\beta$ coefficients of
\beqn\label{eq:341_beta}
&& (  b_{{\rm SU}(3)_c}^{ (1 ) }  \,,b_{{\rm SU}(4)_W}^{ (1 ) }  \,, b_{{\rm U}(1)_{X_1} }^{ (1 ) } ) = (  -\frac{5}{3} \,, -3 \,, + \frac{  247 }{ 18  } ) \,, \non
 && b_{\Gc_{341} }^{ (2 ) } = \left( \ba{ccc}
  226/3  & 75/2 &  97/36 \\
  20 & 571/4  &  187/24 \\
   194/9  & 935/8  & 8639/432  \\  \ea \right)   \,. 
\eeqn
When the RGEs evolve down to $v_{341}$, the gauge couplings should match according to Eq.~\eqref{eq:341_coupMatch}.

\para
Between the $v_{331}\leq \mu\leq v_{341}$, the massless Higgs fields are
\beqn\label{eq:Higgs_331to341}
&&  ( \rep{1} \,, \repb{3} \,, -\frac{1}{3} )_{\rep{H}\,, 3\,, {\rm VI} } \subset ( \rep{1} \,, \repb{4} \,, -\frac{1}{4} )_{\rep{H}\,, 3\,, {\rm VI} }  \subset  \repb{8_H}_{\,, 3\,, {\rm VI}} \,, \non
&&  ( \rep{1} \,, \repb{3} \,, -\frac{1}{3} )_{\rep{H}\,, \dot 2\,, \dot {\rm VIII} }^\prime  \subset  ( \rep{1} \,, \rep{6 } \,, -\frac{1}{2} )_{\rep{H}\,, \dot 2\,, \dot {\rm VIII} }  \subset ( \rep{1 } \,, \rep{6 } \,, -\frac{1}{2} )_{\rep{H}\,, \dot 2\,, \dot {\rm VIII} } \subset \repb{28_H}_{\,, \dot 2\,, \dot {\rm VIII}} \,,\non
&&  ( \rep{1} \,, \repb{3} \,, -\frac{1}{3} )_{\rep{H}\,, \dot 2\,, \dot {\rm IX} }  \subset  ( \rep{1} \,, \repb{4} \,, -\frac{1}{4} )_{\rep{H}\,, \dot 2\,, \dot {\rm IX} }  \subset ( \repb{4 } \,, \repb{4} \,, 0 )_{\rep{H}\,, \dot 2\,, \dot {\rm IX} } \subset \repb{28_H}_{\,,\dot 2\,, \dot {\rm IX}} \,,\non
&& ( \rep{1 } \,, \repb{3} \,, +\frac{2 }{ 3} )_{\rep{H} }^{\prime  \prime \prime } \subset   ( \rep{1 } \,, \repb{4} \,, +\frac{3 }{ 4} )_{\rep{H} }^{\prime  } \subset  ( \rep{4 } \,, \repb{4} \,, +\frac{1 }{ 2 } )_{\rep{H} }  \subset \rep{70_H} \,.
\eeqn
The massless $\Gc_{331}$ fermions are listed as follows 
\beqn\label{eq:331_fermions}
&& \Big[ ( \repb{3}\,, \rep{1}\,, +\frac{1}{3} )_{ \mathbf{F}}^\Omega \oplus ( \rep{1}\,, \rep{1}\,, 0 )_{ \mathbf{F}}^\Omega \Big]  \oplus \Big[ ( \rep{1}\,, \repb{3}\,, -\frac{1}{3} )_{ \mathbf{F}}^\Omega \oplus ( \rep{1}\,, \rep{1}\,, 0 )_{ \mathbf{F}}^{\Omega^{\prime \prime }}  \Big]  \subset \repb{8_F}^\Omega \,, \non
&& \Omega = ( \omega \,, \dot \omega ) \,, \quad \omega = (3\,, {\rm VI})\,,  \quad \dot \omega = (\dot 1\,, \dot 2\,, \dot {\rm VIII}\,, \dot {\rm IX} )\,,\non
&& ( \rep{1}\,, \rep{1}\,, 0)_{ \mathbf{F}}^{{\rm IV} } \subset \repb{8_F}^{{\rm IV} } \,, \quad ( \rep{1}\,, \rep{1}\,, 0)_{ \mathbf{F}}^{{\rm V} \,, {\rm V}^{\prime\prime} } \subset \repb{8_F}^{{\rm V} } \,, \non
&&  ( \rep{1}\,, \rep{1}\,, 0)_{ \mathbf{F}}^{\dot {\rm VII} } \oplus ( \rep{1}\,, \rep{1}\,, 0)_{ \mathbf{F}}^{\dot {\rm VII}^{ \prime \prime} } \subset \repb{8_F}^{\dot {\rm VII} } \,,  \non
&& ( \repb{3}\,, \rep{1}\,, -\frac{2}{3} )_{ \mathbf{F}}  \oplus ( \rep{1}\,, \repb{3}\,, +\frac{2}{3} )_{ \mathbf{F}}  \oplus  ( \rep{3}\,, \rep{3}\,, 0 )_{ \mathbf{F}}   \subset \rep{28_F}\,, \non
&&( \rep{1}\,, \repb{3}\,, +\frac{2}{3} )_{ \mathbf{F}}^\prime \oplus  ( \repb{3}\,, \rep{1}\,, -\frac{2}{3} )_{ \mathbf{F}}^\prime \oplus  ( \rep{3}\,, \rep{3}\,, 0 )_{ \mathbf{F}}^\prime \non
&\oplus& ( \rep{1}\,, \repb{3}\,, +\frac{2}{3})_{ \mathbf{F}}^{\prime \prime}  \oplus ( \rep{3}\,, \rep{3}\,, 0)_{ \mathbf{F}}^{\prime\prime} \oplus  ( \repb{3}\,, \rep{1}\,, -\frac{2}{3})_{ \mathbf{F}}^{\prime\prime\prime}  \subset \rep{56_F}\,.
\eeqn
Correspondingly, we have the $\Gc_{331}$ $\beta$ coefficients of
\beqn\label{eq:331_beta}
&& ( b^{(1)}_{3c} \,, b^{(1)}_{3W } \,, b^{(1)}_{X_2 } ) = ( -5 \,, -  \frac{ 23}{ 6} \,, + \frac{82}{ 9} )  \,, \non
 && b_{\Gc_{331} }^{ (2 ) } = \left( \ba{ccc}
 12   & 12 &  2 \\
 12  & 113/3  & 38/9  \\
   16  & 304/9  & 304/27  \\  \ea \right)   \,. 
\eeqn
When the RGEs evolve down to $v_{331}$, the gauge couplings should match according to Eq.~\eqref{eq:331_coupMatch}.

\para
Between the $v_{\rm EW}\leq \mu\leq v_{331}$, there is only one massless SM Higgs field of
\beqn\label{eq:Higgs_SMto331}
&& ( \rep{1 } \,, \repb{2} \,, +\frac{1 }{ 2} )_{\rep{H} }^{\prime  \prime \prime } \subset ( \rep{1 } \,, \repb{3} \,, +\frac{2 }{ 3} )_{\rep{H} }^{\prime  \prime \prime } \subset   ( \rep{1 } \,, \repb{4} \,, +\frac{3 }{ 4} )_{\rep{H} }^{\prime  } \subset  ( \rep{4 } \,, \repb{4} \,, +\frac{1 }{ 2 } )_{\rep{H} }  \subset \rep{70_H} \,.
\eeqn
The massless SM fermions (together with massless sterile neutrinos) are listed as follows 
\beqn\label{eq:SM_fermions}
&& \Big[ ( \repb{3}\,, \rep{1}\,, +\frac{1}{3} )_{ \mathbf{F}}^\Omega \oplus ( \rep{1}\,, \rep{1}\,, 0 )_{ \mathbf{F}}^\Omega \Big]  \oplus \Big[ ( \rep{1}\,, \repb{2}\,, -\frac{1}{2} )_{ \mathbf{F}}^\Omega \oplus ( \rep{1}\,, \rep{1}\,, 0 )_{ \mathbf{F}}^{\Omega^{\prime}} \oplus ( \rep{1}\,, \rep{1}\,, 0 )_{ \mathbf{F}}^{\Omega^{\prime\prime}}   \Big]  \subset \repb{8_F}^\Omega \,, \quad \Omega = ( \dot 1\,, \dot 2\,, 3 ) \,, \non
&& ( \rep{1}\,, \rep{1}\,, 0)_{ \mathbf{F}}^{{\rm IV} } \oplus ... \oplus ( \rep{1}\,, \rep{1}\,, 0)_{ \mathbf{F}}^{\dot {\rm IX} } \subset \repb{8_F}^{\Omega } \,,  \non
&& ( \rep{1}\,, \rep{1}\,, 0)_{ \mathbf{F}}^{{\rm VI}^\prime } \oplus  ( \rep{1}\,, \rep{1}\,, 0)_{ \mathbf{F}}^{\dot {\rm VIII}^\prime } \oplus ( \rep{1}\,, \rep{1}\,, 0)_{ \mathbf{F}}^{\dot {\rm IX}^\prime } \subset \repb{8_F}^{\Omega^\prime } \,,  \non
&& ( \rep{1}\,, \rep{1}\,, 0)_{ \mathbf{F}}^{{\rm V}^{ \prime \prime} } \oplus ... \oplus ( \rep{1}\,, \rep{1}\,, 0)_{ \mathbf{F}}^{\dot {\rm IX}^{ \prime \prime } } \subset \repb{8_F}^{\Omega^{ \prime\prime} } \,,  \non
&& ( \repb{3}\,, \rep{1}\,, -\frac{2}{3} )_{ \mathbf{F}}   \oplus  ( \rep{1}\,, \rep{1}\,, +1 )_{ \mathbf{F}} \oplus  ( \rep{3}\,, \rep{2}\,, +\frac{1}{6} )_{ \mathbf{F}}   \subset \rep{28_F}\,, \non
&&  ( \rep{1}\,, \rep{1}\,, +1 )_{ \mathbf{F}}^\prime  \oplus ( \repb{3}\,, \rep{1}\,, -\frac{2}{3} )_{ \mathbf{F}}^\prime  \oplus  ( \rep{3}\,, \rep{2}\,, +\frac{1}{6} )_{ \mathbf{F}}^\prime  \non 
&\oplus&  ( \rep{1}\,, \rep{1}\,, +1)_{ \mathbf{F}}^{\prime\prime \prime} \oplus ( \rep{3}\,, \rep{2}\,, +\frac{1}{6} )_{ \mathbf{F}}^{\prime\prime \prime}  \oplus ( \repb{3}\,, \rep{1}\,, -\frac{2}{3})_{ \mathbf{F}}^{\prime\prime\prime}   \subset \rep{56_F} \,.
\eeqn
Correspondingly, we have the usual SM $\beta$ coefficients of
\beqn\label{eq:SM_beta}
&& ( b^{(1)}_{3c} \,,  b^{(1) }_{2W} \,,  b^{ (1)}_{Y } )= ( -7 \,, - \frac{ 19 }{ 6} \,,  + \frac{ 41 }{ 6} ) \,, \non
&& b_{\Gc_{\rm SM} }^{ (2 ) } = \left( \ba{ccc}
 -26  &9/2 & 11/6  \\
 12  & 35 /6 & 3 /2  \\
 44/3  & 9/2 & 199/18  \\
\ea \right)   \,. 
\eeqn

\begin{figure}[htb]
\centering
\includegraphics[height=4.5cm]{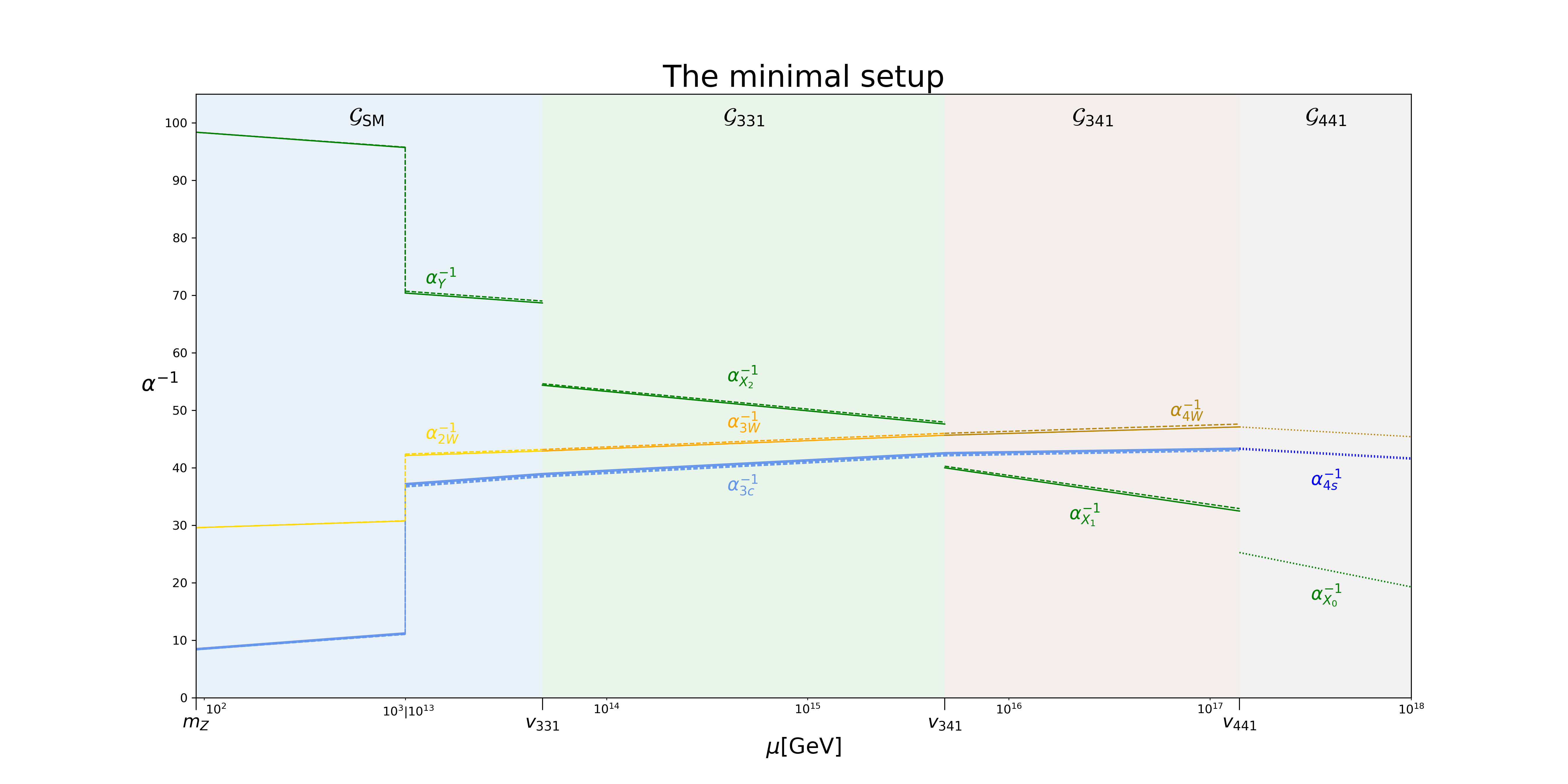}
\caption{The RGEs of the minimal ${\rm SU}(8)$ setup according to the survival hypothesis.
The dashed lines and the solid lines represent the one-loop and two-loop RGEs, respectively.}
\label{fig:RGE_mini} 
\end{figure}

\para
With the one- and two-loop $\beta$ coefficients defined in Eqs.~\eqref{eq:441_beta}, \eqref{eq:341_beta}, \eqref{eq:331_beta}, and \eqref{eq:SM_beta}, we plot the RGEs of the minimal ${\rm SU}(8)$ setup in Fig.~\ref{fig:RGE_mini}.
Three intermediate symmetry breaking scales follow from the benchmark point in Eq.~\eqref{eq:benchmark}.
The discontinuities of the ${\rm U}(1)$ gauge couplings at three intermediate scales follow from their definitions in Eqs.~\eqref{eq:441_coupMatch}, \eqref{eq:341_coupMatch}, \eqref{eq:331_coupMatch}, respectively.
The RG evolutions within $10^{3}\,{\rm GeV} \lesssim \mu \lesssim 10^{13}\,{\rm GeV}$ are zoomed out in order to highlight the behaviors in three intermediate symmetry breaking scales.
The current plot indicates $\alpha_{4S}^{-1}( \mu) \sim \alpha_{4W}^{-1}( \mu) \simeq 40$ and $\alpha_{X_0 }^{-1} ( \mu )\simeq 20$ for $\mu \simeq 5\times 10^{17}\,{\rm GeV}$.
Furthermore, the one-loop $\beta$ coefficients of $ b_{{\rm SU}(4)_s}^{ (1 ) }  = b_{{\rm SU}(4)_W}^{ (1 ) } <  b_{{\rm U}(1)_{X_0} }^{ (1 ) }$ from Eq.~\eqref{eq:441_beta} guarantee that two non-Abelian ${\rm SU}(4)_{S \,, W}$ gauge couplings can never unify with the Abelian ${\rm U}(1)_{X_0}$ gauge coupling in the minimal setup.

\para
The one-loop threshold effect was tentatively neglected in Fig.~\ref{fig:RGE_mini}. 
This was usually considered as a major source to modify the RG behavior within the field theory~\cite{Weinberg:1980wa,Hall:1980kf,Ernst:2018bib}, and it can be generically expressed as
\beqn\label{eq:threshold_match}
 \alpha_{ \rm UV }^{-1} ( \mu )  &=&  \alpha_{ \rm IR }^{-1} ( \mu ) +  \frac{ \Delta_{  \rm IR } ( \mu ) }{ 12\pi} \,, \non
\Delta_{  \rm IR } ( \mu ) &=& C_2 ( \Gc_{  \rm UV } ) - C_2 ( \Gc_{ \rm IR } ) - 21 \sum_{ V_k \in \Gc_{ \rm IR }   }  {\rm dim} ( V_k ) \log \frac{ M_{V_k } }{ \mu } \non
&&+ 2 \sum_{ \Sc_k \in \Gc_{ \rm IR }  } {\rm dim} ( \Sc_k) \log \frac{ M_{ \Sc_k } }{ \mu } + 4 \sum_{ \Fc_k \in \Gc_{ \rm IR }  } {\rm dim}( \Fc_k ) \log \frac{ M_{ \Fc_k} }{ \mu }  \,,
\eeqn
where $V_k$, $S_k$ and $\Fc_k$ represent the vector bosons, complex scalars and Weyl fermions that are integrated out at the threshold scale $\mu$.
Their dimensions are always given in terms of the broken subgroup of $\Gc_{ \rm IR } $.
One may evaluate contributions due to the threshold effects in Eq.~\eqref{eq:threshold_match} to see if they can alter the gauge coupling evolutions significantly.
Let us take the renormalization scale of $\mu = \overline M_{V_k}$ (with $\overline M_{V_k}$ being the average of the gauge boson masses) in Eq.~\eqref{eq:threshold_match}, and list the contributions from the massive scalars and fermions at each symmetry breaking stage.
Given the RG behaviors depicted in Fig.~\ref{fig:RGE_mini}, we will parametrize the differences between the Abelian and the non-Abelian gauge couplings.
For simplicity, we take a common mass for both massive fermions and massive scalar fields at each stage, and we estimate the largest possible threshold effects by assuming these masses to be $10$ times larger than the renormalization scale of $\mu = \overline M_{V_k}$.
The massive vectorlike fermions at each symmetry breaking stage have been summarized in Tab.~\ref{tab:SU8_vectorferm}.

\para
At the $\Gc_{331}$-breaking stage, we find
\beqn\label{eq:331threshold}
&& \{ {\rm massive}~S_k \} = (  \rep{1}\,, \repb{2} \,, - \frac{1}{2} )_{ \rep{H} \,, 3 \,, {\rm VI}\,, \dot 2\,, \dot{\rm IX} } \oplus (  \rep{1}\,, \repb{2} \,, - \frac{1}{2} )_{ \rep{H} \,, \dot 2\,, \dot{\rm VIII} }^\prime \,, \non
&& ( \Delta_{ 3c } \,, \Delta_{ 2W } \,, \Delta_{ Y } ) = ( 0\,, 1\,, 1) +  ( 6\,,  0 \,, 4) \log \frac{ M_{  ( \rep{3} \,, \rep{1}\,, - \frac{1}{3} )_{\rep{F}} } }{ \overline M_{V_{331} }} +   ( 6\,,  0 \,, 4) \log \frac{ M_{  ( \repb{3} \,, \rep{1}\,, + \frac{1}{3} )_{\rep{F}} } }{\overline M_{V_{331} }} \non
&+&  ( 0\,,  6 \,, 6) \log \frac{ M_{  ( \rep{1 } \,, \repb{2 }\,, + \frac{1}{2 } )_{\rep{F}} } }{ \overline M_{V_{331} }} + ( 0\,,  6 \,, 6) \log \frac{ M_{  ( \rep{1 } \,, \repb{2 }\,, - \frac{1}{2 } )_{\rep{F}} } }{ \overline M_{V_{331} }} + ( 0\,,  6 \,, 6) \log \frac{ M_{  ( \rep{1 } \,, \repb{2 }\,, - \frac{1}{2 } )_{\rep{H}} } }{ \overline M_{V_{331} }}  \non
&=& ( 12 \log \frac{ M_{ \Fc } }{ \overline M_{V_{331} }}   \,, 1+ 12  \log \frac{ M_{ \Fc } }{ \overline M_{V_{331} }}  + 6 \log \frac{ M_{ \Sc } }{ \overline M_{V_{331} }}   \,,   1+ 20  \log \frac{ M_{ \Fc } }{ \overline M_{V_{331} }}  + 6 \log \frac{ M_{ \Sc } }{ \overline M_{V_{331} }}  ) \,, \non
&\Rightarrow&   \Delta_Y - \frac{  \Delta_{ 3c } + \Delta_{ 2W } }{2}  =  \hf + 8  \log \frac{ M_{ \Fc } }{ \overline M_{V_{331} }} + 3 \log \frac{ M_{ \Sc } }{ \overline M_{V_{331} }}  \lesssim 25 \,.
\eeqn
At the $\Gc_{341}$-breaking stage, we find
\beqn\label{eq:341threshold}
&& \{ {\rm massive}~S_k \} =  (  \rep{1}\,, \repb{3 } \,, - \frac{1}{3 } )_{ \rep{H}  \,, {\rm V} } \oplus (  \rep{1}\,, \repb{3 } \,, - \frac{1}{3 } )_{ \rep{H} \,, \dot 1 \,,  \dot {\rm VII} } \oplus (  \rep{1}\,, \repb{3 } \,, - \frac{1}{3 } )_{ \rep{H} \,, \dot 1  }^\prime \oplus (  \rep{1}\,, \rep{3 } \,, - \frac{2 }{3 } )_{ \rep{H} \,, \dot 1  }  \,, \non
&& ( \Delta_{ 3c } \,, \Delta_{ 3W } \,, \Delta_{ X_2 } ) = ( 0\,, 1\,, \frac{2}{3} ) +   ( 4\,, 0 \,, \frac{ 8}{ 3} ) \log \frac{ M_{  ( \rep{3} \,, \rep{1}\,, - \frac{1}{3} )_{\rep{F}} } }{ \overline M_{V_{341} }} + ( 4\,, 0 \,, \frac{ 8}{ 3} ) \log \frac{ M_{  ( \repb{3} \,, \rep{1}\,, + \frac{1}{3} )_{\rep{F}} } }{ \overline M_{V_{341} }}  \non
&+& ( 0\,, 4 \,, \frac{ 8}{ 3} ) \log \frac{ M_{  ( \rep{1} \,, \rep{3}\,, + \frac{1}{3} )_{\rep{F}} } }{ \overline M_{V_{341} }} + ( 0\,, 4 \,, \frac{ 8}{ 3} ) \log \frac{ M_{  ( \rep{1} \,, \repb{3}\,, - \frac{1}{3} )_{\rep{F}} } }{ \overline M_{V_{341} }}  +  ( 6 \,, 6\,, 4  ) \log \frac{ M_{  ( \rep{3 } \,, \repb{3}\,, + \frac{1}{3} )_{\rep{F}} } }{ \overline M_{V_{341} }} + ( 6 \,, 6\,, 4  ) \log \frac{ M_{  ( \repb{3 } \,, \rep{3}\,, - \frac{1}{3} )_{\rep{F}} } }{ \overline M_{V_{341} }}  \non
&+& ( 0 \,, 0 \,, 4  ) \log \frac{ M_{  ( \rep{1 } \,, \rep{1}\,, +1 )_{\rep{F}} } }{ \overline M_{V_{341} }}  + ( 0 \,, 0 \,, 4  ) \log \frac{ M_{  ( \rep{1 } \,, \rep{1}\,, -1 )_{\rep{F}} } }{ \overline M_{V_{341} }}  + ( 0 \,, 4 \,, \frac{ 8}{3}  ) \log \frac{ M_{  ( \rep{1 } \,, \repb{3 }\,, -\frac{1 }{ 3} )_{\rep{H}} } }{ \overline M_{V_{341} }} + ( 0 \,, 1 \,, \frac{ 8}{3}  ) \log \frac{ M_{  ( \rep{1 } \,, \rep{3 }\,, -\frac{2 }{ 3} )_{\rep{H}} } }{ \overline M_{V_{341} }}  \non
&=&  (  20 \log \frac{ M_{ \Fc } }{ \overline M_{V_{341} }}   \,,  1+ 20 \log \frac{ M_{ \Fc } }{ \overline M_{V_{341} }}  +  5\log \frac{ M_{ \Sc } }{ \overline M_{V_{341} }}   \,,  \frac{2 }{ 3} + \frac{ 80}{3} \log \frac{ M_{ \Fc } }{ \overline M_{V_{341} }}  + \frac{16}{3} \log \frac{ M_{ \Sc } }{ \overline M_{V_{341} }} ) \,, \non
&\Rightarrow&   \Delta_{X_2 } - \frac{  \Delta_{ 3c } + \Delta_{ 3W } }{2}  = \frac{1}{6} + \frac{20}{3} \log \frac{ M_{ \Fc } }{ \overline M_{V_{341} }}  +  \frac{17}{6} \log \frac{ M_{ \Sc } }{ \overline M_{V_{341} }}  \lesssim 21 \,.
\eeqn
At the $\Gc_{441}$-breaking stage, we find
\beqn\label{eq:441threshold}
&& \{ {\rm massive}~S_k \} =  (  \repb{3 } \,,  \rep{1}\,, - \frac{1}{3 } )_{ \rep{H}  \,, \Omega=( \omega \,, \dot \omega ) } \oplus  (  \rep{1 } \,,  \repb{ 4}\,, - \frac{1}{4 } )_{ \rep{H}  \,,  {\rm IV}  } \oplus  (  \rep{3 } \,,  \rep{1}\,, + \frac{2 }{3 } )_{ \rep{H}  \,, \dot \omega  } \non
&\oplus& (  \rep{1 } \,,  \rep{ 6 }\,, - \frac{1}{2 } )_{ \rep{H}  \,,  \dot {\rm VII} \,, \dot {\rm IX}  }   \oplus (  \rep{1 }\,, \rep{ 4  } \,, - \frac{ 3 }{ 4 }  )_{ \rep{H} }  \non
&\oplus&  (  \rep{1}\,, \rep{1} \,, -1 )_{ \rep{H} } \oplus (  \rep{1}\,, \rep{1} \,, +1 )_{ \rep{H} } \oplus (  \rep{3 }\,, \repb{4 } \,, +\frac{5}{12 }  )_{ \rep{H} } \oplus (  \repb{3 }\,, \rep{4 } \,, - \frac{5}{12 }  )_{ \rep{H} }  \oplus (  \rep{3 }\,, \rep{6 } \,, + \frac{ 1}{ 6 }  )_{ \rep{H} } \oplus (  \repb{3 }\,, \rep{6 } \,, - \frac{ 1}{ 6 }  )_{ \rep{H} }  \non
&& ( \Delta_{ 3c } \,, \Delta_{ 4W } \,, \Delta_{ X_1 } ) = ( 1 \,, 0 \,, \frac{2}{3} )  +   ( 2\,, 0 \,, \frac{ 4}{ 3} ) \log \frac{ M_{  ( \rep{3} \,, \rep{1}\,, - \frac{1}{3} )_{\rep{F}} } }{ \overline M_{V_{441} }} +   ( 2\,, 0 \,, \frac{ 4}{ 3} ) \log \frac{ M_{  ( \repb{3} \,, \rep{1}\,, + \frac{1}{3} )_{\rep{F}} } }{ \overline M_{V_{441} }} \non
&+& ( 0\,, 2 \,, 1 ) \log \frac{ M_{  ( \rep{1 } \,, \rep{4 }\,, + \frac{1}{4} )_{\rep{F}} } }{ \overline M_{V_{441} }} +  ( 0\,, 2 \,, 1 ) \log \frac{ M_{  ( \rep{1 } \,, \repb{4 }\,, - \frac{1}{4} )_{\rep{F}} } }{ \overline M_{V_{441} }}   \non
&+&  ( 9 \,, 0 \,, 6 ) \log \frac{ M_{  ( \repb{3 } \,, \rep{1 }\,, + \frac{1}{3 } )_{\rep{H}} } }{ \overline M_{V_{441} }} +   ( 0 \,, 2 \,, 1  ) \log \frac{ M_{  ( \rep{1 } \,, \repb{4 }\,, - \frac{1}{4 } )_{\rep{H}} } }{ \overline M_{V_{441} }}  +  ( 5 \,, 0 \,, \frac{40}{3}  ) \log \frac{ M_{  ( \rep{3 } \,, \rep{1 }\,, + \frac{2}{3 } )_{\rep{H}} } }{ \overline M_{V_{441} }} \non
&+& ( 0 \,, 4 \,, 6  ) \log \frac{ M_{  ( \rep{1 } \,, \rep{6 }\,, - \frac{1 }{2 } )_{\rep{H}} } }{ \overline M_{V_{441} }} +  ( 20 \,, 15 \,, \frac{5}{6} ) \log \frac{ M_{  ( \repb{3 } \,, \repb{4 }\,, + \frac{1 }{ 12 } )_{\rep{H}} } }{ \overline M_{V_{441} }}  \non
&+& ( 4 \,, 3 \,, \frac{25}{6} ) \log \frac{ M_{  ( \rep{3 } \,, \repb{4 }\,, + \frac{5 }{12 } )_{\rep{H}} } }{ \overline M_{V_{441} }}  + ( 4 \,, 3 \,, \frac{25}{6} ) \log \frac{ M_{  ( \repb{3 } \,, \rep{4 }\,, - \frac{5 }{12 } )_{\rep{H}} } }{ \overline M_{V_{441} }}  \non
&+&  ( 0 \,, 0 \,, 2  ) \log \frac{ M_{  ( \rep{1 } \,, \rep{1 }\,, +1 )_{\rep{H}} } }{ \overline M_{V_{441} }} +  ( 0 \,, 0 \,, 2  ) \log \frac{ M_{  ( \rep{1 } \,, \rep{1 }\,, -1 )_{\rep{H}} } }{ \overline M_{V_{441} }}   \non
&+& ( 6 \,, 6 \,, 1 ) \log \frac{ M_{  ( \rep{3 } \,, \rep{6 }\,, + \frac{1}{6 } )_{\rep{H}} } }{ \overline M_{V_{441} }} +  ( 6 \,, 6 \,, 1 ) \log \frac{ M_{  ( \repb{3 } \,, \rep{6 }\,, - \frac{1}{6 } )_{\rep{H}} } }{ \overline M_{V_{441} }} + ( 0 \,, 1 \,, \frac{9}{2} ) \log \frac{ M_{  ( \rep{1 } \,, \rep{4 }\,, - \frac{3 }{4 } )_{\rep{H}} } }{ \overline M_{V_{441} }}  \non
&=&  (  1+ 4 \log \frac{ M_{ \Fc } }{ \overline M_{V_{441} }}  + 54 \log \frac{ M_{ \Sc } }{ \overline M_{V_{441} }}   \,,  4 \log \frac{ M_{ \Fc } }{ \overline M_{V_{441} }}  +  40 \log \frac{ M_{ \Sc } }{ \overline M_{V_{441} }}   \,,   \frac{2}{3 } + \frac{14 }{ 3 } \log \frac{ M_{ \Fc } }{ \overline M_{V_{441} }}  +  46 \log \frac{ M_{ \Sc } }{ \overline M_{V_{441} }} )  \,, \non 
&\Rightarrow&  \Delta_{X_1 } - \frac{  \Delta_{ 3c } + \Delta_{ 4W } }{2}   = \frac{1}{6} + \frac{2}{3} \log \frac{ M_{ \Fc } }{ \overline M_{V_{441} }}  -   \log \frac{ M_{ \Sc } }{ \overline M_{V_{441} }} \lesssim -0.6 \,.
\eeqn
By collecting all threshold effects from Eqs.~\eqref{eq:331threshold}, \eqref{eq:341threshold}, and \eqref{eq:441threshold} into Eq.~\eqref{eq:threshold_match}, we find the overall threshold shift of $\Delta_{ {\rm U}(1)}/(12 \pi)$ relative to the threshold effects of two non-Abelian couplings to be at most $\sim \Oc(1)$.
Hence, the one-loop threshold effect cannot significantly modify the RG evolutions in Fig.~\ref{fig:RGE_mini} such that the large discrepancy of $| \alpha_{X_0}^{-1} - ( \alpha_{4S}^{-1} + \alpha_{ 4W}^{-1} )/2 | \sim 20$ can be eliminated.

\subsection{The gauge coupling unification with additional adjoint Higgs fields}
\label{section:SU8_RGEs_adj}

\para
The RG evolutions of the non-Abelian and Abelian gauge couplings with the minimal ${\rm SU}(8)$ setup as depicted in Fig.~\ref{fig:RGE_mini} do not achieve the unification, since the ${\rm U}(1)$ gauge couplings get enhanced much faster than two non-Abelian couplings with the current field contents.
A tentative solution within the field theoretical framework is to further introduce adjoint Higgs fields of $\rep{63_H}$'s.
If one assumes the $( \rep{15} \,, \rep{1} \,, 0 )_{ \mathbf{H}} \oplus ( \rep{1} \,, \rep{15} \,, 0 )_{ \mathbf{H}}$ components in Eq.~\eqref{eq:63H_decomp} remain massless after the GUT-scale symmetry breaking, they bring additional positive contributions to the $\beta$ coefficients of two non-Abelian gauge couplings.
Naturally, one expects the unification scale to locate below the reduced Planck scale.
We display two examples of this possibility in Fig.~\ref{fig:RGE_addition}, while they both require as many as $150-200$ adjoint massless Higgs fields of $( \rep{15} \,, \rep{1} \,, 0 )_{ \mathbf{H}} \oplus ( \rep{1} \,, \rep{15} \,, 0 )_{ \mathbf{H}}$ within the range of $v_{441}\lesssim \mu \lesssim v_U$ such that $v_U\simeq 5\times 10^{17}\,{\rm GeV}$.

\begin{figure}[htb]
\centering
\includegraphics[height=4.2cm]{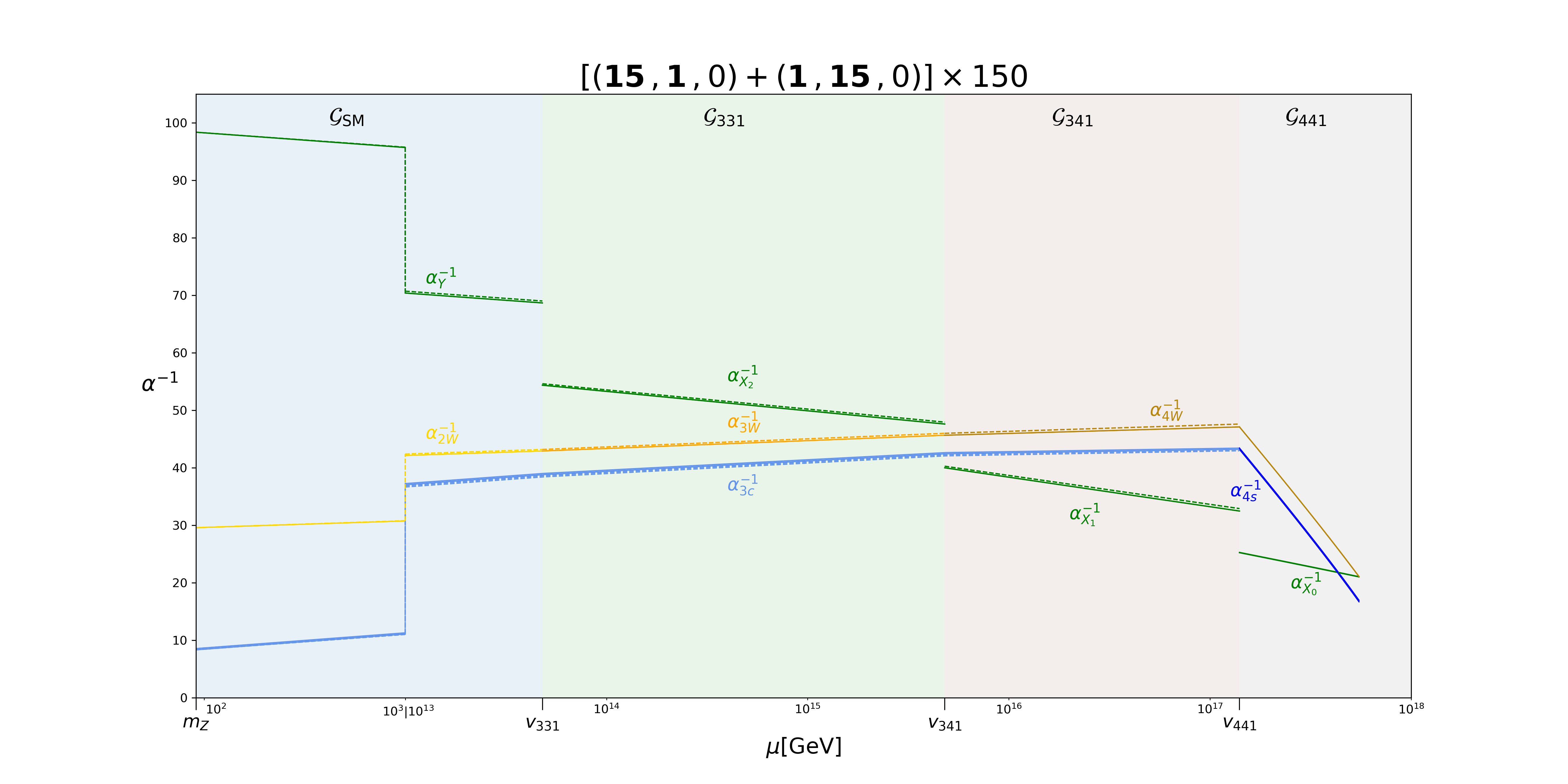}\\
\includegraphics[height=4.2cm]{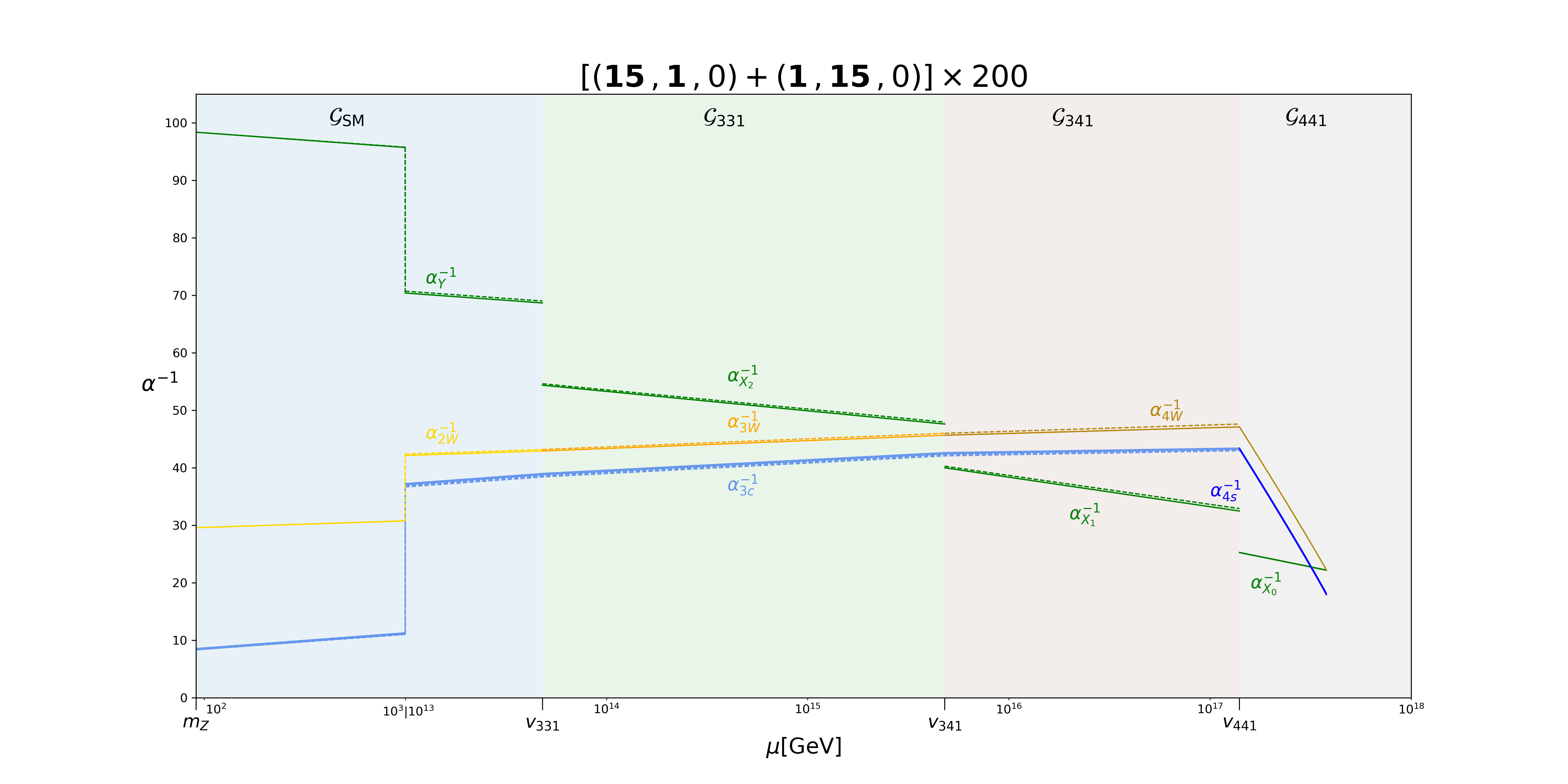}
\caption{The RGEs of the ${\rm SU}(8)$ theory with 150 additional Higgs fields of $( \rep{15} \,, \rep{1} \,, 0 )_{ \mathbf{H}} \oplus ( \rep{1} \,, \rep{15} \,, 0 )_{ \mathbf{H}}$ (left panel), and 200 additional Higgs fields of $( \rep{15} \,, \rep{1} \,, 0 )_{ \mathbf{H}} \oplus ( \rep{1} \,, \rep{15} \,, 0 )_{ \mathbf{H}}$ (right panel).
The dashed lines and the solid lines represent the one-loop and two-loop RGEs, respectively.}
\label{fig:RGE_addition} 
\end{figure}

\para
Given that the unification scale is precious close to the Planck scale, there were also consideration of the gravitational effects to the gauge coupling unification, such as in the context of the ${\rm SU}(5)$ and ${\rm SO}(10)$ GUTs~\cite{Hill:1983xh,Shafi:1983gz,Parida:1989kg,Datta:1994yz}.
At the leading order, we consider the following $d=5$ HSW operator in the ${\rm SU}(8)$ theory
\beqn\label{eq:O_HSW}
\Oc_{\rm HSW}&\equiv& - \hf \frac{ c_{\rm HSW} }{ M_{\rm pl}} \Tr[ \rep{63_H} U^{ \mu \nu } U_{\mu \nu } ] \,,
\eeqn
with $U_{\mu \nu } = U_{\mu \nu }^{\AG} T^{\AG}$ being the ${\rm SU}(8)$ field strength.
After the GUT-scale symmetry breaking with the VEV in Eq.~\eqref{eq:63H_VEV}, this HSW operator shifts two ${\rm SU}(4)_{S/W}$ field strengths of $G_{ \mu \nu }=G_{ \mu \nu }^{\bar A} T^{\bar A}$ and $A_{ \mu \nu }= A_{ \mu \nu }^{\bar I} T^{\bar I}$ as follows
\beqn
&& - \frac{1 }{2 } ( 1 - \frac{ 1  }{4  } c_{\rm HSW} \zeta_0 ) \Tr ( G_{ \mu \nu } G^{ \mu \nu } ) - \frac{1 }{2 } ( 1 + \frac{ 1  }{4  } c_{\rm HSW} \zeta_0 ) \Tr ( A_{ \mu \nu } A^{ \mu \nu } )\,,
\eeqn
with $\zeta_0\equiv v_U/ M_{\rm pl}$.
The gauge coupling unification is modified into
\beqn\label{eq:unification_HSW}
&& ( 1 - \frac{ 1  }{4  } c_{\rm HSW} \zeta_0 ) \alpha_{4S} ( v_U ) =  ( 1 + \frac{ 1  }{4  } c_{\rm HSW} \zeta_0 ) \alpha_{4W} ( v_U ) = \alpha_{X_0 } ( v_U )  = \alpha_{U } ( v_U ) \,.
\eeqn
Therefore, we find the following results
\beqn
n_{\rm Adj}=150~&:&~ c_{\rm HSW}=1.91 \,, \quad v_U \simeq 4.9 \times 10^{17}\,{\rm GeV} \,, \quad \alpha_U^{-1} (v_U) \simeq 21.4 \,, \non
n_{\rm Adj}=200~&:&~c_{\rm HSW}= 2.54 \,,  \quad v_U \simeq 3.5 \times 10^{17}\,{\rm GeV} \,, \quad \alpha_U^{-1} (v_U) \simeq 22.5  \,,
\eeqn
in accordance to two plots in Fig.~\ref{fig:RGE_addition}.

\subsection{RGEs of the minimal setup with the lighter Higgs fields assumptions}
\label{section:SU8_RGEs_lightH}

\para
The derivations in Sec.~\ref{section:SU8_RGEs_survival} relies on the minimal set of massless Higgs fields that can develop the VEVs according to Eqs.~\eqref{eqs:SU8_Higgs_VEVs_mini}.
Next, we consider the RGEs of the minimal setup by assuming more lighter Higgs fields relative to the survival hypothesis.
In this scenario, the Higgs fields from the $\repb{8_H}_{\,,\omega} \oplus \repb{28_H}_{\,,\dot \omega}$ remain massless until certain components obtained VEVs according to Eqs.~\eqref{eqs:SU8_Higgs_VEVs_mini}.
Meanwhile, the massless fermions at each symmetry breaking stage have been fixed according to the anomaly-free conditions~\cite{Chen:2023qxi,Chen:2024cht}.

\para
Between the $v_{441}\leq \mu\leq v_U$, we consider the identical massless Higgs fields in Eq.~\eqref{eq:Higgs_441to8}.
Between the $v_{341}\leq \mu\leq v_{441}$, we assume the massless Higgs fields of
\beqn\label{eq:HiggsB_341to441}
&& (  \repb{3} \,, \rep{1} \,, + \frac{1}{3} )_{\rep{H}\,, 3\,, {\rm V} \,, {\rm VI}}  \oplus ( \rep{1} \,, \repb{4} \,, -\frac{1}{4} )_{\rep{H}\,, 3\,, {\rm V} \,, {\rm VI}}  \subset  \repb{8_H}_{\,, 3\,, {\rm V} \,, {\rm VI}} \,, \non
&&   (\rep{3 } \,,   \rep{1 } \,, +\frac{2 }{ 3 } )_{\rep{H}\,, \dot \omega } \oplus (\repb{3 } \,,   \rep{1 } \,, +\frac{1}{3 } )_{\rep{H}\,, \dot \omega } \oplus  ( \rep{1 } \,, \rep{6 } \,, -\frac{1}{2} )_{\rep{H}\,, \dot \omega } \non
&\oplus& ( \repb{3 } \,, \repb{4 } \,, +\frac{1}{12} )_{\rep{H}\,, \dot \omega } \oplus ( \rep{1 } \,, \repb{4 } \,, -\frac{1 }{4} )_{\rep{H}\,, \dot \omega }  \subset   \repb{28_H}_{\,, \dot \omega } \,, \quad \dot \omega = ( \dot 1\,, \dot 2\,, \dot{\rm VII}\,, \dot {\rm VIII}\,, \dot {\rm IX} )\,,\non
&&  ( \rep{1 } \,, \repb{4} \,, +\frac{3 }{ 4} )_{\rep{H} }^{\prime  } \oplus ( \rep{1 } \,, \rep{4} \,, - \frac{3 }{ 4} )_{\rep{H} }^{\prime  } \subset ( \rep{4 } \,, \repb{4} \,, +\frac{1 }{ 2 } )_{\rep{H} }  \oplus ( \repb{4 } \,, \rep{4 } \,, - \frac{1 }{ 2 } )_{\rep{H} }  \subset \rep{70_H} \,,
\eeqn
which leads to the $\Gc_{341}$ $\beta$ coefficients of
\beqn\label{eq:341B_beta}
&& (  b_{{\rm SU}(3)_c}^{ (1 ) }  \,,b_{{\rm SU}(4)_W}^{ (1 ) }  \,, b_{{\rm U}(1)_{X_1} }^{ (1 ) } ) = (   +\frac{23 }{6 } \,,  + \frac{1 }{2} \,,  + \frac{677 }{ 36 } ) \,, \non
 && b_{\Gc_{341} }^{ (2 ) } = \left( \ba{ccc}
 589/3   & 225/2  &  331/36 \\
  60  & 509/2  & 45/4  \\
 662/9  &    675/4 & 8893/216  \\  \ea \right)   \,. 
\eeqn
Between the $v_{331}\leq \mu\leq v_{341}$, we assume the massless Higgs fields of
\beqn\label{eq:HiggsB_331to341}
&& (  \repb{3} \,, \rep{1} \,, + \frac{1}{3} )_{\rep{H}\,, 3 \,, {\rm VI}}  \oplus ( \rep{1} \,, \repb{3 } \,, -\frac{1}{3 } )_{\rep{H}\,, 3 \,, {\rm VI}}  \subset  \repb{8_H}_{\,, 3 \,, {\rm VI}} \,, \non
&&  \[ (\rep{3 } \,,   \rep{1 } \,, +\frac{2 }{ 3 } )_{\rep{H}\,, \dot \omega } \oplus (\repb{3 } \,,   \rep{1 } \,, +\frac{1}{3 } )_{\rep{H}\,, \dot \omega }  \]\oplus  \[ ( \rep{1 } \,, \repb{3 } \,, -\frac{1}{3 } )_{\rep{H}\,, \dot \omega }^\prime \oplus ( \rep{1 } \,, \rep{3 } \,, -\frac{2 }{ 3 } )_{\rep{H}\,, \dot \omega } \] \non
&\oplus& \[ ( \repb{3 } \,, \repb{3 } \,, 0  )_{\rep{H}\,, \dot \omega } \oplus ( \repb{3 } \,, \rep{1 } \,, +\frac{1}{ 3 } )_{\rep{H}\,, \dot \omega }  \oplus ( \rep{1 } \,, \repb{3 } \,, -\frac{1 }{3 } )_{\rep{H}\,, \dot \omega }  \] \subset   \repb{28_H}_{\,, \dot \omega } \,, \quad \dot \omega = (  \dot 2\,,  \dot {\rm VIII}\,, \dot {\rm IX} )  \,, \non
&& ( \rep{1 } \,, \repb{3} \,, +\frac{2 }{ 3} )_{\rep{H} }^{\prime  \prime \prime } \subset   ( \rep{1 } \,, \repb{4} \,, +\frac{3 }{ 4} )_{\rep{H} }^{\prime  } \subset  ( \rep{4 } \,, \repb{4} \,, +\frac{1 }{ 2 } )_{\rep{H} }  \subset \rep{70_H} \,.
\eeqn
which leads to the $\Gc_{331}$ $\beta$ coefficients of
\beqn\label{eq:331B_beta}
&& (  b_{{\rm SU}(3)_c}^{ (1 ) }  \,,b_{{\rm SU}(3 )_W}^{ (1 ) }  \,, b_{{\rm U}(1)_{X_2 } }^{ (1 ) } ) = (  - \frac{ 5}{ 3}  \,,  - \frac{ 3 }{  2 }  \,,  + \frac{ 116}{ 9 }  ) \,, \non
 && b_{\Gc_{331} }^{ (2 ) } = \left( \ba{ccc}
  256/3  & 36  & 58/9  \\
  36  & 89  &  22/3 \\
 464/9   & 176/3 &   728/27 \\  \ea \right)   \,. 
\eeqn
Between the $v_{\rm EW}\leq \mu\leq v_{331}$, the massless Higgs field is identical to Eq.~\eqref{eq:Higgs_SMto331}.

\begin{figure}[htb]
\centering
\includegraphics[height=4.5cm]{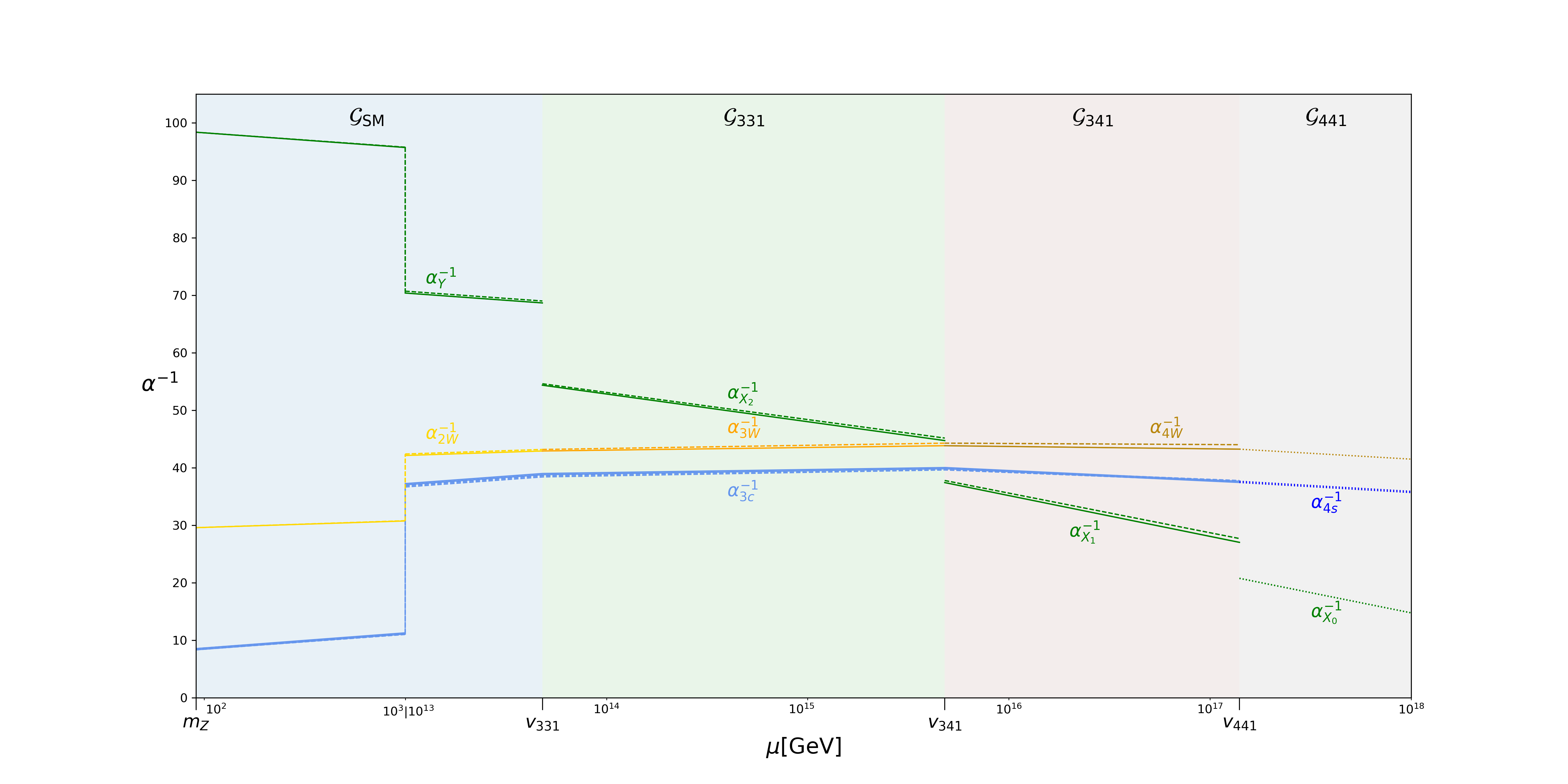}
\caption{The RGEs of the minimal ${\rm SU}(8)$ setup with the lighter Higgs fields assumptions.
The dashed lines and the solid lines represent the one-loop and two-loop RGEs, respectively.}
\label{fig:RGE_lightH} 
\end{figure}

\para
Accordingly, we plot the RGEs with more lighter Higgs fields in Fig.~\ref{fig:RGE_lightH}.
A very similar behavior can be found by comparing to the RGEs displayed in Fig.~\ref{fig:RGE_mini} with the survival hypothesis.
The large discrepancy of $| \alpha_{X_0}^{-1} - ( \alpha_{4S}^{-1} + \alpha_{ 4W}^{-1} )/2 | \sim 20$ beyond the $v_{441}$ can not be eliminated by considering the threshold effects as well.
Without the inclusion of additional adjoint Higgs fields, the minimal ${\rm SU}(8)$ theory can not achieve the gauge coupling unification with two different assumptions of the Higgs fields masses.

\section{Summary and outlook}
\label{section:conclusion}

\subsection{The main results}

\para
We study the RGEs of the gauge couplings in an ${\rm SU}(8)$ theory through its SWW symmetry breaking pattern in Eq.~\eqref{eq:Pattern}.
Three intermediate symmetry breaking scales follow from the previous Ref.~\cite{Chen:2024cht}, which were obtained such that three generational SM quark/lepton masses and the CKM mixing pattern are consistently predicted.
The gauge couplings at each symmetry breaking stage are defined in Sec.~\ref{section:SU8_gauge}.
Since all three generational SM fermions are non-trivially embedded in the ${\rm SU}(8)$ theory, the flavor non-universality in terms of the neutral currents in the $\Gc_{441}$ and $\Gc_{341}$ effective theories are explicitly displayed in Tab.~\ref{tab:Maon_VA} and \ref{tab:Zpp_VA}.
Together with the symmetry breaking scales determined in Eq.~\eqref{eq:benchmark}, we obtain the RG evolution of three gauge couplings in Fig.~\ref{fig:RGE_mini} for the minimal setup.
The gauge coupling unification cannot be achieved even when we include the one-loop threshold effects into account.
The only possible solution within the field theory is to include addition $150 - 200 $ adjoint massless Higgs fields of $( \rep{15} \,, \rep{1} \,, 0 )_{ \mathbf{H}} \oplus ( \rep{1} \,, \rep{15} \,, 0 )_{ \mathbf{H}}$ within the range of $v_{441}\lesssim \mu \lesssim v_U$.

\para
We wish to mention that the examples shown in Fig.~\ref{fig:RGE_addition} are only possible accommodation when the current SWW symmetry breaking pattern is considered.
Within the field theoretical framework, alternative symmetry breaking patterns other than the one in Eq.~\eqref{eq:Pattern} are possible, such as
\beqs
\beqn
{\rm WSW}~&:&~ {\rm SU}(8) \xrightarrow{ v_U } \Gc_{441} \xrightarrow{ v_{441} } \Gc_{431} \xrightarrow{v_{431} } \Gc_{331} \xrightarrow{ v_{331} } \Gc_{\rm SM} \xrightarrow{ v_{\rm EW} } {\rm SU}(3)_{c}  \otimes  {\rm U}(1)_{\rm EM} \,, \\[1mm]
{\rm WWS}~&:&~ {\rm SU}(8) \xrightarrow{ v_U } \Gc_{441} \xrightarrow{ v_{441} } \Gc_{431} \xrightarrow{v_{431} } \Gc_{ 421} \xrightarrow{ v_{421} } \Gc_{\rm SM} \xrightarrow{ v_{\rm EW} } {\rm SU}(3)_{c}  \otimes  {\rm U}(1)_{\rm EM} \,,
\eeqn
\eeqs
or even the one where the ${\rm SU}(8)$ was not maximally broken, such as in Ref.~\cite{Ma:1981pr}.

\subsection{On the SUSY extension and the Ka{\v{c}}-Moody extension}

\para
More broadly speaking, the unified theories were extended to an $\Nc=1$ SUSY theory, such as the SUSY ${\rm SU}(5)$ theory~\cite{Dimopoulos:1981zb}.
The gauge coupling unification was achieved at $v_U\simeq 2\times 10^{16}\,{\rm GeV}$, with the sparticle masses assumed to be $\sim \Oc(1)\,{\rm TeV}$ therein.
A SUSY ${\rm SU}(8)$ theory with non-trivially embedded SM fermions was not known before, and it is useful to sketch some underlying issues below for the future work.
\begin{itemize}

\item The current minimal ${\rm SU}(8)$ theory is anomalous, when one promotes the fields in Eqs.~\eqref{eq:SU8_3gen_fermions} and \eqref{eq:SU8_Higgs} into chiral superfields.
At least two types of SUSY extensions to the minimal ${\rm SU}(8)$ theory are possible, with the additional chiral superfields of
\beqs
\beqn
 \{ H \}_{ {\rm I}} &=&\rep{8_H}^{  \omega}  \oplus \rep{28_H}^{ \dot \omega} \,,  ~ \omega = ( 3\,, {\rm IV}\,, {\rm V}\,, {\rm VI}) \,, ~  \dot \omega = (\dot 1\,, \dot 2\,, \dot {\rm VII}\,, \dot {\rm VIII}\,, \dot {\rm IX} )   \,, \\[1mm]
 \{ H \}_{ {\rm II}} &=& \rep{ 36_H} \oplus \rep{36_H}^\prime \,.
\eeqn
\eeqs

\item With the superpotential of 
\beqn
W&=& \hf M_U \Tr \rep{\Phi}_{\rm Adj}^2 + \frac{ \lambda_U  }{3} \Tr \rep{\Phi}_{\rm Adj}^3 \,, \quad \rep{\Phi}_{\rm Adj}\equiv \rep{24_H} \,,
\eeqn
in the SUSY ${\rm SU}(5)$ theory, Witten has shown that two distinctive symmetry breaking patterns of ${\rm SU}(5) \to \Gc_{\rm SM}$ and ${\rm SU}(5) \to \Gc_{41}$ are degenerate~\cite{Witten:1981nf}.
For the ${\rm SU}(8)$ theory with $\rep{\Phi}_{\rm Adj}\equiv \rep{63_H}$, the symmetry breaking patterns of
\beqn
&& {\rm SU}(8) \to \Gc_{441} \,,  \quad {\rm SU}(8) \to \Gc_{351}/ \Gc_{531} \,, \quad  {\rm SU}(8) \to  \Gc_{ 621} \,, \quad {\rm SU}(8) \to \Gc_{71}\,,
\eeqn
at the GUT scale all lead to degenerate minima.
The subgroups of all four distinctive symmetry breaking patterns contain the $\Gc_{\rm SM}$ as their subgroup, hence they are phenomenologically reasonable.
Correspondingly, the gauge couplings will be defined differently from the maximally breaking pattern of ${\rm SU}(8) \to \Gc_{441}$.

\item The other issue in the SUSY ${\rm SU}(8)$ theory is the sparticle masses, which are determined by the SUSY breaking mechanisms.
The modifications of the RGEs between the non-SUSY sector and the SUSY sector will be determined accordingly.
Additional contributions to the one-loop threshold effect in Eq.~\eqref{eq:threshold_match} are necessary to account for the difference between the dimensional reduction scheme in the SUSY sector and the $\overline{\rm MS}$ scheme in the non-SUSY sector~\cite{Langacker:1992rq}.

\end{itemize}

\para
Lastly, we briefly comment on a possible unification scenario within the string theoretical framework, which was previously studied in the ${\rm SU}(5)/{\rm SO}(10)$ theories and their SUSY extensions~\cite{Font:1990uw,Dienes:1996du}.
In the heterotic string theory, Ginsparg has shown in Ref.~\cite{Ginsparg:1987ee} that the gauge and gravitational couplings must unify according to
\beqn
&& k_i g_i^2 = g_{\rm string}^2 = \frac{ 8 \pi G_N }{ \alpha^\prime} \,,
\eeqn
with $k_i$ being the Ka{\v{c}}-Moody levels of each gauge couplings, and $\alpha^\prime$ being the Regge slope.
The string unification scale was previously given by~\cite{Kaplunovsky:1987rp}
\beqn
&& M_{\rm string} \approx g_{\rm string} \times 5 \times 10^{17}\, {\rm GeV}\,,
\eeqn
which is quite close to the $v_{441}$ in Eq.~\eqref{eq:benchmark}.
It is notable that our benchmark scales given in Ref.~\cite{Chen:2024cht} were estimated by fitting the SM flavor parameters within the ${\rm SU}(8)$ framework, and are independent of the derivation of the string scale in Ref.~\cite{Ginsparg:1987ee,Kaplunovsky:1987rp}.
To achieve the string unification based on some extensions from the minimal setup of the ${\rm SU}(8)$ theory, one has to fully study all possible symmetry breaking patterns under the SUSY extensions described above, as well as to look for the constraints of the unitarity and conformal invariance from the Ka{\v{c}}-Moody Lie algebra~\cite{Dienes:1996du}.

\section*{Acknowledgements}
%
%
\para
We would like to thank Tianjun Li, Kaiwen Sun, Yuan Sun, Yinan Wang, Zhong-Zhi Xianyu, and Wenbin Yan for very enlightening discussions and communications. 
N.C. would like to thank Southeast University, Nanjing University, Institute of Theoretical Physics CAS for hospitality when preparing this work.
N.C. is partially supported by the National Natural Science Foundation of China (under Grants No. 12035008 and No. 12275140) and Nankai University.
Y.N.M. is partially supported by the National Natural Science Foundation of China (under Grant No. 12205227), the Fundamental Research Funds for the Central Universities (WUT: 2022IVA052), and Wuhan University of Technology.

\appendix

\section{Conventions, decomposition rules and charge quantizations in the ${\rm SU}(8)$ theory}
\label{section:Br}

\begin{table}[htp]
\begin{center}
\begin{tabular}{c|ccc}
\hline \hline
Indices   &  group  & irrep &  range    \\
\hline
$\aG\,,\bG\,,\cG$   &  ${\rm SU}(8)$   &  fundamental  & $1\,,...\,,8$   \\
   &      & anti-fundamental  &   \\
 $\AG\,,\BG\,,\CG$   &     ${\rm SU}(8)$  & adjoint  & $1\,,...\,,63$  \\ \hline
$\bar a\,, \bar b\,, \bar c$   &  ${\rm SU}(4)_s$   &  fundamental  & $\clubsuit\,,\diamondsuit\,, \heartsuit\,, \spadesuit$   \\
   &      & anti-fundamental  &   \\
 $\bar A\,,\bar B\,,\bar C$   &     ${\rm SU}(4)_s$  & adjoint  & $1\,,...\,,15$  \\ \hline
$a\,,b\,,c$   &  ${\rm SU}(3)_c$   &  fundamental  & $\clubsuit\,,\diamondsuit\,, \heartsuit$   \\
   &      & anti-fundamental  &   \\
 $A\,,B\,,C$   &     ${\rm SU}(3)_c$  & adjoint  & $1\,,...\,,8$  \\  \hline
 $\bar i\,,\bar j\,, \bar k$   &  ${\rm SU}(4)_W$   &  fundamental  & $1\,,2\,,3\,,4$   \\
   &      & anti-fundamental  &   \\
  $\bar I\,,\bar J\,, \bar K$   &     ${\rm SU}(4)_W$  & adjoint  & $1\,,...\,,15$  \\ \hline
 $\tilde i \,, \tilde j \,, \tilde k $   &  ${\rm SU}(3)_W$   &  fundamental  & $1\,,2\,,3$   \\
   &      & anti-fundamental  &   \\
  $ \tilde I \,, \tilde J \,, \tilde K $   &     ${\rm SU}(3)_W$  & adjoint  & $1\,,...\,,8$  \\ \hline
 $ i\,, j$   &  ${\rm SU}(2)_W$   &  fundamental  & $1\,,2$   \\
   &      & anti-fundamental  &   \\
  $ I\,, J\,, K$   &     ${\rm SU}(2)_W$  & adjoint  & $1\,,2\,,3$  \\ 
\hline\hline
\end{tabular}
\end{center}
\caption{
Definition of indices for various gauge groups from the maximal ${\rm SU}(8)$ symmetry breaking pattern.
The fundamental and anti-fundamental indices will be distinguished by superscripts and subscripts, respectively.
}
\label{tab:notations}
\end{table}%

%
\para
According to the symmetry breaking pattern in Eq.~\eqref{eq:Pattern}, we define the conventions of gauge group indices in Tab.~\ref{tab:notations}.
The fundamental and anti-fundamental representations will be denoted by superscripts and subscripts, respectively.
We also define the decomposition rules in the ${\rm SU}(8)$ theory.
At the zeroth GUT-scale symmetry breaking, we define the ${\rm U}(1)_{X_0}$ charges for the ${\rm SU}(8)$ fundamental representation as follows
\beqn\label{eq:X0charge}
\hat X_0( \rep{8} ) &\equiv& {\rm diag} ( \underbrace{ - \frac{1}{4}  \mathbb{I}_{4 \times 4}  }_{ \rep{4_s} }\,, \underbrace{ +\frac{1}{4} \mathbb{I}_{4 \times 4}   }_{ \rep{4_W} } )\,.
\eeqn
%
%
%
%
%
%
Sequentially, the ${\rm U}(1)_{X_1}$, ${\rm U}(1)_{X_2}$, and ${\rm U}(1)_{Y}$ charges are defined according to the ${\rm SU}(4)_s$ and the ${\rm SU}(4)_W$ fundamental representations as follows
\beqs\label{eq:U1charges_fund}
\beqn
{\rm first ~stage}~&:&~\hat X_1(\rep{4_s}) \equiv  {\rm diag} \, \Big( \underbrace{  (- \frac{1}{12}+ \Xc_0 ) \mathbb{I}_{3\times 3} }_{ \rep{3_c} } \,, \frac{1}{4}+ \Xc_0 \Big) \,,\label{eq:X1charge_4sfund}\\[1mm]
{\rm second ~stage}~&:&~ \hat X_2 ( \rep{4_W} ) \equiv {\rm diag} \, \Big( \underbrace{  ( \frac{1}{12} + \Xc_1 ) \mathbb{I}_{3\times 3} }_{ \rep{3_W} } \,, -\frac{1}{4} + \Xc_1 \Big)  \,, \label{eq:X2charge_4Wfund} \\[1mm]
{\rm third ~stage}~&:&~ \hat Y ( \rep{4_W} ) \equiv  {\rm diag} \, \Big(  ( \frac{1}{6}+ \Xc_2 ) \mathbb{I}_{2\times 2} \,,- \frac{1}{3}+ \Xc_2 \,, \Xc_2 \Big) \non
&=& {\rm diag} \, \Big(  \underbrace{ ( \frac{1}{4} + \Xc_1 ) \mathbb{I}_{2\times 2} }_{ \rep{2_W} } \,,  ( - \frac{1}{4} + \Xc_1  ) \mathbb{I}_{2\times 2} \Big) \,, \label{eq:Ycharge_4Wfund} \\[1mm]
{\rm fourth ~stage}~&:&~ \hat Q_e ( \rep{4_W} ) \equiv T_{ {\rm SU}(4) }^3 +  \hat Y ( \rep{4_W} ) = {\rm diag} \, \Big( \frac{3}{4} + \Xc_1  \,, ( - \frac{1}{4} + \Xc_1  ) \mathbb{I}_{3\times 3}  \Big) \,. \label{eq:Qcharge_4Wfund}
\eeqn
\eeqs
Based on the above definitions for the fundamental representations, the rules for other higher rank anti-symmetric representations can be derived by tensor productions.
For adjoint representations of the ${\rm SU}(4)_s$ and the ${\rm SU}(4)_W$ groups, these charges are defined by setting all relevant ${\rm U}(1)$ charges to zero in Eqs.~\eqref{eq:U1charges_fund}
\beqs\label{eq:U1charges_adj}
\beqn
\hat X_1(\rep{15_s}) &\equiv&  {\rm diag} ( - \frac{1}{12} \mathbb{I}_{3\times 3} \,,  \frac{1}{4} ) \,,\label{eq:X1charge_4sadj}\\[1mm]
\hat X_2 ( \rep{15_W} )&\equiv& {\rm diag} ( \frac{1}{12} \mathbb{I}_{3\times 3} \,, -\frac{1}{4}   )  \,, \label{eq:X2charge_4Wadj} \\[1mm]
\hat Y ( \rep{15_W} )&\equiv&  {\rm diag}(  \frac{1}{4} \mathbb{I}_{2\times 2}  \,,- \frac{1}{4} \mathbb{I}_{2\times 2} )  \,, \label{eq:Ycharge_4Wadj} \\[1mm]
\hat Q_e ( \rep{15_W} )&\equiv&  {\rm diag} \, \Big( \frac{3}{4}   \,, - \frac{1}{4}   \mathbb{I}_{3\times 3}  \Big) \,. \label{eq:Qcharge_4Wadj} 
\eeqn
\eeqs

\section{Note added in proof of the ${\rm SU}(8)$ maximally breaking pattern}
\label{section:proof}

\para
In the maximally symmetry breaking pattern of the Georgi-Glashow theory
\beqn
&& {\rm SU}(5) \to  {\rm SU}(3)_c  \otimes  {\rm SU}(2)_W \otimes  {\rm U}(1)_Y  \,,
\eeqn
the gauge coupling of the ${\rm U}(1)_Y$ is normalized as 
\beqn\label{eq:SU5_U1norm}
&& g_Y {\rm diag} (- \frac{1}{3} \mathbb{I}_{3\times 3}  \,, + \frac{1}{2} \mathbb{I}_{ 2 \times 2 }  ) = g_1 \frac{1 }{ \sqrt{15 } } { \rm diag} ( - \mathbb{I}_{3\times 3} \,, +\frac{3}{2} \mathbb{I}_{ 2 \times 2 } )  \Rightarrow  g_Y^2 = \frac{3}{5} g_1^2 \,.
\eeqn
Let us generalize it to the maximally symmetry breaking pattern of a generic ${\rm SU}(N)$ theory as follows
\beqn
&& {\rm SU}(N) \to {\rm SU}(n_S ) \otimes {\rm SU}(n_W ) \otimes {\rm U}(1) \,,
\eeqn
with $n_S = n_W =N/2$ for the even-$N$, or $n_{S\,,W}=(N \pm 1)/2$ for the odd-$N$.
The ${\rm U}(1)$ charges of the ${\rm SU}(N)$ fundamental representation after the symmetry breaking are given by
\beqn 
&& {\rm diag} \, ( \underbrace{ - \frac{1}{n_S }  \,, ... }_{ \times n_S }\,, \underbrace{ + \frac{1}{n_W } \,, ... }_{ \times n_W } )\,.
\eeqn
Meanwhile, the VEV of the adjoint Higgs for the symmetry-breaking pattern is given by
\beqn
\langle \rep{\Phi}_{\rm Adj} \rangle &=& \frac{1}{ \sqrt{ 2 n_S n_W ( n_S + n_W ) } } {\rm diag}\,( \underbrace{ - n_W\,, ... }_{ \times n_S } \,, \underbrace{ + n_S \,, ... }_{ \times n_W } ) v_U \,,
\eeqn
with the normalization of $\Tr\rep{\Phi}_{\rm Adj}^2 = \hf v_U^2$.
If the ${\rm U}(1)$ coupling does not need to be normalized, one has the following relations
\beqn
&& \frac{1}{n_S } = \frac{ n_W }{ \sqrt{ 2 n_S n_W ( n_S + n_W ) }  } \Rightarrow n_S n_W = 2 ( n_S + n_W ) = 2N \,, \non
&&{\rm even}-N~:~n_S = n_W=N/2 \Rightarrow N=8 \,, \non
&&{\rm odd}-N~:~n_S = (N+1)/2\,, n_W =(N-1)/2 \Rightarrow  N^2 -1 = 8N \,.
\eeqn
Apparently, the condition for the odd-$N$ case has no integer solution. 
Thus, the unique gauge group with its unique maximally symmetry breaking pattern where the ${\rm U}(1)$ charge does not need additional normalization factor as in Eq.~\eqref{eq:SU5_U1norm} is the ${\rm SU}(8) \to \Gc_{441}$.
If the maximally symmetry breaking pattern was relaxed, the breaking of ${\rm SU}(8) \to \Gc_{441}$ is not the unique one where the ${\rm U}(1)$ charge does not need additional normalization.
One such example is the breaking of ${\rm SU}(9) \to \Gc_{631}/ \Gc_{361}$, with the ${\rm U}(1)$ charges of the ${\rm SU}(9)$ fundamental representation given by
\beqn
&& {\rm diag} \, ( - \frac{1}{6 } \mathbb{I}_{6\times 6} \,,  + \frac{1}{3 } \mathbb{I}_{ 3\times 3 } ) \,, ~~ {\rm or}~~ {\rm diag} \, ( - \frac{1}{3 } \mathbb{I}_{3 \times 3 } \,,  + \frac{1}{ 6 } \mathbb{I}_{ 6 \times 6 } ) \,.
\eeqn


\providecommand{\href}[2]{#2}\begingroup\raggedright\endgroup

\end{document}